\documentclass[amsmath,amssymb,reprint,aps,prd,eqsecnum]{revtex4-2}

\usepackage{graphicx}
\usepackage{dcolumn}
\usepackage{bm}
\usepackage{mathrsfs}
\usepackage{hyperref}
\usepackage[dvipsnames]{xcolor}
\usepackage{comment}

\allowdisplaybreaks

\begin{document}

\title{Renormalized charged scalar current on a Reissner-Nordstr\"om black hole \\ in the presence of charge superradiance}

\author{George Montagnon}

\email{GJMontagnon1@sheffield.ac.uk}

\affiliation{School of Mathematical and Physical Sciences,
The University of Sheffield,
Hicks Building,
Hounsfield Road,
Sheffield. S3 7RH United Kingdom}

\author{Elizabeth Winstanley}

\email{E.Winstanley@sheffield.ac.uk}

\affiliation{School of Mathematical and Physical Sciences,
The University of Sheffield,
Hicks Building,
Hounsfield Road,
Sheffield. S3 7RH United Kingdom}

\date{\today}

\begin{abstract}
We compute the renormalized charge current for a massless, minimally coupled, charged quantum scalar field on a charged
Reissner-Nordstr\"om black hole space-time, using the method of pragmatic mode-sum renormalization.
Since the field exhibits superradiance, we consider the past Unruh, Boulware and Candelas-Chrzanowski-Howard states,
and study how the renormalized current in these states depends on the charges of the black hole and scalar field.
We also study the backreaction of the charge current on the electromagnetic field.
\end{abstract}

\maketitle

\section{Introduction}
\label{sec:intro}

Isolated, nonextremal black holes formed by gravitational collapse emit Hawking radiation \cite{Hawking:1974rv,Hawking:1975vcx}, a thermal flux of quantum particles.
If the black hole is static and uncharged (or we consider just the emission of neutral particles from a charged black hole), the Hawking radiation switches off in the extremal limit as the black hole temperature goes to zero.
This is not the case for a rotating black hole: in addition to the Hawking radiation, the black hole also emits Unruh-Starobinskii radiation \cite{Starobinsky:1973aij,Unruh:1974bw} which persists even if the temperature of the black hole vanishes.
The Unruh-Starobinskii radiation is the quantum analogue of the classical phenomenon of superradiance \cite{Brito:2015oca}, whereby low-frequency modes of a bosonic field are amplified on scattering by the black hole.
Even in the absence of Hawking radiation, a rotating black hole spontaneously emits particles in precisely those field modes which exhibit classical superradiance.

Similar classical superradiance phenomena occur for charged bosonic fields interacting with a static, charged black hole \cite{Brito:2015oca,Bekenstein:1973mi,Benone:2015bst,DiMenza:2014vpa}. 
While charged fermionic fields do not experience classical superradiance \cite{Maeda:1976tm}, the corresponding quantum process occurs for both bosonic \cite{Gibbons:1975kk,Balakumar:2020gli} and fermionic \cite{Alvarez-Dominguez:2024ahv} charged fields. 

A classical static, electrically charged black hole has, in addition to a background space-time metric, a nonzero background electrostatic field with which the quantum field interacts.  
The expectation value of the charge current operator $\langle {\hat {J}}^{\mu } \rangle $ of the charged quantum field will, in turn, act as a source in the semiclassical Maxwell equations (we use Gaussian units):
\begin{equation}
    \nabla _{\mu }F^{\mu \nu } = 4\pi \langle {\hat {J}}^{\nu } \rangle ,
    \label{eq:SCmaxwell}
\end{equation}
where $F^{\mu \nu }$ is the electromagnetic field strength tensor, and thus (\ref{eq:SCmaxwell}) governs the backreaction of a charged quantum field on the electromagnetic field.

In this paper we compute the renormalized charge current for a massless, charged, quantum scalar field propagating on an electrically charged, nonextremal Reissner-Nordstr\"om (RN) black hole.
We treat both the space-time metric and the corresponding electrostatic field as fixed, classical, background quantities. 
Only the charged scalar field is quantized.  

The first step in finding the renormalized current is to define a suitable quantum state for the scalar field.
This is complicated by the presence of classically superradiant scalar field modes \cite{Balakumar:2022yvx}, and a plethora of putative quantum states can be defined. 
In this paper we focus on three (uncontroversial) states. 
All three states are ``past'' states in the nomenclature of \cite{Balakumar:2022yvx}, in that they are defined by considering the charged scalar field modes in the distant past. 
Specifically, we consider the following states \cite{Balakumar:2022yvx}:
\begin{description}
    \item[Boulware state $|{\mathrm{B}}\rangle $] 
    This state is as empty as possible far from the black hole in the distant past; in the distant future, far from the black hole, there is an outgoing flux of particles in the classically superradiant modes only.
    \item[Unruh state $|{\mathrm{U}}\rangle $] 
    This state is also as empty as possible far from the black hole in the distant past; however, in the distant future it contains the outgoing thermal Hawking radiation flux (in addition to the superradiant particle flux). 
    \item[Candelas-Chrzanowski-Howard state $|{\mathrm{CCH}}\rangle $] 
    This state is not empty far from the black hole in the distant past; it contains an ingoing flux of thermally-distributed particles. In the distant future, far from the black hole, there is also an outgoing thermal flux of particles, as well as the superradiant emission.
    \end{description}
All three states also have analogues for a neutral quantum scalar field (respectively, the Boulware \cite{Boulware:1974dm}, Unruh \cite{Unruh:1976db} and Hartle-Hawking \cite{Hartle:1976tp} states),
which have been comprehensively studied and whose properties for neutral scalar fields are well-understood.
For a neutral scalar field, the Boulware state \cite{Boulware:1974dm} corresponds to a vacuum state for a static observer far from the black hole, in both the distant past and distant future, whilst, for a charged scalar field on a charged black hole, the state $|{\mathrm{B}}\rangle $ is no longer a vacuum state in the far future \cite{Gibbons:1975kk,Balakumar:2020gli}.
In contrast, the properties of the Unruh state $|{\mathrm{U}}\rangle $ are very similar for both neutral and charged fields on a charged black hole; both contain the thermal Hawking radiation.
In the limit in which the charged black hole becomes extremal, the Hawking radiation disappears, and the Unruh state $|{\mathrm{U}}\rangle $ becomes the Boulware state $|{\mathrm{B}}\rangle $.
The Hartle-Hawking state \cite{Hartle:1976tp} for a neutral scalar field is not a vacuum state far from the black hole; it describes a thermal state at the Hawking temperature of the black hole.
Due to charge superradiance, it does not seem to be possible, at least within a conventional quantization scheme, to define a corresponding thermal equilibrium state for a charged scalar field on a charged black hole space-time (the ``Hartle-Hawking''-like state defined in \cite{Balakumar:2022yvx} relies on a nonstandard treatment of the superradiant modes). 
While the Candelas-Chrzanowski-Howard (CCH) state $|{\mathrm{CCH}}\rangle $ contains both ingoing and outgoing thermal radiation, this radiation is not in thermal equilibrium \cite{Balakumar:2022yvx}, due to the different thermal factors for the ingoing and outgoing scalar field modes. 

Having selected suitable quantum states, we can proceed to the computation of the renormalized expectation value of the charged scalar current operator ${\hat {J}}^{\mu }$.
Formally, the divergences which arise due to ${\hat {J}}^{\mu }$ involving products of operators at the same space-time point can be regularized by point-splitting \cite{Christensen:1976vb,Christensen:1978yd}, which involves applying an appropriate differential operator to the point-split Green function. 
The divergences which then arise in the limit in which the points are brought together are removed by subtracting from the Green function a suitable singular parametrix, then applying the differential operator, and finally the coincidence limit can be taken, resulting in a finite expectation value.
In this work we use the Hadamard parametrix \cite{Fulling:1978ht,Balakumar:2019djw} to regularize the Green function.
This parametrix is independent of the quantum state under consideration, and depends only on the details of the space-time geometry and background electromagnetic field.
For a general background and two points lying in a normal neighbourhood (so that they are connected by a unique geodesic), the Hadamard parametrix can be written as a covariant series expansion in the separation of the points; the coefficients in this expansion depending on the metric, curvature tensors, electrostatic potential and derivatives of these quantities \cite{Balakumar:2019djw}. 
As an alternative, one can consider the DeWitt-Schwinger expansion of the Green function \cite{Herman:1995hm}, which has, by necessity, the same singular terms as the Hadamard parametrix, but contains different finite terms \cite{Pla:2022spt}. 

The remaining technical challenge is then to subtract the chosen parametrix from the point-split Green function, which, for the quantum states of interest, is given as an infinite sum over scalar field modes, each of which can only be computed numerically.
Practical implementations, which enable the resulting renormalized expectation values to be computed efficiently, have been developed recently for a charged scalar field (see \cite{Herman:1995hm,Herman:1998dz} for early work in this direction). 
In the literature to date, there are two main approaches to the computation of the renormalized current for a charged scalar field on a charged black hole.
First, one can perform a Wick rotation of the static black hole metric and work on a Euclideanized space-time.
This has the advantage that there are no null geodesics, and therefore the singularities in the Green function arise only in the coincidence limit.
In contrast, the Green function on the original Lorentzian space-time is singular also when the points are connected by a null geodesic. 
The Euclidean approach naturally applies to a quantum field in the thermal equilibrium Hartle-Hawking state, due to the periodicity of such a state in imaginary time.
Recently, the ``extended coordinates'' methodology developed in \cite{Taylor:2016edd,Taylor:2017sux,Taylor:2022sly,Arrechea:2023fas} for a neutral scalar field has been generalized to the case of a massive charged quantum scalar field on a charged black hole background \cite{Breen:2024ggu}.
If the scalar field is sufficiently massive that classical superradiance is absent \cite{DiMenza:2014vpa}, the Hartle-Hawking state can be defined, and in \cite{Breen:2024ggu}, the renormalized expectation value of the charge current (and also the stress-energy tensor) is computed in this state. 
Since differences in expectation values between two states do not require renormalization (as the Hadamard parametrix is state-independent), it is comparatively straightforward to then find the renormalized current in other quantum states (see also \cite{Arrechea:2024cnv} for a direct computation for a neutral scalar field in the Boulware state using the extended coordinates method). 

The second possibility is to work on the original Lorentzian space-time, which avoids the requirement to perform a Wick rotation and circumvents potential difficulties in situations where the thermal equilibrium Hartle-Hawking state does not exist. 
The ``pragmatic mode-sum'' approach \cite{Levi:2015eea,Levi:2016esr,Levi:2016quh,Levi:2016paz,Levi:2016exv}, developed for a neutral scalar field, has enabled the renormalized stress-energy tensor to be computed on a variety of black hole backgrounds, notably including the rotating Kerr black hole \cite{Levi:2016exv}.
In this paper we extend the ``pragmatic mode-sum'' methodology to the computation of the renormalized current for a massless charged scalar field on a charged RN black hole background. 
Previous computations of the renormalized charged scalar current \cite{Klein:2021ctt,Klein:2021les} have considered a Reissner-Nordstr\"om-de Sitter black hole, with the renormalized current on the outer and inner horizons of an RN black hole considered only very recently \cite{Alberti:2025mpg}.
Central to the method employed in \cite{Klein:2021ctt,Klein:2021les} is the choice of gauge for the electrostatic potential, as a result of which only finite renormalization terms are required to find the charged scalar current.
In the approach of \cite{Klein:2021ctt,Klein:2021les}, a different choice of gauge is required at each space-time point exterior to the event horizon.
In this work, we use the same gauge throughout space-time, and therefore have to perform a renormalization procedure 
involving singular renormalization terms.

The outline of the paper is as follows.
In Sec.~\ref{sec:chargedScalar} we review the salient features of a classical charged scalar field on a nonextremal RN black hole, before outlining the canonical quantization of the scalar field and defining the three quantum states of interest. 
In Sec.~\ref{sec:expvals} we extend the $t$-splitting variant of the ``pragmatic mode-sum'' methodology \cite{Levi:2015eea} to find the renormalized charged scalar current (and the square of the scalar field operator).
Our numerical results are presented in Sec.~\ref{sec:results}, and we study the backreaction of the charged quantum scalar field on the electromagnetic field by solving the backreaction equations (\ref{eq:SCmaxwell}) in Sec.~\ref{sec:backreaction}. 
Finally, our conclusions can be found in Sec.~\ref{sec:conc}.

\section{Charged quantum scalar field \\ on an RN black hole}
\label{sec:chargedScalar}

In this paper we are concerned with the behaviour of a massless, minimally coupled, charged quantum scalar field on the subextremal RN space-time, as described by the line element (in the usual spherical polar coordinates)
\begin{equation}
    ds^{2}=-f(r) \, dt^{2}+\frac{1}{f(r)}dr^{2}+r^{2} \, d\Omega^{2},
    \label{eq:metric}
\end{equation}
where $d\Omega^{2}$ denotes the line element on $\mathbb{S}^{2}$
\begin{equation}
    d\Omega ^{2} = d\theta ^{2} + \sin ^{2} \theta \, d\varphi ^{2},
\end{equation}
and the metric function $f(r)$ takes the form
\begin{equation}
    f(r)=1-\frac{2M}{r}+\frac{Q^{2}}{r^{2}}. 
    \label{eq:frRN}
\end{equation}
The parameters $M$ and $Q$ are the black hole mass and charge respectively, for which we have $Q<M$ in the subextremal case. 
The zeros of the metric function $f(r)$ are
\begin{equation}
    r_{\pm}=M\pm\sqrt{M^{2}-Q^{2}},
    \label{eq:rpm}
\end{equation}
and correspond to the event ($r_{+}$) and inner ($r_{-}$) horizons. 
Here, and throughout this paper, the metric has signature $(-,+,+,+)$ and we use units in which $G = \hbar = c = k_{\rm {B}}=1$.

The coordinates in which the line element (\ref{eq:metric}) is expressed cover the exterior region of the black hole ($r>r_{+}$), however we will later require a coordinate system that is regular across the event horizon. 
For this purpose, we introduce ingoing and outgoing Eddington-Finkelstein coordinates
\begin{equation}
    v=t+r_{*},\qquad u=t-r_{*},
    \label{eq:EF}
\end{equation}
where $r_{*}$ is the usual tortoise coordinate defined by
\begin{equation}
    \frac{dr_{*}}{dr}=\frac{1}{f(r)}.
    \label{eq:tortoise}
\end{equation}
The coordinate systems $(v,r,\theta,\varphi)$ and $(u,r,\theta,\varphi)$ are regular across the future and past event horizons, respectively.
We then define the coordinates 
\begin{equation}
    U=-\frac{1}{\kappa}e^{-\kappa u},\qquad  V=\frac{1}{\kappa}e^{\kappa v},
    \label{eq:Kruskal}
\end{equation}
where
\begin{equation}
    \kappa=\frac{1}{2}f'(r_{+})
    \label{eq:kappa}
\end{equation}
is the surface gravity of the event horizon. 
The coordinates $(U,V,\theta, \varphi )$ are regular across both the future ${\mathcal {H}}^{+}$ and past ${\mathcal {H}}^{-}$ event horizons.

\begin{figure}
    \centering
    \includegraphics[scale=0.65]{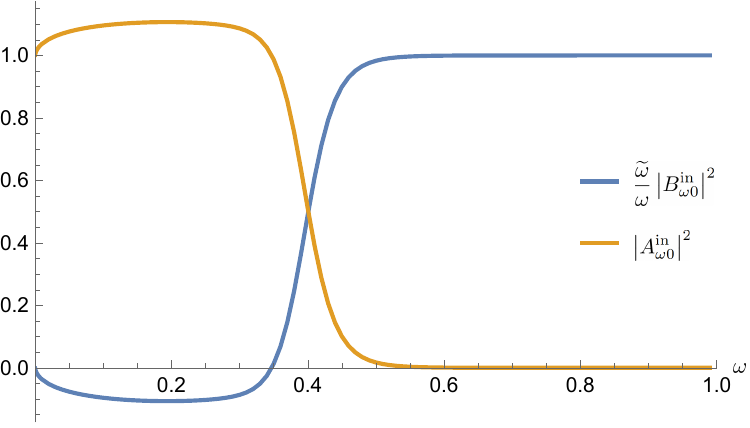}
    \caption{Reflection $|A^{{\mathrm {in}}}_{\omega\ell}|^{2}$ and transmission $({\widetilde {\omega }}/\omega )|B^{{\mathrm {in}}}_{\omega\ell}|^{2}$ coefficients for an ``in'' charged scalar field mode with $\ell = 0$. The scalar field has charge $qM=0.4$ and the black hole parameters are $Q/M=0.99$.
    The reflection coefficient is greater than unity for small positive frequency $\omega $ satisfying (\ref{eq:SR}), and in this regime classical charge superradiance occurs.}
    \label{fig:classicalSup}
\end{figure}

The massless, minimally coupled, charged scalar field satisfies the equation of motion
\begin{equation}
    D_{\mu }D^{\mu }\Phi=0,
    \label{eq:fieldeqn}
\end{equation}
where $D_{\mu}=\nabla_{\mu}-iqA_{\mu}$ is the gauge covariant derivative, $q$ is the field charge and $A_{\mu}$ is the gauge field, which, via a choice of gauge,  we  take to have the form
\begin{equation}
    A=-\frac{Q}{r}dt.
    \label{eqn:gaugeField}
\end{equation}
The scalar field equation (\ref{eq:fieldeqn}) admits an orthonormal basis of ``in" and ``up" mode solutions of the form \cite{Balakumar:2022yvx}
\begin{subequations}
\label{eq:modes}
    \begin{align}
    \phi^{{\mathrm {in}}}_{\omega\ell m}(x) & =  ~ 
    \frac{e^{-i\omega t}}{r{\sqrt {4\pi |\omega| }}} X^{\rm {in}}_{\omega \ell }(r) Y_{\ell m}(\theta , \varphi ),
\\
    \phi^{{\mathrm {up}}}_{\omega\ell m}(x)  & = ~ 
    \frac{e^{-i\omega t}}{r{\sqrt {4\pi \left|{\widetilde {\omega }}\right| }}} X^{\rm {up}}_{\omega \ell }(r) Y_{\ell m}(\theta , \varphi ),
\end{align}
\end{subequations}
where 
\begin{equation}
    {\widetilde {\omega }} = \omega - \frac{qQ}{r_{+}},
    \label{eq:wtilde}
\end{equation}
and $Y_{\ell m}(\theta, \varphi )$ is a spherical harmonic.
The quantum number $\ell =0,1,2\ldots $ and $m=-\ell, -\ell+ 1, \ldots, \ell - 1, \ell$.
The radial functions $X^{{\mathrm {in/up}}}_{\omega\ell}(r)$ satisfy the one-dimensional scattering equation
\begin{equation}
    \left[ \frac{d^{2}}{dr_{*}^{2}}+V(r)\right] X^{{\rm {in/up}}}_{\omega\ell}(r)=0,
    \label{eq:radial}
\end{equation}
where the effective potential is given by
\begin{equation}
     V(r)=\left(\omega-\frac{qQ}{r}\right)^{2}-\frac{f(r)}{r^{2}}\left[ \ell(\ell+1)+rf'(r)\right] .
     \label{eq:potential}
\end{equation}
In the asymptotic regions close to the event horizon ($r\rightarrow r_{+}$, $r_{*}\rightarrow - \infty $) and at infinity ($r\rightarrow \infty $, $r_{*} \rightarrow \infty $), the radial functions take the form
\begin{subequations}
\label{eq:inupmodes}
\begin{align}
    X^{\rm {in}}_{\omega \ell }(r) = & ~
    \begin{cases}
        B^{\rm {in}}_{\omega \ell } e^{-i\widetilde{\omega} r_{*}}, & r_{*}\rightarrow - \infty ,
        \\
        e^{-i\omega r_{*}} + A^{\rm {in}}_{\omega \ell }e^{i\omega r_{*}},
        & r_{*}\rightarrow \infty ,
    \end{cases}
\\
    X^{\rm {up}}_{\omega \ell }(r) = & ~
    \begin{cases}
        e^{i\widetilde{\omega}r_{*}} + A^{\rm {up}}_{\omega \ell }e^{-i\widetilde{\omega} r_{*}},
        & r_{*}\rightarrow -\infty ,
        \\
        B^{\rm {up}}_{\omega \ell } e^{i\omega r_{*}}, & r_{*}\rightarrow  \infty.
    \end{cases}
\end{align}
\end{subequations}
The complex coefficients $A_{\omega\ell}^{{\mathrm {in/up}}}$, $B_{\omega\ell}^{{\mathrm {in/up}}}$  satisfy the relations \cite{Balakumar:2022yvx}
\begin{subequations}
 \label{eqn:coeffsRels}
\begin{align}
    {\widetilde {\omega }}\left|B^{{\mathrm {in}}}_{\omega\ell}\right|^{2} = & ~
    \omega\left(1-\left|A^{{\mathrm {in}}}_{\omega\ell}\right|^{2}\right),
    \\
     \omega\left|B^{{\mathrm{up}}}_{\omega\ell}\right|^{2} = & ~ 
     {\widetilde {\omega }}\left(1-\left|A^{{\mathrm {up}}}_{\omega\ell}\right|^{2}\right),
     \\
    {\widetilde {\omega }}B^{{\mathrm {in}}}_{\omega\ell}= & ~ \omega B^{{\mathrm {up}}}_{\omega\ell} ,
    \\
    {\widetilde {\omega }}A^{{\mathrm {up}}*}_{\omega\ell}B^{{\mathrm {in}}}_{\omega\ell}= & ~
    -\omega A^{{\mathrm {in}}}_{\omega\ell}B^{{\mathrm {up}}*}_{\omega\ell},
\end{align}
\end{subequations}
from which we observe that for values of $\omega$ for which
\begin{equation}
    \omega{\widetilde {\omega }}<0,
    \label{eq:SR}
\end{equation}
we have $|A^{{\mathrm {in/up}}}_{\omega\ell}|>1$. 
This is the classical phenomenon of charge superradiance: a charged scalar field wave is scattered off the potential (\ref{eq:potential}) in such a way that the reflection coefficient is greater than unity, in other words the reflected part of the wave has a larger amplitude than the incident part of the wave, see 
Fig.~\ref{fig:classicalSup} for an example illustrating this effect.

The canonical quantization of a massless charged scalar field, and the corresponding construction of states, was investigated in detail in \cite{Balakumar:2022yvx}. 
Here we consider three of the states constructed in \cite{Balakumar:2022yvx}, namely the ``past'' Boulware $|{\mathrm{B}}\rangle $, Unruh $|{\mathrm{U}}\rangle $ and CCH $|{\mathrm{CCH}}\rangle $ states.
Hereafter, we drop the ``past'' prefix in discussing the states.
The full construction of these states can be found in \cite{Balakumar:2022yvx}; here we briefly review the key aspects.

To construct the Boulware state $|{\mathrm{B}}\rangle $, the quantum charged scalar field operator ${\hat {\Phi}}$ is expanded in terms of the basis modes (\ref{eq:modes}) as
\begin{multline}
    \hat{\Phi}(x)
    = \sum_{\ell=0}^{\infty}\sum_{m=-\ell}^{\ell}\Bigg\{\int_{0}^{\infty}d\omega \,
    \hat{a}^{{\mathrm {in}}}_{\omega\ell m}\phi^{{\mathrm {in}}}_{\omega\ell m}(x)
    \\ + \int _{-\infty }^{0} d\omega  \,
    \hat{b}^{{\mathrm {in}}\dagger}_{\omega\ell m}\phi^{{\mathrm {in}}}_{\omega\ell m}(x)
    +\int_{0}^{\infty}d{\widetilde {\omega}} \, \hat{a}^{{\mathrm {up}}}_{\omega\ell m}\phi^{{\mathrm {up}}}_{\omega\ell m}(x)
    \\ +\int_{-\infty }^{0}d{\widetilde {\omega}} \, \hat{b}^{{\mathrm {up}}\dagger}_{\omega\ell m}\phi^{{\mathrm {up}}}_{\omega\ell m}(x)
    \Bigg\},
    \label{eq:Bexpansion}
\end{multline}
where the operators satisfy the standard commutation relations (all commutation relations not given explicitly below vanish identically)
\begin{align}
    \left[ {\hat{a}}^{{\mathrm {in}}}_{\omega\ell m}, {\hat{a}}^{{\mathrm {in}}\dagger}_{\omega'\ell' m'} \right]
    & = \delta_{\ell\ell'}\delta_{mm'}\delta(\omega-\omega'), \qquad \omega>0  ,
    \nonumber \\
    \left[ {\hat{b}}^{{\mathrm {in}}}_{\omega\ell m}, {\hat{b}}^{{\mathrm {in}}\dagger}_{\omega'\ell' m'} \right]
     & = \delta_{\ell\ell'} \delta_{mm'} \delta(\omega-\omega'), \qquad \omega<0,
    \nonumber \\
    \left[ {\hat{a}}^{{\mathrm {up}}}_{\omega\ell m}, {\hat{a}}^{{\mathrm {up}}\dagger}_{\omega'\ell' m'} \right] & = \delta_{\ell\ell'}\delta_{mm'}\delta(\omega-\omega'), \qquad {\widetilde {\omega}}  > 0,
    \nonumber \\
    \left[ {\hat{b}}^{{\mathrm {up}}}_{\omega\ell m}, {\hat{b}}^{{\mathrm {up}}\dagger}_{\omega'\ell' m'}\right]
    & = \delta_{\ell\ell'}\delta_{mm'}\delta(\omega-\omega'), \qquad {\widetilde {\omega}} <0 .
\end{align}
The Boulware state $|{\mathrm{B}}\rangle $ is annihilated by all the annihilation operators:
\begin{align}
    {\hat{a}}^{{\mathrm {in}}}_{\omega\ell m}|{\mathrm{B}}\rangle  &  = 0, \qquad \omega >0,
    \nonumber \\
    {\hat{b}}^{{\mathrm {in}}}_{\omega\ell m}|{\mathrm{B}}\rangle & =0, \qquad \omega <0 ,
    \nonumber\\
    {\hat{a}}^{{\mathrm {up}}}_{\omega\ell m}|{\mathrm{B}}\rangle & =0, \qquad {\widetilde {\omega}}>0 ,
    \nonumber\\
   {\hat{b}}^{{\mathrm {up}}}_{\omega\ell m}|{\mathrm{B}}\rangle  & =0,  \qquad {\widetilde {\omega}}<0. 
\end{align}
We interpret the Boulware state $|{\mathrm{B}}\rangle $ as being devoid of particles in either the ``in'' or ``up'' modes in the distant past.
However, this state is not empty at future null infinity; it contains an outgoing flux of ``up'' mode particles in the superradiant regime \cite{Balakumar:2020gli}.

To construct the Unruh state $|{\mathrm{U}}\rangle $, the quantum scalar field is expanded in an orthonormal basis of field modes consisting of the same ``in'' modes as for the Boulware state, but an alternative set of ``up'' modes.
These are defined to have positive/negative frequency with respect to the Kruskal coordinate $U$ (\ref{eq:Kruskal}) in the vicinity of the past event horizon (see \cite{Balakumar:2022yvx} for details).
The resulting field expansion is, in the region exterior to the event horizon, 
\begin{multline}
    {\hat{\Phi}}(x)= 
    \sum_{\ell=0}^{\infty}\sum_{m=-\ell}^{\ell}\Bigg\{ 
    \int_{0}^{\infty}d\omega \, {\hat{c}}^{{\mathrm {in}}}_{\omega\ell m}\phi^{{\mathrm {in}}}_{\omega\ell m}(x)
    \\+ \int _{-\infty }^{0} d\omega \, {\hat{d}}^{{\mathrm {in}}\dagger}_{\omega\ell m}\phi^{{\mathrm {in}}}_{\omega\ell m}(x)
    \\
    + \int _{-\infty }^{\infty } d{\widetilde {\omega }} \, 
    \frac{\phi^{{\mathrm {up}}}_{\omega\ell m }(x)}{\sqrt{2\left|\sinh\left(\frac{\pi {\widetilde {\omega}}}{\kappa }\right) \right| }} 
    \left[ e^{\frac{\pi {\widetilde {\omega }}}{2\kappa}} {\hat{c}}^{{\mathrm {up}}}_{\omega\ell m}
    +e^{-\frac{\pi {\widetilde {\omega }}}{2\kappa}}{\hat{d}}^{{\mathrm {up}}\dagger}_{\omega\ell m}\right]
    \Bigg\} .
    \label{eq:Uexpansion}
\end{multline}
The nonzero commutators between the operators appearing in the expansion (\ref{eq:Uexpansion}) are
\begin{align}
    \left[ {\hat{c}}^{{\mathrm {in}}}_{\omega\ell m}, {\hat{c}}^{{\mathrm {in}}\dagger}_{\omega'\ell' m'}\right] 
    & = \delta_{\ell\ell'}\delta_{mm'}\delta(\omega-\omega'), \qquad \omega>0 ,
    \nonumber \\
    \left[ {\hat{d}}^{{\mathrm {in}}}_{\omega\ell m}, {\hat{d}}^{{\mathrm {in}}\dagger}_{\omega'\ell' m'}\right] 
    & = \delta_{\ell\ell'}\delta_{mm'}\delta(\omega-\omega'), \qquad \omega <0 ,
    \nonumber\\
    \left[ {\hat{c}}^{{\mathrm {up}}}_{\omega\ell m}, {\hat{c}}^{{\mathrm {up}}\dagger}_{\omega'\ell' m'}\right]
    & = \delta_{\ell\ell'}\delta_{mm'}\delta(\omega-\omega'), \qquad \forall \, {\widetilde {\omega }},
    \nonumber \\
    \left[ {\hat{d}}^{{\mathrm {up}}}_{\omega\ell m}, {\hat{d}}^{{\mathrm {up}}\dagger}_{\omega'\ell' m'}\right]
    & = \delta_{\ell\ell'}\delta_{mm'}\delta(\omega-\omega'), \qquad \forall \, {\widetilde {\omega }}, 
\end{align}
and the Unruh state $|{\mathrm{U}}\rangle $ is annihilated by all the annihilation operators:
\begin{align}
    {\hat{c}}^{{\mathrm {in}}}_{\omega\ell m}|{\mathrm{U}}\rangle  &  = 0, \qquad \omega >0,
    \nonumber \\
    {\hat{d}}^{{\mathrm {in}}}_{\omega\ell m}|{\mathrm{U}}\rangle & =0, \qquad \omega <0 ,
    \nonumber\\
    {\hat{c}}^{{\mathrm {up}}}_{\omega\ell m}|{\mathrm{U}}\rangle & =0, \qquad \forall \, {\widetilde {\omega }} ,
    \nonumber\\
   {\hat{d}}^{{\mathrm {up}}}_{\omega\ell m}|{\mathrm{U}}\rangle  & =0,  \qquad \forall \, {\widetilde {\omega }} . 
\end{align}
Like the Boulware state, the Unruh state $|{\mathrm{U}}\rangle $, in the distant past, contains no incoming particles in the ``in'' modes. 
The ``up'' modes are thermally populated, corresponding to the Hawking radiation at temperature $T=\kappa /2\pi $, where $\kappa $ is the surface gravity of the black hole (\ref{eq:kappa}). 
Note that the frequency appearing in the argument of the $\sinh $ function in  (\ref{eq:Uexpansion}) (which will give a thermal factor for the ``up'' modes in the corresponding expectation values) is the shifted frequency ${\widetilde {\omega }}$ (\ref{eq:wtilde}). 

Finally, we construct the CCH state $|{\mathrm{CCH}}\rangle $ using the same set of ``up'' modes as for the Unruh state, but a new set of ``in'' modes, which have positive/negative frequency with respect to the Kruskal coordinate $V$ (\ref{eq:Kruskal}) in the vicinity of the future event horizon. 
In the region exterior to the event horizon, the field expansion takes the form
\begin{multline}
    \hat{\Phi}(x) = \sum_{\ell=0}^{\infty}\sum_{m=-\ell}^{\ell}
    \\ \Bigg\{
    \int_{-\infty }^{\infty} d\omega \, \frac{\phi^{{\mathrm {in}}}_{\omega\ell m}(x)}{\sqrt{2\left|\sinh\left(\frac{\pi\omega}{\kappa}\right)\right|}}
    \left[ e^{\frac{\pi\omega}{2\kappa}}{\hat{f}}^{\mathrm {in}}_{\omega\ell m}
    +e^{-\frac{\pi\omega}{2\kappa}}{\hat{g}}^{{\mathrm {in}}\dagger}_{\omega\ell m} \right]
    \\
     + \int _{-\infty }^{\infty } d{\widetilde {\omega }} \, 
    \frac{\phi^{{\mathrm {up}}}_{\omega\ell m}(x)}{\sqrt{2\left|\sinh\left(\frac{\pi{\widetilde {\omega }}}{\kappa}\right)\right|}}
    \left[ e^{\frac{\pi {\widetilde {\omega }}}{2\kappa}} {\hat{f}}^{{\mathrm {up}}}_{\omega\ell m}
    +e^{-\frac{\pi {\widetilde {\omega }}}{2\kappa}} {\hat{g}}^{{\mathrm {up}}\dagger}_{\omega\ell m}\right] \Bigg\},
    \label{eq:CCHexpansion}
\end{multline}
with the operators in the expansion satisfying the nonzero commutation relations
\begin{align}
    \left[ {\hat{f}}^{{\mathrm {in}}}_{\omega\ell m}, {\hat{f}}^{{\mathrm {in}}\dagger}_{\omega'\ell' m'}\right] 
    & = \delta_{\ell\ell'}\delta_{mm'}\delta(\omega-\omega'), \qquad \forall \, \omega ,
    \nonumber \\
    \left[ {\hat{g}}^{{\mathrm {in}}}_{\omega\ell m}, {\hat{g}}^{{\mathrm {in}}\dagger}_{\omega'\ell' m'}\right] 
    & = \delta_{\ell\ell'}\delta_{mm'}\delta(\omega-\omega'), \qquad  \forall \, \omega ,
    \nonumber\\
    \left[ {\hat{f}}^{{\mathrm {up}}}_{\omega\ell m}, {\hat{f}}^{{\mathrm {up}}\dagger}_{\omega'\ell' m'}\right]
    & = \delta_{\ell\ell'}\delta_{mm'}\delta(\omega-\omega'), \qquad \forall \, {\widetilde {\omega }},
    \nonumber \\
    \left[ {\hat{g}}^{{\mathrm {up}}}_{\omega\ell m}, {\hat{g}}^{{\mathrm {up}}\dagger}_{\omega'\ell' m'}\right]
    & = \delta_{\ell\ell'}\delta_{mm'}\delta(\omega-\omega'), \qquad \forall \, {\widetilde {\omega }}, 
\end{align}
and the CCH state $|{\mathrm{CCH}}\rangle $ being annihilated by all the annihilation operators:
\begin{align}
    {\hat{f}}^{{\mathrm {in}}}_{\omega\ell m}|{\mathrm{CCH}}\rangle  &  = 0, \qquad \forall \, \omega ,
    \nonumber \\
    {\hat{g}}^{{\mathrm {in}}}_{\omega\ell m}|{\mathrm{CCH}}\rangle & =0, \qquad  \forall \, \omega ,
    \nonumber\\
    {\hat{f}}^{{\mathrm {up}}}_{\omega\ell m}|{\mathrm{CCH}}\rangle & =0, \qquad \forall \, {\widetilde {\omega }} ,
    \nonumber\\
   {\hat{g}}^{{\mathrm {up}}}_{\omega\ell m}|{\mathrm{CCH}}\rangle  & =0,  \qquad \forall \, {\widetilde {\omega }} . 
\end{align}
The expansion (\ref{eq:CCHexpansion}) will result in thermal factors for both the ``in'' and ``up'' modes in the expectation values we study in the remainder of this paper.
We note however that the arguments of the $\sinh $ functions in (\ref{eq:CCHexpansion}) for the ``in'' and ``up'' modes are not the same: for the ``in'' modes the argument involves the frequency $\omega $, while, for the ``up'' modes, as in the expansion (\ref{eq:Uexpansion}) leading to the Unruh state, we have the shifted frequency ${\widetilde {\omega }}$ (\ref{eq:wtilde}). 
As a result, the CCH state will not be a thermal equilibrium state. 

\section{Renormalized expectation values}
\label{sec:expvals}

In this section we outline our method, utilising the $t$-splitting variant of ``pragmatic mode-sum'' regularization \cite{Levi:2015eea} to obtain the renormalized expectation value of the charged scalar current (and also the square of the scalar field operator) for a massless charged quantum scalar field in the three states reviewed in the previous section.
The classical expression for the current is given by 
(in our conventions this has the opposite sign to the conventions of \cite{Klein:2021ctt,Klein:2021les})
\begin{equation}
J_{\mu}=-\frac{q}{4\pi}\text{Im}\left(\Phi^{*}D_{\mu}\Phi\right),
\end{equation}
and therefore, due to the field products, naive calculations of the expectation value of its corresponding operator in the quantum theory will be ill-defined. 
It is therefore necessary to employ a renormalization procedure in order to remove these divergences and obtain meaningful results. 

\subsection{Hadamard renormalization}
\label{sec:Hadamard}

In this paper, we adopt the Hadamard renormalization scheme which involves working with Hadamard states -- the defining feature of which is that the short-distance singular behaviour of their two-point functions is captured by the Hadamard parametrix. 
Let the two space-time points $x, x'$ lie in a normal neighbourhood (so that they are connected by a unique geodesic).
Then, for a Hadamard state $| \Psi \rangle$, one has 
\begin{equation}
    \langle \Psi | \hat{\Phi}(x)\hat{\Phi}^{\dagger}(x') | \Psi \rangle=K(x,x')+W_{\Psi }(x,x'),
\end{equation}
where $W_{\Psi }$ is a regular biscalar which depends on the details of the state $|\Psi \rangle $, and $K(x,x')$ denotes the Hadamard parametrix in four space-time dimensions \cite{Decanini:2005eg,Balakumar:2019djw}:
\begin{equation}
    K(x,x')=\frac{1}{8\pi^2}\left[ \frac{U(x,x')}{\sigma_{\varepsilon}(x,x')}+V(x,x')\ln\left(\frac{\sigma_{\varepsilon}(x,x')}{L^2}\right) \right].
    \label{eq:HadK}
\end{equation}
Here, $L^{2}$ is an arbitrary renormalization length scale, $U(x,x')$ and $V(x,x')$ are regular biscalars, and $\sigma_{\varepsilon}(x,x')$ is the Synge world function, defined by
\begin{equation}
    2\sigma = \sigma^{\mu}\sigma_{\mu},
\end{equation}
where $\sigma _{\mu } = \partial _{\mu } \sigma $,
and equipped with the usual $i\varepsilon$ prescription to account for points connected via null geodesics:
\begin{equation}
    \sigma_{\varepsilon}(x,x')=\sigma(x,x')+2i\varepsilon(t-t')+\varepsilon^{2}.
\end{equation}

The biscalars $U(x,x')$ and $V(x,x')$ are independent of the quantum state under consideration; they depend only on the background metric, curvature, electromagnetic potential and derivatives of these quantities.
The biscalar $V(x,x')$ is expanded as a series in the Synge world function:
\begin{equation}
    V(x,x')=\sum_{k}V_{k}(x,x')\sigma(x,x')^{k} .
\end{equation}
The coefficients of this expansion, and the biscalar $U(x,x')$ are subsequently expanded as covariant Taylor series:
\begin{subequations}
    \begin{align}
    U(x,x')= & ~ \sum_{j}U_{\alpha_{1}...\alpha_{j}}(x)\sigma^{\alpha_{1}}(x,x')...\sigma^{\alpha_{j}}(x,x'),
\\
    V_{k}(x,x')= & ~\sum_{j}V_{kj\alpha_{1}...\alpha_{j}}(x)\sigma^{\alpha_{1}}(x,x')...\sigma^{\alpha_{j}}(x,x').
\end{align}
\end{subequations}
From the equation of motion for the charged scalar field (\ref{eq:fieldeqn}),
the expansion coefficients $U_{\alpha_{1}...\alpha_{j}}(x)$, $V_{kj\alpha_{1}...\alpha_{j}}(x)$ can be determined from the transport equations \cite{Balakumar:2019djw} 
\begin{subequations}
\begin{align}
    0 & = \left[ 2\sigma^{\alpha }D_{\alpha }+\Box\sigma-4\right] U(x,x') ,
\\
0 &  = 
\left[ 2\sigma^{\alpha }D_{\alpha }+\Box\sigma-2 \right] V_{0}(x,x') 
   +D_{\alpha }D^{\alpha }U(x,x') ,
\\
0 & = 
    (k+1)\left[2\sigma^{\alpha }D_{\alpha }+\Box\sigma+2k\right]V_{k+1}(x,x') 
    \nonumber \\ & \qquad 
+ D_{\alpha }D^{\alpha }V_{k}(x,x') ,
\end{align}
together with the boundary condition
\begin{equation}
    U(x,x)=1.
\end{equation}
\end{subequations}
Expressions for the first few coefficients, sufficient for the renormalization of the stress-energy tensor operator, can be found in \cite{Balakumar:2019djw}. 

To perform Hadamard renormalization of the charged scalar field current, we first consider the unrenormalized 
expectation value with the two space-time points separated, namely:
\begin{multline}
    \langle \Psi | \hat{J}_{\mu'} | \Psi \rangle_{{\mathrm {unren}}}=\frac{iq}{8\pi}
    \\ \times
    \lim_{x\to x'}\langle \Psi | \{\hat{\Phi}^{\dagger}(x),D_{\mu'}\hat{\Phi}(x')\}  
    -\{\hat{\Phi}(x),\left[D_{\mu'}\hat{\Phi}(x')\right]^{\dagger}\} | \Psi \rangle,
\end{multline}
where $\{A,B\}=\frac{1}{2}\left( AB + BA \right)$ denotes the symmetrized product. Expanding the gauge derivative, this may be written as 
\begin{multline}
    \langle \Psi | \hat{J}_{\mu'} | \Psi \rangle_{{\mathrm {unren}}}=\frac{iq}{8\pi} \\
    \times \lim_{x\to x'}\Big( \langle \Psi | \{\hat{\Phi}^{\dagger}(x),\left[\partial_{\mu'}\hat{\Phi}(x')\right]\}-\{\hat{\Phi}(x),\left[\partial_{\mu'}\hat{\Phi}(x')\right]^{\dagger}\} | \Psi \rangle
    \\
    -iqA_{\mu'}(x')\langle \Psi | \{\hat{\Phi}^{\dagger}(x),\hat{\Phi}(x')\}+\{\hat{\Phi}(x),\hat{\Phi}^{\dagger}(x')\}| \Psi \rangle\Big).
    \label{eqn:currentDefn}
\end{multline}
Therefore, in order to calculate the renormalized expectation value of the charged scalar current, we must renormalize the quantities  
\begin{subequations}
\label{eq:unren}
    \begin{equation}
    \langle\Psi | \{\hat{\Phi}^{\dagger}(x),\left[\partial_{\mu'}\hat{\Phi}(x')\right]\}-\{\hat{\Phi}(x),\left[\partial_{\mu'}\hat{\Phi}(x')\right]^{\dagger}\} | \Psi \rangle
    \label{eq:currentDerivPart}
\end{equation}
and 
\begin{equation}
    \langle \Psi | \{\hat{\Phi}^{\dagger}(x),\hat{\Phi}(x')\}+\{\hat{\Phi}(x),\hat{\Phi}^{\dagger}(x')\} | \Psi \rangle.
    \label{eq:SC}
\end{equation}
\end{subequations}
The second of these expressions (\ref{eq:SC}) is proportional to the squared magnitude of the scalar field operator, which from now on we call the ``vacuum polarization'', and which is of interest in its own right.
In the following two subsections, we first describe our methodology for computing the vacuum polarization (that is, renormalizing (\ref{eq:SC})), and then the charged scalar current (that is, renormalizing (\ref{eq:currentDerivPart})). 

In outline, for both quantities the basic strategy is the same. 
Following \cite{Levi:2015eea}, we choose space-time points which are separated in the $t$-direction only.
We subtract from the unrenormalized expectation values (\ref{eq:unren}) a set of suitable counterterms which includes all local divergences. 
In Hadamard renormalization, these counterterms are constructed from $K(x,x')$ (\ref{eq:HadK}). 
Having fixed our point-splitting, we write the required counterterms as expansions in the time separation $(t-t')$.

The challenge is then to write the resulting renormalized expectation values in a form which can be computed numerically. 
The unrenormalized expectation values (\ref{eq:unren}) will be written as sums over the charged scalar field modes (\ref{eq:modes}), from which we need to subtract the counterterm expansions in $(t-t')$.
We tackle this problem by extending, to a charged scalar field, the $t$-splitting variant of the ``pragmatic mode-sum'' method developed for a neutral scalar field \cite{Levi:2015eea}.
In \cite{Levi:2015eea} deWitt-Schwinger rather than Hadamard renormalization is employed.
The relationship between deWitt-Schwinger and Hadamard renormalization for a charged quantum scalar field is elucidated in Ref.~\cite{Pla:2022spt}. 
In particular, the counterterms in deWitt-Schwinger and Hadamard renormalization have the same divergences (as they must, in order to yield finite renormalized expectation values), and hence both give physically meaningful results. 
We have chosen Hadamard rather than deWitt-Schwinger counterterms as we can use the readily available expressions for the former in Ref.~\cite{Balakumar:2019djw}, which were also employed in the recent work \cite{Breen:2024ggu}.

Before describing the details of our methodology, we note that we have chosen to fix the gauge potential (\ref{eqn:gaugeField}) from the outset, unlike the approach in \cite{Klein:2021ctt}, where a judicious choice of gauge means that only finite renormalization terms are required to find the expectation value of the charged scalar current. 
The latter choice of gauge corresponds, at each space-time point where the expectation value is to be computed, to setting the gauge potential $A_{\mu }$ to vanish.
Since the gauge potential appears in the radial equation (\ref{eq:radial}), this entails computing a separate set of radial functions appearing in the charged scalar field modes (\ref{eq:modes}) for each value of the radial coordinate $r$ under consideration.
Given that finding the radial functions is the most computationally-intensive part of the method, and we are interested in a large number of values of the radial coordinate, we have found it more palatable to carry out the renormalization.  

\subsection{Vacuum polarization}
\label{sec:SC}

The expression for the unrenormalized expectation value of the vacuum polarization $\langle   |{\hat {\Phi }} |^{2}  \rangle $ may be written as
\begin{equation}
    \langle |{\hat {\Phi }} |^{2} \rangle = \lim_{x\to x'}\text{Re}\left[\langle\{\hat{\Phi}(x),\hat{\Phi}^{\dagger}(x')\}\rangle\right] .
    \label{eq:SCunren}
\end{equation}
Following \cite{Levi:2015eea}, we wish to write this as a mode sum over the quantum numbers $\ell $ and $m$, and 
and an integral over {\em {positive}} values of the frequency $\omega $. 
Considering the quantum field in the Boulware, Unruh and CCH states, and our choice of point splitting in the $t$-direction, the unrenormalized expectation values (\ref{eq:SCunren}) take the form
\begin{widetext}
\begin{subequations}
\label{eq:SCunrenstates}
\begin{align}
    \langle {\mathrm{B}} | \,|{\hat {\Phi }} |^{2} \, |{\mathrm{B}}\rangle = & 
    ~ \frac{1}{2}\lim_{x\to x'} 
    \int _{0}^{\infty } d\omega\cos\left(\omega\epsilon\right)
    \left[ F^{{\mathrm {in}}}(\omega,r)+F^{{\mathrm {in}}}(-\omega,r)
    +F^{{\mathrm {up}}}(\omega,r)+F^{{\mathrm {up}}}(-\omega,r)\right],
    \label{eq:SCunrenB}
\\
    \langle {\mathrm{U}} | \,|{\hat {\Phi }} |^{2} \, |{\mathrm{U}}\rangle = & 
    ~ \frac{1}{2}\lim_{x\to x'} 
    \int _{0}^{\infty } d\omega \cos\left(\omega\epsilon\right)
    \Bigg[ F^{{\mathrm {in}}}(\omega,r)+F^{{\mathrm {in}}}(-\omega,r)
    + F^{{\mathrm {up}}}(\omega,r)\coth\left|\frac{\pi{\widetilde {\omega }}}{\kappa}\right|
    \nonumber \\ & \qquad \qquad 
    + F^{{\mathrm {up}}}(-\omega,r) \coth\left|\frac{\pi{\overline {\omega }}}{\kappa}\right|
    \Bigg] ,
    \label{eq:SCunrenU}
\\
    \langle {\mathrm{CCH}} | \,|{\hat {\Phi }} |^{2} \, | {\mathrm{CCH}} \rangle= & ~
    \frac{1}{2}\lim_{x\to x'} 
    \int _{0}^{\infty } d\omega \cos\left(\omega\epsilon\right)
    \Bigg[ \left\{ F^{{\mathrm {in}}}(\omega,r)+F^{{\mathrm {in}}}(-\omega,r)\right\} 
    \coth\left|\frac{\pi\omega}{\kappa}\right|
   \nonumber \\ & \qquad  \qquad 
    + F^{{\mathrm {up}}}(\omega,r) \coth\left|\frac{\pi{\widetilde {\omega }}}{\kappa}\right|
    + F^{{\mathrm {up}}}(-\omega,r) \coth\left|\frac{\pi{\overline {\omega }}}{\kappa}\right| 
    \Bigg],
    \label{eq:SCunrenCCH}
\end{align}
\end{subequations}
\end{widetext}
where we have defined (cf.~(\ref{eq:wtilde}))
\begin{equation}
    {\overline {\omega }} = \omega + \frac{qQ}{r_{+}}.
    \label{eq:wbar}
\end{equation}
In (\ref{eq:SCunrenstates}) we have defined the following functions, employing similar notation to that in  Ref.~\cite{Levi:2015eea},
\begin{subequations}
\label{eq:FmodeSC}
    \begin{align}
    F^{{\mathrm {in}}}(\omega,r)= & ~
    \frac{1}{16\pi ^{2}}\sum _{\ell =0}^{\infty }\frac{\left(2\ell+1\right)}{|\omega|r^{2}} 
    |X^{\mathrm {in}}_{\omega\ell}(r)|^{2} ,
    \\
    F^{{\mathrm {up}}}(\omega,r)= & ~
    \frac{1}{16\pi ^{2}}\sum _{\ell =0}^{\infty }\frac{\left(2\ell+1\right)}{|{\widetilde {\omega}} |r^{2}} 
    |X^{\mathrm {up}}_{\omega\ell}(r)|^{2} .
\end{align}
\end{subequations}
For a neutral scalar field, we have $F^{\mathrm {in/up}}(-\omega ,r)=F^{\mathrm {in/up}}(\omega ,r)$ because in that case the radial functions satisfy $X_{-\omega \ell }(r) = X_{\omega \ell }^{*}(r)$, where a star $*$ denotes the complex conjugate.
However, this does not hold for the charged scalar field modes, because the radial equation (\ref{eq:radial}) is not invariant under the transformation $\omega \rightarrow -\omega $.
Therefore our mode sums in (\ref{eq:SCunrenstates}) have separate contributions from the positive and negative frequency modes, and we have to compute almost twice as many scalar field modes as in the neutral case. 
The unrenormalized expectation values $\langle {\mathrm{U}} | \,|{\hat {\Phi }} |^{2} \, |{\mathrm{U}}\rangle$ and $\langle {\mathrm {CCH}} | \,|{\hat {\Phi }} |^{2} \, |{\mathrm{CCH}}\rangle$ contain thermal $\coth $ factors.
In the Unruh state, this thermal factor is present only for the ``up'' modes.
For the positive frequency modes, the thermal factor depends on the shifted frequency ${\widetilde {\omega }}$ (\ref{eq:wtilde}), corresponding to thermal emission of particles with a chemical potential equal to $qQ/r_{+}$ \cite{Gibbons:1975kk}.
The thermal factor for the negative frequency modes depends, instead, on ${\overline {\omega }}$ (\ref{eq:wbar}), corresponding to a change in sign of the chemical potential \cite{Gibbons:1975kk,Balakumar:2022yvx}. 
The CCH state also contains a thermal factor for the ``in'' modes, but this depends only on the frequency $\omega $ and the chemical potential is absent.
 
The renormalization terms required for the vacuum polarization are simply given by taking the real part of the Hadamard parametrix (\ref{eq:HadK}).
To renormalize the vacuum polarization we need to expand (\ref{eq:HadK}) up to (and including) terms which are finite in the coincidence limit.  
For a massless, minimally coupled scalar field, we see from the expressions provided in \cite{Balakumar:2019djw} that the Hadamard coefficients we require, for a massless and minimally-coupled charged scalar field, are 
\begin{subequations}
\label{eq:realHadC}
\begin{align}
    \Re \left[U_{0}(x)\right] & = 1,  \\ 
    \Re \left[U_{1\mu}(x)\right] & =0,  \\ 
    \Re \left[ U_{2\mu\nu}(x)\right] & = \frac{1}{12}R_{\mu\nu}-\frac{q^{2}}{2}A_{(\mu}A_{\nu)}, \\
    \Re \left[ V_{00}(x)\right] & = 0,  \\
   \Re \left[ V_{01\mu}(x)\right] & = 0,
\end{align}
\end{subequations}
where $\Re $ denotes the real part.
We also need to expand the Synge world function and its derivatives in powers of the time separation $(t-t')=\epsilon $.
For points separated in the time direction, from \cite{Anderson:1993if} we have
\begin{subequations}
\label{eq:sigma}
\begin{align}
    \sigma= & ~ -\frac{1}{2}f(r)\epsilon^{2}-\frac{1}{96}f(r)f'(r)^{2}\epsilon^{4}+\mathcal{O}(\epsilon^{6}),
\\
    \sigma^{t}  = &  ~ \epsilon+\frac{f'(r)^{2}}{24}\epsilon^{3}+\mathcal{O}(\epsilon^{5}),
\\
    \sigma^{r}  = & ~ -\frac{1}{4}f(r)f'(r)\epsilon^{2}+\mathcal{O}(\epsilon^{4}),
\\
    \sigma^{\theta}  = & ~ \sigma^{\varphi}=0.
\end{align}
\end{subequations}
Combining (\ref{eq:realHadC}, \ref{eq:sigma}), the Hadamard counterterms that we need to subtract to renormalize the vacuum polarization are
\begin{multline}
    \Re \left[K(x,x')\right]=-\frac{1}{4\pi^{2}\epsilon^{2}f(r)}\\
    +\frac{24q^{2}rA_{t}(r)^{2}-4f(r)f'(r)+rf'(r)^{2}-2rf(r)f''(r)}{192\pi^{2}rf(r)}.
    \label{eq:SCHad}
\end{multline}
We note that (\ref{eq:SCHad}) takes a very similar form to that for a neutral scalar field, with the gauge potential contributing only to the first term in the numerator on the second line.

To find the renormalized vacuum polarization, we need to subtract (\ref{eq:SCHad}) from the mode sums in (\ref{eq:SCunrenstates}). 
To achieve this subtraction, following \cite{Levi:2015eea}, we write the singular part of (\ref{eq:SCHad}) as an integral over frequency, using the identity
\begin{equation}
    \frac{1}{\epsilon^{2}}=-\int_{0}^{\infty}\omega\cos{\left(\omega\epsilon\right)} \, d\omega,
    \label{eqn:Id1}
\end{equation}
which gives
\begin{equation}
    -\frac{1}{4\pi^{2}\epsilon^{2}f(r)}=\int_{0}^{\infty}F^{{\mathrm{VP}}}_{{\mathrm {sing}}}(\omega,r)\cos{\left(\omega\epsilon\right)} \, d\omega,
    \label{eq:singSCint}
\end{equation}
where we have defined 
\begin{equation}
    F^{\mathrm{VP}}_{{\mathrm {sing}}}(\omega,r)=\frac{\omega }{4\pi^{2}f(r)},
    \label{eq:FsingSC}
\end{equation}
and ``VP'' stands for ``vacuum polarization''. 

We now demonstrate the complete procedure for computing the expectation value of the vacuum polarization for the charged scalar field in the Boulware state $|{\mathrm{B}}\rangle $, at a single radial point. 
The calculation in the other two states, $|{\mathrm{U}}\rangle$ and $|{\mathrm{CCH}}\rangle$, and at other values of the radial coordinate follows in exactly the same way. 
In the following we set $M=1$, $Q=4M/5$ and $qM=8/25$, and take $|\omega| \in[0,4]$, with a grid spacing $\delta \omega = 1/300$. 
We set the radial point to be $r_{0}\approx5.96158M$. 

Subtracting the renormalization terms (\ref{eq:SCHad}) (with the first term written as the integral (\ref{eq:singSCint})) from the unrenormalized expectation value (\ref{eq:SCunrenB}), the quantity we seek to compute is
\begin{multline}
    \langle {\mathrm{B}} |\,|{\hat {\Phi }} |^{2} \, | {\mathrm{B}}\rangle=
    \int _{0}^{\infty } d\omega\,  F^{\mathrm{VP}}_{{\mathrm {reg}}}(\omega,r)\\
    -\frac{24q^{2}rA_{t}(r)^{2}-4f(r)f'(r)+rf'(r)^{2}-2rf(r)f''(r)}{192\pi^{2}rf(r)},
    \label{eq:SCrenB}
\end{multline}
where we have defined
\begin{equation}
    F^{\mathrm{VP}}_{{\mathrm {reg}}}(\omega,r)=F^{\mathrm{VP}}_{{\rm {total}}}(\omega,r)-F^{\mathrm {VP}}_{{\mathrm {sing}}}(\omega,r),
    \label{eq:FregSC}
\end{equation}
with
\begin{multline}
    F^{\mathrm{VP}}_{{\mathrm {total}}}(\omega,r)=\\
    \frac{1}{2}\left[ F^{{\mathrm {in}}}(\omega,r)+F^{{\mathrm {in}}}(-\omega,r)
    +F^{{\mathrm {up}}}(\omega,r)+F^{{\mathrm {up}}}(-\omega,r) \right] ,
    \label{eq:FtotalSC}
\end{multline}
and $F^{\mathrm {in/up}}(\omega ,r)$ are given in (\ref{eq:FmodeSC}). 

The first step is to compute $F^{{\mathrm {in/up}}}(\pm \omega,r_{0})$. 
Since these are given as mode sums, we first integrate the radial equation (\ref{eq:radial}) to obtain the ``in'' and ``up'' modes. 
In this section we focus on the renormalization procedure, so we postpone discussion of the details of this integration to Sec.~\ref{sec:radial}.
We find the radial mode functions for all values of $\omega$ in the frequency grid and for $0\leq\ell\leq\ell_{\mathrm{max}}$, where $\ell_{\mathrm{max}}$ is the value of $\ell$ at which we truncate the sums in $F^{{\mathrm {in/up}}}(\pm \omega ,r_{0})$ (\ref{eq:FmodeSC}). 
Following \cite{Levi:2015eea}, we truncate the sums over $\ell $ at a value $\ell_{\mathrm {max}}$, beyond which the contribution to each sum is less than $10^{-10}$.

\begin{figure}
    \centering
    \includegraphics[scale=0.65]{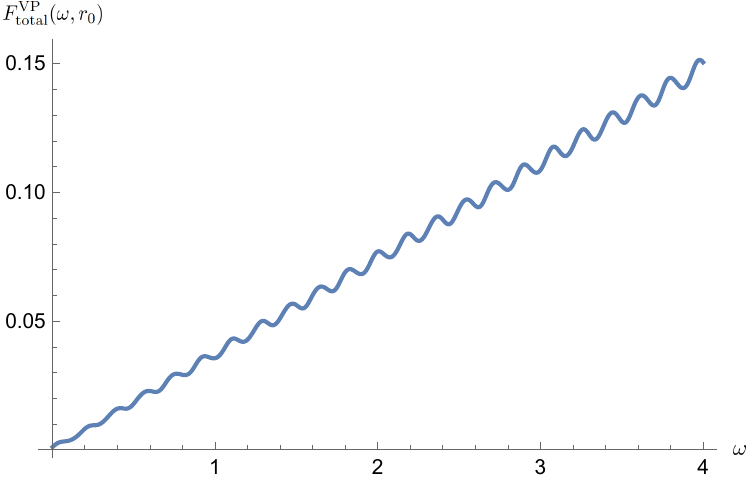}
    \caption{Mode-sum contribution $F^{\mathrm{VP}}_{{\mathrm {total}}}(\omega,r_{0})$ (\ref{eq:FtotalSC}) to the vacuum polarization for the Boulware state $|{\mathrm{B}}\rangle $ for a charged scalar field on an RN black hole background, as a function of frequency $\omega $ at fixed radial coordinate $r_{0}\approx5.96158M$.
    The RN black hole has mass $M=1$ and charge $Q=4M/5$, while the scalar field has charge $qM=8/25$.
    }
    \label{fig:FtotSC}
\end{figure}

\begin{figure}
    \centering
    \includegraphics[scale=0.65]{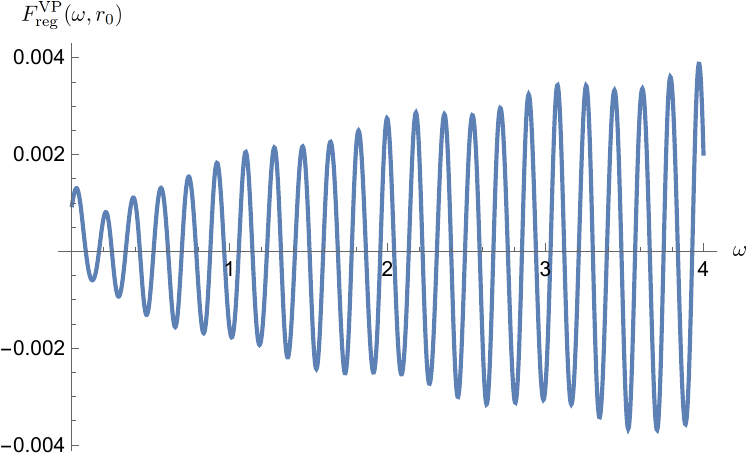}
    \caption{Regularized mode-sum contribution $F^{\mathrm{VP}}_{{\mathrm {reg}}}(\omega,r_{0})$ (\ref{eq:FregSC}) to the vacuum polarization for the Boulware state $|{\mathrm{B}}\rangle $ for a charged scalar field on an RN black hole background, as a function of frequency $\omega $ at fixed radial coordinate $r_{0}\approx5.96158M$.
    The black hole and scalar field parameters are as in Fig.~\ref{fig:FtotSC}.}
    \label{fig:FregSC}
\end{figure}

Fig.~\ref{fig:FtotSC} shows $F^{\mathrm{VP}}_{{\mathrm {total}}}(\omega,r_{0})$ (\ref{eq:FtotalSC}) as a function of $\omega $, for the particular value of the radial coordinate we are using to illustrate our procedure.
As is the case for a neutral scalar field on a Schwarzschild black hole background \cite{Levi:2015eea}, we see that $F^{\mathrm{VP}}_{{\mathrm {total}}}(\omega,r_{0})$ grows linearly as $\omega $ increases, with oscillations around this linear growth.
The linear divergence is removed by subtracting from $F^{\mathrm{VP}}_{{\mathrm {total}}}(\omega,r_{0})$ the singular term $F^{\mathrm{VP}}_{{\mathrm {sing}}}(\omega,r_{0})$ (\ref{eq:FsingSC}), as can be seen in Fig.~\ref{fig:FregSC}.

However, on subtracting the linear divergence, in Fig.~\ref{fig:FregSC} we see that $F^{\mathrm{VP}}_{{\mathrm {reg}}}(\omega,r_{0})$ is oscillating as a function of the frequency $\omega $, with the amplitude of the oscillations increasing as $\omega $ increases.
Thus the integral in (\ref{eq:SCrenB}) is still divergent. 
As discussed in \cite{Levi:2015eea}, the origin of these oscillations is the existence of an additional singularity in the unrenormalized vacuum polarization (\ref{eq:SCunrenB}), that is, a singularity which arises while the points are still separated. 
This singularity arises when there are pairs of space-time points $(t,\bf{x})$ and $(t',\bf{x})$ which are connected by a null geodesic which orbits the event horizon one (or more) times. 
For black hole space-times, there will be an infinite number of such geodesics, orbiting the event horizon an integer number of times, before returning to the same point in space (but at a different coordinate time). 
The Hadamard parametrix (\ref{eq:HadK}) only captures the local singular behaviour, and not the nonlocal singularities resulting from these null geodesics. 

Following the method of \cite{Levi:2015eea}, we evaluate the integral in (\ref{eq:SCrenB}) as a generalized, ``self-cancellation'' integral, defined as follows.
Let
\begin{equation}
    H(\omega)=\int_{0}^{\omega}h(x) \, dx,
    \label{eq:Hint}
\end{equation}
denote the integral we wish to evaluate in the limit $\omega \rightarrow \infty $.
Then, if the integrand $h(x)$ oscillates with a wavelength $\lambda $, we define the ``self-cancellation'' integral to be
\begin{equation}
    \int_{0}^{\infty({\rm {sc}})}h(\omega) \, d\omega=\lim_{\omega\to\infty} \frac{1}{2}\left[H(\omega)+H\left(\omega+\frac{\lambda}{2}\right)\right], 
\end{equation}
where we use lower case ${\mathrm {sc}}$ to denote ``self-cancellation''. 
This procedure will completely cancel constant amplitude oscillations with wavelength $\lambda $, enabling the limit on the right-hand-side to be taken, yielding a finite result.
In our case, each null geodesic wrapping round the black hole and connecting space-time points with the same spatial components will contribute oscillations in $F^{\mathrm{VP}}_{{\mathrm {reg}}}(\omega,r_{0})$ having a wavelength equal to the orbital period of the geodesic. 

Given that there is a countably infinite number of null geodesics orbiting the black hole, each with a different wavelength, in principle the self-cancellation procedure would need to be performed separately for each wavelength.
However, the amplitude of the oscillations resulting from a null geodesic decreases rapidly as the number of times the geodesic orbits the black hole increases. 
As in \cite{Levi:2015eea}, we find that performing the self-cancellation procedure for the first four wavelengths is sufficient for our purposes.
 
The next step in the computation is to apply self-cancellation to the integral in (\ref{eq:SCrenB}). 
For this, we need to find the relevant wavelengths $\lambda _{i}$, for $i=1,2,3,4$.
These are given by 
\begin{equation}
    \lambda _{i}= \frac{2\pi}{\varepsilon_{i}} ,
\end{equation}
where ${\varepsilon _{i}}$ is the coordinate time taken for a null geodesic to complete $i$ closed orbits of the black hole round the black hole and return to the same spatial location.
We find these orbital periods by numerically integrating the geodesic equations. 
For our representative radial coordinate $r_{0}\approx 5.96158M$, we find that the first four wavelengths are given approximately by
\begin{align}
    \lambda_{1}\approx 0.1795,\qquad & \lambda_{2}\approx 0.0988,
    \nonumber \\ 
    \lambda_{3}\approx  0.0682,\qquad & \lambda_{4}\approx 0.0493.
\end{align}

If the amplitude of the oscillations in the integrand in (\ref{eq:SCrenB}) were constant, then, for each wavelength, it would be necessary to apply the self-cancellation procedure just once. 
However, as can be seen in Fig.~\ref{fig:FregSC}, the amplitude of the oscillations is in fact growing as $\omega $ increases. 
Therefore, in order to cancel the oscillations, it is necessary to apply the self-cancellation procedure multiple times for each wavelength.
As in \cite{Levi:2015eea}, we find that we need to apply self-cancellation four times for each wavelength in order to cancel the oscillations. 

The final answer for the renormalized vacuum polarization in the Boulware state is then written in terms of a self-cancellation generalized integral. 
We define the self-cancellation operator $T_{\lambda }$, for an oscillation having wavelength $\lambda $, applied to the integral $H(\omega )$ (\ref{eq:Hint}) to be \cite{Levi:2015eea}
\begin{equation}
    T_{\lambda}\left[H(\omega)\right]=\frac {1}{2} \left[ H(\omega)+H\left(\omega+\frac{\lambda}{2}\right)\right] , 
    \label{eq:selfcancelT}
\end{equation} 
and the regularized quantity resulting from applying this operator four times for each wavelength to be
\begin{equation}
     H_{*}(\omega,r):=(T_{\lambda_{1}})^{4}(T_{\lambda_{2}})^{4}(T_{\lambda_{3}})^{4}(T_{\lambda_{4}})^{4}\left[H(\omega,r)\right].
     \label{eq:Hstar}
\end{equation}
With this notation, the renormalized vacuum polarization in the Boulware state is
\begin{multline}
    \langle {\mathrm{B}}| \,|{\hat {\Phi }} |^{2} \, |{\mathrm{B}}\rangle=
    \int _{0}^{\infty({\mathrm {sc}}) }  F^{\mathrm {VP}}_{{\mathrm {reg}}}(\omega,r) \, d\omega\\
    -\frac{24q^{2}rA_{t}(r)^{2}-4f(r)f'(r)+rf'(r)^{2}-2rf(r)f''(r)}{192\pi^{2}rf(r)},
\end{multline}
where now 
\begin{equation}
    \int _{0}^{\infty({\mathrm {sc}}) }  F^{\mathrm{VP}}_{{\mathrm {reg}}}(\omega,r) \, d\omega = \lim_{\omega\to\infty}H_{*}^{\mathrm {VP}}(\omega,r),
    \label{eq:FSCSC}
\end{equation}
with $H_{*}^{\mathrm {VP}}(\omega,r)$ given by (\ref{eq:Hstar}) for $H(\omega ,r)=H^{\mathrm{{VP}}}(\omega,r)$ and 
\begin{equation}
    H^{\mathrm{{VP}}}(\omega,r)=\int_{0}^{\omega}F^{\mathrm{VP}}_{{\mathrm {reg}}}(\omega',r) \, d\omega'.
    \label{eq:HSC}
\end{equation}

\begin{figure}
    \centering
    \includegraphics[scale=0.65]{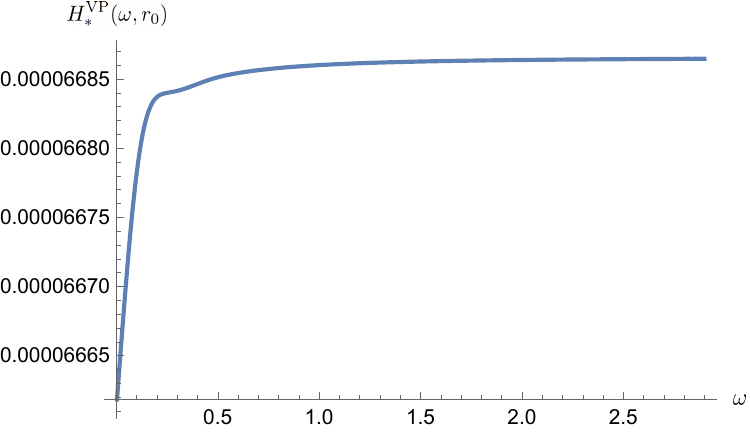}
    \caption{Self-cancellation generalized integral $H_{*}^{\mathrm {VP}}(\omega,r)$ (\ref{eq:Hstar}, \ref{eq:HSC}) for the charged vacuum polarization in the Boulware state $|{\mathrm{B}}\rangle $, as a function of frequency $\omega $ at fixed radial coordinate $r_{0}\approx5.96158M$.
    The black hole and charged scalar field parameters are as in Fig.~\ref{fig:FtotSC}.
    }
    \label{fig:HstarSC}
\end{figure}

Figure \ref{fig:HstarSC} shows the result of performing the self-cancellation integral for our example value of the radial coordinate, $r=r_{0}$. 
It can be seen that we are left with a function devoid of oscillations, that is rapidly convergent as $\omega\to\infty$. 
From this, one can readily deduce the value of the renormalized vacuum polarization for the particular state and radial coordinate under consideration.

\subsection{Renormalized charge current}
\label{sec:Current}

As a result of the spherical symmetry inherited by our three quantum states of interest, the angular components of the expectation value of the charge current vanish identically \cite{Balakumar:2022yvx}:
\begin{subequations}
\label{eq:Jangular}
\begin{align}
  0 & =  \langle {\mathrm{B}}| {\hat{J}}_{\theta}| {\mathrm{B}} \rangle
  = \langle {\mathrm{U}} | {\hat{J}}_{\theta}| {\mathrm{U}}\rangle
  =\langle {\mathrm{CCH}} | {\hat{J}}_{\theta}| {\mathrm{CCH}} \rangle ,
\\ 
  0 & =  \langle {\mathrm{B}} | {\hat{J}}_{\varphi} | {\mathrm{B}}\rangle
  =\langle {\mathrm{U}} | {\hat{J}}_{\varphi} | {\mathrm{U}}\rangle
  =\langle {\mathrm{CCH}} | {\hat{J}}_{\varphi} | {\mathrm{CCH}} \rangle .
\end{align}
\end{subequations}
The radial component of the charge current does not require renormalization \cite{Balakumar:2022yvx}, and is thus given by the mode sums
\begin{widetext}
\begin{subequations}
\label{eq:Jrmode}
\begin{align}    
    \langle {\mathrm{B}}| {\hat{J}}_{r}| {\mathrm{B}}\rangle
    = & ~ -\frac{q}{8\pi}\lim_{x\to x'} 
    \int _{0}^{\infty } d\omega
    \left[ G^{{\mathrm {in}}}(\omega,r)+G^{{\mathrm {in}}}(-\omega,r)+G^{{\mathrm {up}}}(\omega,r)
    +G^{{\mathrm {up}}}(-\omega,r) \right] ,
\\
    \langle {\mathrm{U}} | {\hat{{J}}}_{r}| {\mathrm{U}} \rangle
    = & ~ -\frac{q}{8\pi}\lim_{x\to x'} 
    \int _{0}^{\infty } d\omega 
    \Bigg[ G^{{\mathrm {in}}}(\omega,r)+G^{{\mathrm {in}}}(-\omega,r)
    +G^{{\mathrm {up}}}(\omega ,r) \coth\left|\frac{\pi{\widetilde {\omega }}}{\kappa}\right| 
    +G^{\mathrm {up}}(-\omega,r)\coth \left|\frac{\pi {\overline {\omega }}}{\kappa}\right| 
     \Bigg]  ,
\\
    \langle {\mathrm{CCH}} | {\hat{J}}_{r}| {\mathrm{CCH}} \rangle
    = & ~ -\frac{q}{8\pi}\lim_{x\to x'} 
    \int _{0}^{\infty } d\omega
    \Bigg[ 
    \left\{ G^{{\mathrm {in}}}(\omega,r)+G^{{\mathrm {in}}}(-\omega,r)\right\}  \coth \left|\frac{\pi\omega}{\kappa}\right| 
    +G^{{\mathrm {up}}}(\omega,r) \coth \left|\frac{\pi{\widetilde {\omega }}}{\kappa}\right| 
    \nonumber \\ & \qquad \qquad 
    +G^{{\mathrm {up}}}(-\omega,r)\coth \left|\frac{\pi {\overline {\omega }}}{\kappa}\right|  
    \Bigg] ,
\end{align}
\end{subequations}
\end{widetext}
where, analogously to (\ref{eq:FmodeSC}), we have defined
\begin{subequations}
    \begin{align}
    G^{{\mathrm {in}}}(\omega,r) & =
    \sum _{\ell =0}^{\infty }\frac{\left(2\ell+1\right)}{16\pi ^{2}\left| \omega \right|}\Im \left[ \frac{X^{\mathrm {in}*}_{\omega\ell}(r)}{r}\frac{d}{dr}\left(  \frac{X^{\mathrm {in}}_{\omega\ell}(r)}{r} \right)
    \right] ,
\\
    G^{{\mathrm {up}}}(\omega,r) & =
    \sum _{\ell =0}^{\infty }\frac{\left(2\ell+1\right)}{16\pi^{2}\left| {\widetilde {\omega }}\right|}\Im \left[
    \frac{X^{\mathrm {up}*}_{\omega\ell}(r)}{r}\frac{d}{dr}\left( \frac{X^{\mathrm {up}}_{\omega\ell}(r)}{r} \right)
    \right].
\end{align}
\end{subequations}
$\Im $ denotes the imaginary part, and the shifted frequencies ${\widetilde {\omega }}$ and ${\overline {\omega }}$ are given in (\ref{eq:wtilde}, \ref{eq:wbar}) respectively.
From the semiclassical Maxwell equations (\ref{eq:SCmaxwell}), it must be the case that the expectation value of the charge current in any quantum state satisfies the conservation equation
\begin{equation}
    \nabla_{\mu}\langle \hat{{J}}^{\mu}\rangle =0.
    \label{eq:conservation}
\end{equation}
Since the charge current expectation value is independent of time and the angular components vanish (\ref{eq:Jangular}), solving the (\ref{eq:conservation}) gives that 
the radial component of the charge current expectation value takes the form \cite{Balakumar:2022yvx}
\begin{equation}
    \langle \hat{{J}}^{r}\rangle=-\frac{\mathcal{K}}{r^{2}},
    \label{eq:Jr}
\end{equation}
where $\mathcal{K}$ is some state-dependent constant, interpreted as the flux of charge from the black hole in the particular state under consideration.
We could use (\ref{eq:Jr}) to simplify the computations and find $\langle \hat{{J}}^{r}\rangle$ at just one value of the radial coordinate $r$ (which would be sufficient to determine ${\mathcal {K}}$).
However, since we have already generated the charged scalar field modes, we find $\langle \hat{{J}}^{r}\rangle$ by evaluating the mode sums (\ref{eq:Jrmode}) at each value of the radial coordinate under consideration, in which case (\ref{eq:Jr}) provides a useful check on the accuracy of our numerical results.

Thus, to obtain the renormalized expectation value of the current in its entirety, it suffices to renormalize the $t$-component. 
We now extend the renormalization procedure outlined in Sec.~\ref{sec:SC} for the vacuum polarization, to renormalize the  $t$-component of (\ref{eq:currentDerivPart}). 
Including the factor of $i$ from (\ref{eqn:currentDefn}), we write the quantity we seek to renormalize as 
\begin{multline}
    i\left(\langle\{\hat{\Phi}^{\dagger}(x),\left[ \partial_{t'}\hat{\Phi}(x')\right] \}-\{\hat{\Phi}(x),\left[\partial_{t'}\hat{\Phi}(x')\right]^{\dagger}\}\rangle\right)\\
    =2 \, \Im \langle\{\hat{\Phi}(x),\left[\partial_{t'}\hat{\Phi}(x')\right]^{\dagger}\}\rangle .
\end{multline}
In the three  quantum states of interest, the unrenormalized expectation values of this quantity can be written as the following mode sums (analogous to (\ref{eq:SCunrenstates})):
\begin{widetext}
\begin{subequations}
\label{eq:Jtunrenstates}
    \begin{align}
    2\, \Im \langle {\mathrm{B}} |\{\hat{\Phi}(x),\left[\partial_{t}\hat{\Phi}(x)\right]^{\dagger}\} | {\mathrm{B}}\rangle 
    & = \lim_{x\to x'} 
    \int _{0}^{\infty } d\omega \, \omega 
     \cos\left(\omega\epsilon\right) \left[ F^{{\mathrm {in}}}(\omega,r)-F^{{\mathrm {in}}}(-\omega,r)
     +F^{{\mathrm {up}}}(\omega,r)-F^{{\mathrm {up}}}(-\omega,r)\right],
     \label{eq:JtunrenB}
\\ 
   2\, \Im \langle {\mathrm{U}}|\{ \hat{\Phi}(x),\left[\partial_{t}\hat{\Phi}(x)\right]^{\dagger}\}| {\mathrm{U}}\rangle
   & = \lim_{x\to x'} 
    \int _{0}^{\infty } d\omega \, \omega 
 \cos\left(\omega\epsilon\right)
 \Bigg[ F^{{\mathrm {in}}}(\omega,r)-F^{{\mathrm {in}}}(-\omega,r)
    +F^{{\mathrm {up}}}(\omega,r)\coth\left|\frac{\pi{\widetilde {\omega }}}{\kappa}\right|
    \nonumber \\ & \qquad \qquad 
    -F^{{\mathrm {up}}}(-\omega,r)\coth\left|\frac{\pi{\overline {\omega }}}{\kappa}\right| 
    \Bigg] ,
\\ 
    2\, \Im \langle {\mathrm{CCH}} |\{\hat{\Phi}(x),\left[\partial_{t}\hat{\Phi}(x)\right]^{\dagger}\}| {\mathrm{CCH}} \rangle
    & =\lim_{x\to x'} 
    \int _{0}^{\infty } d\omega \, \omega 
    \cos\left(\omega\epsilon\right)
    \Bigg[ 
    \left\{ F^{{\mathrm {in}}}(\omega,r)-F^{{\mathrm {in}}}(-\omega,r)\right\}  
    \coth\left|\frac{\pi\omega}{\kappa}\right|
   \nonumber \\ &  \qquad \qquad 
    +F^{{\mathrm {up}}}(\omega,r) \coth\left|\frac{\pi{\widetilde {\omega }}}{\kappa}\right|
    -F^{{\mathrm {up}}}(-\omega,r) \coth \left|\frac{\pi{\overline {\omega }}}{\kappa}\right| 
    \Bigg], 
\end{align}
\end{subequations}
\end{widetext}
where the functions $F^{\mathrm {in/up}}(\omega ,r)$ are those appearing in the vacuum polarization (\ref{eq:FmodeSC}).
Notice that the mode-sum expressions (\ref{eq:Jtunrenstates}) involve the difference in contributions from the positive and negative frequency modes.
Unlike the situation for a neutral scalar field, these do not cancel for a charged scalar field, although we may anticipate that the renormalized current will be smaller than the renormalized vacuum polarization. 

The relevant renormalization counterterms can be found by taking the $t'$ derivative of the imaginary part of the Hadamard parametrix (\ref{eq:HadK}). 
Using \cite{Balakumar:2019djw,Klein:2021les}, we find, for a massless and minimally-coupled charged scalar field,
\begin{subequations}
\label{eq:imHadC}
\begin{align}
    \Im \left[U_{0}(x)\right] & =0,
    \\
    \Im \left[ U_{1\mu}(x)\right] &  =qA_{\mu},
    \\
   \Im \left[U_{2\mu\nu}(x)\right] & =-\frac{q}{2}\nabla_{(\mu}A_{\nu)},
   \\
   \Im \left[U_{3\mu\nu\rho}(x)\right] & = \frac{q}{6}\nabla_{(\mu}\nabla_{\nu}A_{\rho)}+\frac{q}{12}R_{(\mu\nu}A_{\rho)}
   \nonumber \\ & \qquad 
   -\frac{q^{3}}{6}A_{\mu}A_{\nu}A_{\rho},
   \\
    \Im \left[V_{00}(x)\right] &=0,
    \\
    \Im \left[ V_{01\mu}(x)\right] & =0,
    \\
    \Im \left[V_{02\mu\nu}(x)\right] &=0, 
    \\
    \Im \left[ V_{10}(x)\right] & =0.
\end{align}
\end{subequations}
Combining (\ref{eq:imHadC}) with our expansions for $\sigma$ and its derivatives (\ref{eq:sigma}), the required 
renormalization counterterm is
\begin{multline}
\label{eq:ImKt}
    2 \, \partial_{t'}\text{Im}\left[K(x,x')\right] = 
    -\frac{qA_{t}(r)}{2\pi^{2}f(r)\epsilon^{2}}
    \\
    -\frac{q}{96\pi^{2}rf(r)}\Bigg[ 8q^{2}rA_{t}(r)^{3}-4qA_{t}(r)f(r)f'(r)
    \\ 
    -2rA_{t}'(r) f(r)f'(r)+rA_{t}(r)f'(r)^{2} 
    \\
    -2rA_{t}(r)f(r)f''(r)\Bigg]  .
\end{multline}
As expected,  (\ref{eq:ImKt}) is proportional to the scalar field charge $q$ and vanishes when either this or the background electrostatic potential $A_{t}(x)$ vanishes.
As we found for the renormalization of the vacuum polarization (\ref{eq:SCHad}), the singular part of the renormalization terms (\ref{eq:ImKt}) only involves a quadratic divergence in the time separation $\epsilon $. 

To subtract (\ref{eq:ImKt}) from the mode sums in (\ref{eq:Jtunrenstates}), we again employ the identity (\ref{eqn:Id1}) to give 
\begin{equation}
    -\frac{qA_{t}(r)}{2\pi^{2}f(r)\epsilon^{2}}=\int_{0}^{\infty}F^{\mathrm {CD}}_{{\mathrm {sing}}}(\omega,r)\cos{\left(\omega\epsilon\right)}\, d\omega,
    \label{eq:tderivsubtract}
\end{equation}
where we have defined 
\begin{equation}
    F^{{\mathrm {CD}}}_{{\mathrm {sing}}}(\omega,r)=\frac{qA_{t}(r)}{2\pi^{2}f(r)}\omega ,
\end{equation}
and ${\mathrm {CD}}$ denotes ``current derivative''. 
To demonstrate the procedure for computing the renormalized quantity (and thereby the charge current expectation value), as in Sec.~\ref{sec:SC}, we follow in detail an example, using the Boulware state, at a single radial point (whose value is the same as in Sec.~\ref{sec:SC}). 
The procedure works in the same way for the other states and other values of the radial coordinate. 
We also fix the black hole mass and charge and the scalar field charge to be the same as in Sec.~\ref{sec:SC}.

Subtracting (\ref{eq:tderivsubtract}) from the mode sum (\ref{eq:JtunrenB}) gives the renormalized expectation value
\begin{widetext}
\begin{multline}
    2\, \Im \langle {\mathrm{B}}|\{\hat{\Phi}(x),\left[\partial_{t}\hat{\Phi}(x)\right]^{\dagger}\}| {\mathrm{B}}\rangle
    =
    \int _{0}^{\infty } d\omega \, F^{{\mathrm {CD}}}_{{\mathrm {reg}}}(\omega,r)
    \\
    +\frac{q}{96\pi^{2}rf(r)}\Bigg[ 8q^{2}rA_{t}(r)^{3}-4A_{t}(r)f(r)f'(r)
    -2rA_{t}'(r)f(r)f'(r)+rA_{t}(r)f'(r)^{2} 
    -2rA_{t}(r)f(r)f''(r)\Bigg],
\end{multline}
\end{widetext}
where we have defined
\begin{equation}
    F^{{\mathrm {CD}}}_{{\mathrm {reg}}}(\omega,r)=F^{{\mathrm {CD}}}_{{\mathrm {total}}}(\omega,r)-F^{{\mathrm {CD}}}_{{\mathrm {sing}}}(\omega,r),
    \label{eq:Fregt'}
\end{equation}
with 
\begin{multline}
    F^{{\mathrm {CD}}}_{{\mathrm {total}}}(\omega,r)=\\
    \omega\left[ F^{{\mathrm {in}}}(\omega,r)-F^{{\mathrm {in}}}(-\omega,r)+F^{{\mathrm {up}}}(\omega,r)
    -F^{{\mathrm {up}}}(-\omega,r)\right] .
    \label{eq:Ftotalt'}
\end{multline}

\begin{figure}
    \centering
    \includegraphics[scale=0.65]{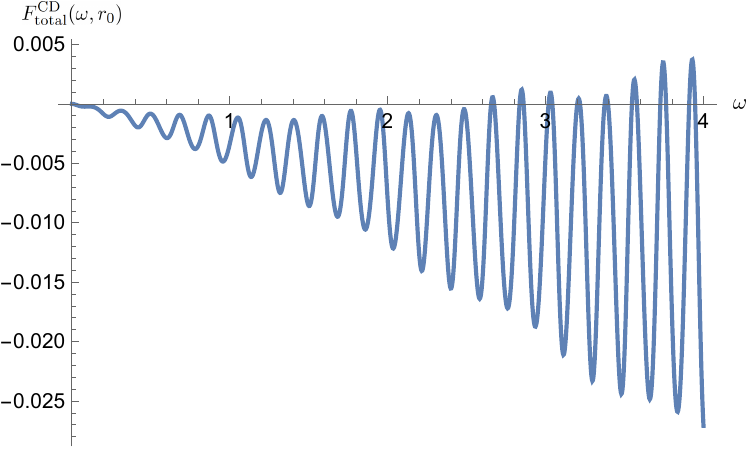}
    \caption{Mode-sum contribution $F^{{\mathrm {CD}}}_{{\mathrm {total}}}(\omega,r)$ (\ref{eq:Ftotalt'}) required for the renormalized charge current expectation value $\langle {\mathrm{B}} | {\hat {J}}_{t} | {\mathrm{B}} \rangle $ in the Boulware state, as a function of frequency $\omega $ at fixed radial coordinate $r_{0}\approx 5.96158M$. The black hole and scalar field parameters are as in Fig.~\ref{fig:FtotSC}.}
    \label{fig:FtotDt}
\end{figure}

\begin{figure}
    \centering
    \includegraphics[scale=0.65]{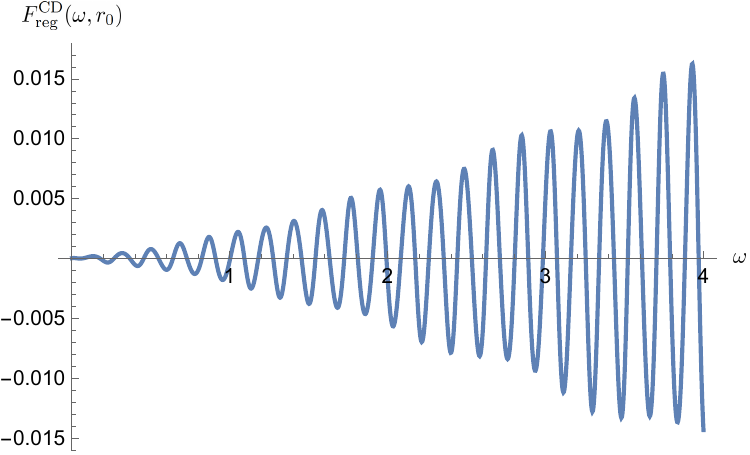}
    \caption{Regularized mode-sum contribution $F^{{\mathrm {CD}}}_{{\mathrm {reg}}}(\omega,r)$ (\ref{eq:Fregt'}) required for the renormalized charge current expectation value $\langle {\mathrm{B}} | {\hat {J}}_{t} | {\mathrm{B}} \rangle $ in the Boulware state, as a function of frequency $\omega $ at fixed radial coordinate $r_{0}\approx 5.96158M$. The black hole and scalar field parameters are as in Fig.~\ref{fig:FtotSC}.}
    \label{fig:FregDt}
\end{figure}

\begin{figure}
    \centering
    \includegraphics[scale=0.65]{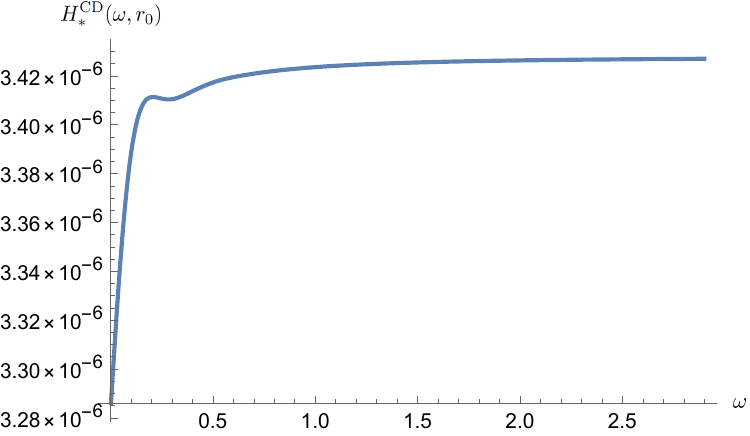}
    \caption{Self-cancellation generalized integral $H_{*}^{{\mathrm {CD}}}(\omega,r)$ (\ref{eq:Hstart'}, \ref{eq:Ht'}) required for the renormalized charge current expectation value $\langle {\mathrm{B}} | {\hat {J}}_{t} | {\mathrm{B}} \rangle $ in the Boulware state $|{\mathrm{B}}\rangle $, as a function of frequency $\omega $ at fixed radial coordinate $r_{0}\approx5.96158M$.
    The black hole and charged scalar field parameters are as in Fig.~\ref{fig:FtotSC}.}
    \label{fig:HstarDt}
\end{figure}

Fig.~\ref{fig:FtotDt} shows $F^{{\mathrm {CD}}}_{{\mathrm {total}}}(\omega,r)$ as a function of the frequency $\omega $, computed using the same method as in Sec.~\ref{sec:SC}. 
We see similar behaviour to the quantity $F^{{\mathrm {VP}}}_{{\mathrm {total}}}(\omega,r)$ (\ref{eq:FtotalSC}) appearing in the renormalized vacuum polarization (see Fig.~\ref{fig:FtotSC}), in particular $F^{{\mathrm {CD}}}_{{\mathrm {total}}}(\omega,r)$ diverges linearly as $\omega $ increases, with oscillations about this linear growth. 
As before, the linear growth is removed by subtracting $F^{{\mathrm {CD}}}_{{\mathrm {sing}}}(\omega,r)$, see Fig.~\ref{fig:FregDt} for the resulting quantity $F^{{\mathrm {CD}}}_{{\mathrm {reg}}}(\omega,r)$ for our example.
As expected, while the linear growth is now absent, there are oscillations in $F^{{\mathrm {CD}}}_{{\mathrm {reg}}}(\omega,r)$, whose amplitude increases as the frequency $\omega $ increases (cf.~Fig.~\ref{fig:FregSC}).

These oscillations are dealt with in exactly the same way as for the vacuum polarization, by a self-cancellation procedure, giving
\begin{widetext}
\begin{multline}
    2\, \Im \langle {\mathrm{B}}|\{\hat{\Phi}(x),\left[\partial_{t}\hat{\Phi}(x)\right]^{\dagger}\}| {\mathrm{B}}\rangle
    =
    \int _{0}^{\infty (\mathrm {sc})} d\omega \, F^{{\mathrm {CD}}}_{{\mathrm {reg}}}(\omega,r)
    \\
    +\frac{q}{96\pi^{2}rf(r)}\Bigg[ 8q^{2}rA_{t}(r)^{3}-4A_{t}(r)f(r)f'(r)
    -2rA_{t}'(r)f(r)f'(r)+rA_{t}(r)f'(r)^{2} 
    -2rA_{t}(r)f(r)f''(r)\Bigg],
\end{multline}
\end{widetext}
where,  in analogy with the vacuum polarization (\ref{eq:FSCSC}),  we have defined the self-cancellation integral
\begin{equation}
    \int _{0}^{\infty({\mathrm {sc}}) }  F^{{\mathrm {CD}}}_{{\mathrm {reg}}}(\omega,r) \, d\omega = \lim_{\omega\to\infty}H_{*}^{{\mathrm {CD}}}(\omega,r),
\end{equation}
with
\begin{equation}
    H_{*}^{{\mathrm {CD}}}(\omega,r):=(T_{\lambda_{1}})^{4}(T_{\lambda_{2}})^{4}(T_{\lambda_{3}})^{4}(T_{\lambda_{4}})^{4}\left[H^{{\mathrm {CD}}}(\omega,r)\right],
    \label{eq:Hstart'}
\end{equation}
and 
\begin{equation}
    H^{{\mathrm {CD}}}(\omega,r)=\int_{0}^{\omega}F^{{\mathrm {CD}}}_{{\mathrm {reg}}}(\omega',r) \, d\omega' ,
    \label{eq:Ht'}
\end{equation}
where the self-cancellation operator $T_{\lambda }[H(\omega )]$ is defined in (\ref{eq:selfcancelT}).
Fig.~\ref{fig:HstarDt} shows the result of performing the self-cancellation integral $H_{*}^{{\mathrm {CD}}}(\omega,r)$. Again we see that the oscillatory behaviour has been quashed, and the resultant function of the frequency $\omega $ is rapidly convergent as $\omega  \rightarrow \infty $. 

From this one can obtain the result for the renormalized expectation value of 
$2 \, \Im \langle\{{\hat {\Phi}}(x),\left[\partial_{t}{\hat {\Phi(x)}}\right]^{\dagger}\}\rangle $.
Using (\ref{eqn:currentDefn}), to find the renormalized expectation value of $\langle  {\hat {J}}_{t} \rangle $, we combine $2 \, \Im \langle\{{\hat {\Phi}}(x),\left[\partial_{t}{\hat {\Phi}}(x)\right]^{\dagger}\}\rangle $ with $-iqA_{t}/2$ times the renormalized vacuum polarization, computed in Sec.~\ref{sec:SC}.

The mode sums required for the radial component $\langle  {\hat {J}}_{r} \rangle $ (\ref{eq:Jrmode}) do not require renormalization and there are no subtleties involved in their direct computation.

\section{Numerical results}
\label{sec:results}

In this section we present our results for the renormalized expectation values of the vacuum polarization, and  the $t$ and $r$ components of the renormalized charged scalar current. 
We set $M=1$ and consider two values of the black hole charge, $Q=0.8M$ and $Q=0.99M$, the motivation for the latter is to allow us to investigate the effects close to the extremal limit. 
For $Q=0.8M$, the event horizon radius is $r_{+}=8M/5$ (\ref{eq:rpm}) and we consider the scalar field charges $qM\in\{0.32, 0.48, 0.64, 0.8\}$, whilst for $Q=0.99M$ we have $r_{+}\approx 1.141M$ and we consider $qM\in\{0.594, 0.792, 0.99\}$.

\begin{figure*}
     \centering
         \includegraphics[width=0.47\textwidth]{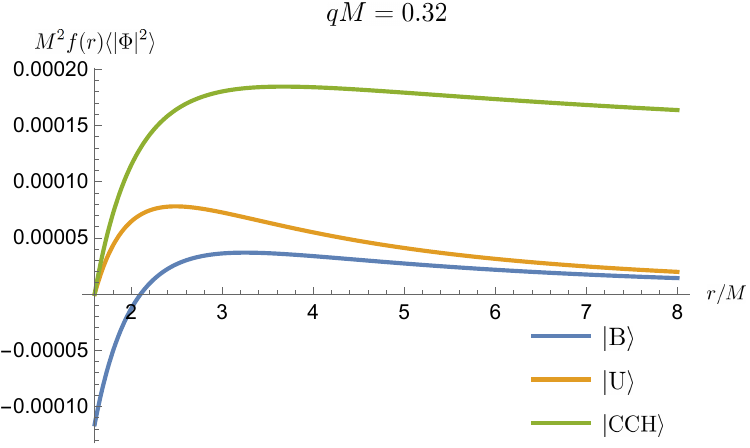}
         \includegraphics[width=0.47\textwidth]{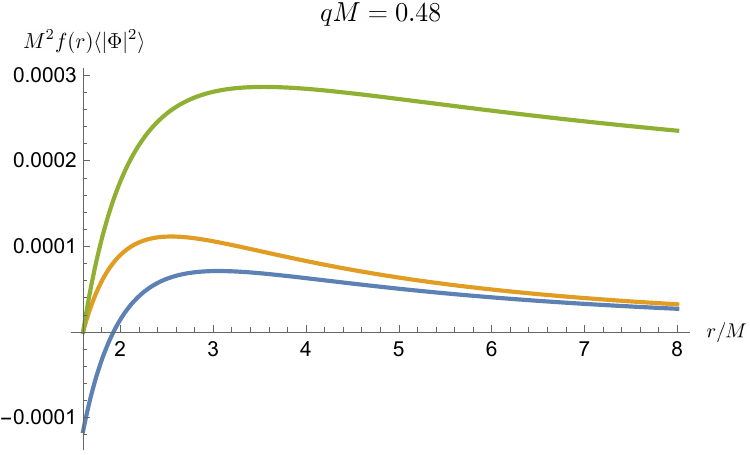}
         \includegraphics[width=0.47\textwidth]{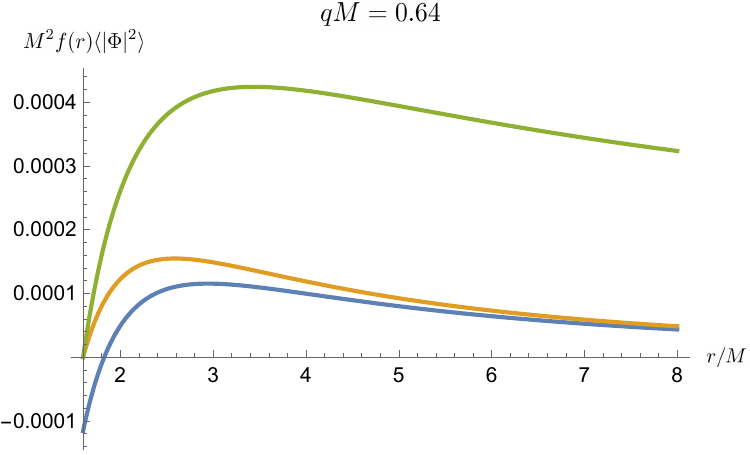}
         \includegraphics[width=0.47\textwidth]{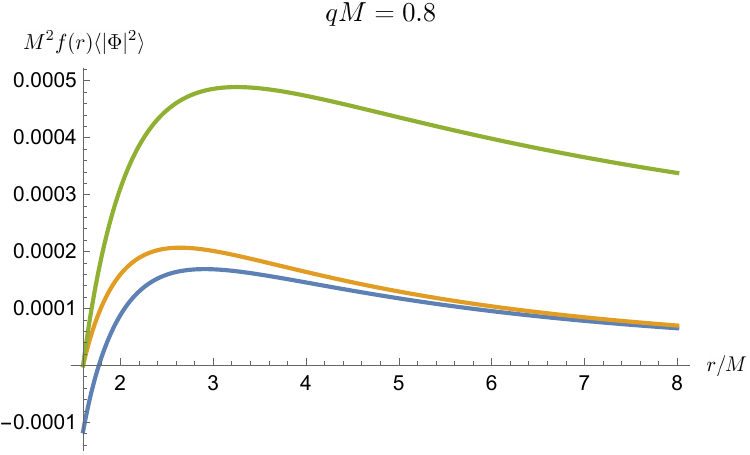}
        \caption{Renormalized vacuum polarization $\langle| {\hat {\Phi }}|^{2} \rangle $, multiplied by the metric function $f(r)$ (\ref{eq:frRN}) for a quantum charged scalar field on an RN black hole with $Q=0.8M$ and a selection of values of the scalar field charge $q$. 
        Three quantum states for the scalar field are considered, namely the Boulware $|{\mathrm{B}}\rangle $, Unruh $|{\mathrm{U}}\rangle $ and CCH $|{\mathrm{CCH}}\rangle $ states.}
        \label{fig:SC1}
\end{figure*}

\subsection{Radial modes}
\label{sec:radial}

We first outline our method for computing the scalar field modes. 
As discussed in \cite{Breen:2024ggu}, the radial equation (\ref{eq:radial}) can be transformed into the Confluent Heun equation \cite{Ronveaux:1995}.

The ``in'' radial functions $X^{\mathrm {in}}_{\omega \ell }(r)/r$ may be written in terms of confluent Heun functions which are built-in to {\tt {Mathematica}}.
Denoting these by ${\widetilde {X}}^{\mathrm {in}}_{\omega \ell }(r)/r$ since they do not (yet) have the required normalization, these are
\begin{multline}
    \frac{1}{r} {\widetilde{X}}^{{\mathrm {in}}}_{\omega\ell}(r) = 
    e^{-i{\widetilde {\omega}}r_{*}}
   e^{
    -\frac{iqQ(r-r_{+})}{r_{+}}} 
    \left(\frac{r-r_{-}}{r_{-}-r_{+}}\right)^{-\frac{iqQr_{-}}{r_{+}}}
    \\
    \times 
    \text{HeunC}\left(a_1,a_2,a_3,a_4,a_5, \frac{r-r_{+}}{r_{-}-r_{+}}\right),
    \label{eq:InHeun}
\end{multline}
where $\textsc{HeunC}\left(...\right)$ denotes the confluent Heun function satisfying the equation
\begin{multline}
    \Big[ z(z-1)\frac{d^{2}}{dz^{2}}+\left\{ a_{3}(z-1)+a_{4}z+z(z-1)a_{5}\right\} \frac{d}{dz} 
    \\
    +\left(a_{2}z-a_{1}\right) \Big]\textsc{HeunC}\left(a_{1},a_{2},a_{3},a_{4},a_{5};z\right) =0,
\end{multline}
with $z=(r-r_{+})/(r_{-}-r_{+})$, 
such that 
\begin{equation}
\textsc{HeunC}\left( a_{1},a_{2},a_{3},a_{4},a_{5};0\right)=1.
\end{equation} 
The constants in (\ref{eq:InHeun}) are given by
\begin{subequations}
    \begin{align}
    a_{1}&  = \ell(\ell+1)-iqQ+2ir_{+}\omega
    ,
    \\
    a_{2} &  = 2i\omega(r_{+}-r_{-})
    ,
    \\
    a_{3} & = \frac{r_{-}+ir_{+}(i-2qQ+2r_{+}\omega)}{r_{-}-r_{+}}
    ,
    \\
    a_{4}  & =\frac{r_{-}-r_{+}+2iqQr_{-}-2ir_{-}^{2}\omega}{r_{-}-r_{+}}
    ,
    \\
    a_{5} &  =2i\omega(r_{+}-r_{-})
    .
\end{align}
\end{subequations}

The ``up'' radial functions ${\widetilde {X}}^{\mathrm {up}}(r)/r$ (again, these do not yet have the required normalization) are obtained by numerically integrating the radial equation (\ref{eq:radial}) via a procedure analogous to that in Ref.~\cite{Breen:2024ggu}. 
We modify the numerical integration {\tt {Mathematica}} notebook of the
{\tt {ReggeWheeler}} package of the Black Hole Perturbation Toolkit \cite{BHPToolkit}, in which the integration is performed using the {\tt {Mathematica}} function {\tt {NDSolve}}.  
The radial equation is integrated inwards from large $r$, and at large $r$ the solution of the radial equation is 
approximated by an expansion of the form 
\begin{equation}
    \frac{1}{r}\widetilde{X}^{{\mathrm {up}}}_{\omega\ell}(r) = \frac{e^{i\omega r_{*}}}{r^{1+iqQ}}\sum_{k}a_{k}r^{-k},
\end{equation}
including sufficient terms in the sum that the solution has converged to the required precision.

To normalize the radial functions as required in (\ref{eq:inupmodes}), we follow the procedure employed in \cite{Arrechea:2023fas}.
We match the asymptotic behaviour of the numerically generated solutions  $\widetilde{X}^{{\mathrm {in/up}}}_{\omega\ell}(r) $, with those given in (\ref{eq:inupmodes}) in terms of reflection and transmission coefficients. 
We obtain 
\begin{subequations}
\begin{align}
    X^{{\mathrm {in}}}_{\omega\ell}(r) &  = \frac{B^{{\mathrm {in}}}_{\omega\ell}}{r_{+}}\left(\frac{r_{+}-r_{-}}{r_{-}-r_{+}}\right)^{\frac{iqQr_{-}}{r_{+}}}{\widetilde{X}}^{{\mathrm {in}}}_{\omega\ell}(r), 
\\
    X^{{\mathrm {up}}}_{\omega\ell}(r) & = B^{{\mathrm {up}}}_{\omega\ell}\widetilde{X}^{{\mathrm {up}}}_{\omega\ell}(r).
\end{align}
\end{subequations}
The transmission coefficients $B^{{\mathrm {in/up}}}_{\omega \ell }$ are found using the Wronskian ${\mathcal{W}_{\omega\ell}}$: 
\begin{equation}
    B^{{\mathrm {in}}}_{\omega\ell}=-\frac{2i\omega r_{+}}{\mathcal{W}_{\omega\ell}}\left(\frac{r_{+}-r_{-}}{r_{-}-r_{+}}\right)^{-\frac{iqQr_{-}}{r_{+}}},
\end{equation}
where  
\begin{equation}
    {\mathcal{W}_{\omega\ell}} = \widetilde{X}^{\mathrm {up}}_{\omega \ell }(r) \frac{d}{dr_{*}}\widetilde{X}^{\mathrm {in}}_{\omega \ell }(r)
    - \widetilde{X}^{\mathrm {in}}_{\omega \ell }(r) \frac{d}{dr_{*}}\widetilde{X}^{\mathrm {up}}_{\omega \ell }(r) ,
\end{equation}
together with the relations (\ref{eqn:coeffsRels}).

The modes are numerically calculated using a working precision of 64 digits, over the frequency grid $|\omega|\in[0,4]$, with spacing of $\delta\omega=1/300$.
We find modes for $0\le \ell \le \ell_{\text{max}}$, where $\ell _{\text{max}}$ is the value of $\ell $ beyond which  the mode contributions to the sums required in Sec.~\ref{sec:expvals} are $<10^{-10}$, following the example of \cite{Levi:2015eea}.

\subsection{Renormalized vacuum polarization}
\label{sec:SCres}

\begin{figure}
     \centering
         \includegraphics[width=0.47\textwidth]{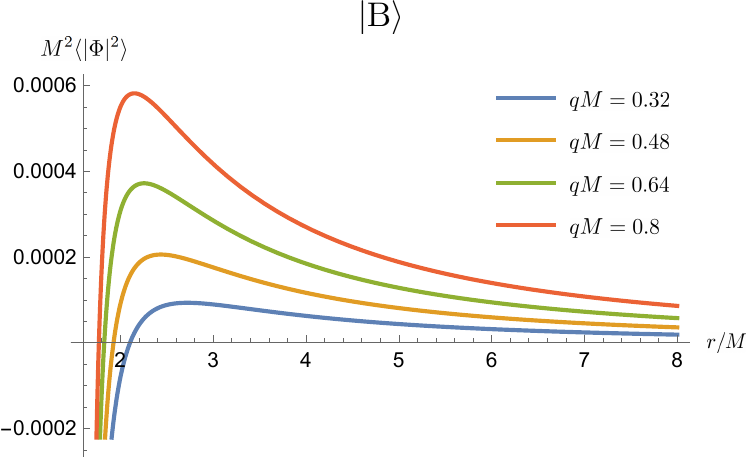}
         \includegraphics[width=0.47\textwidth]{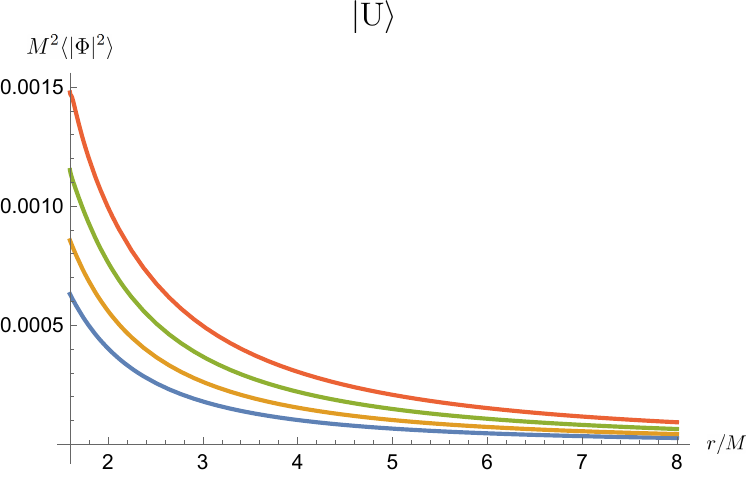} 
         \includegraphics[width=0.47\textwidth]{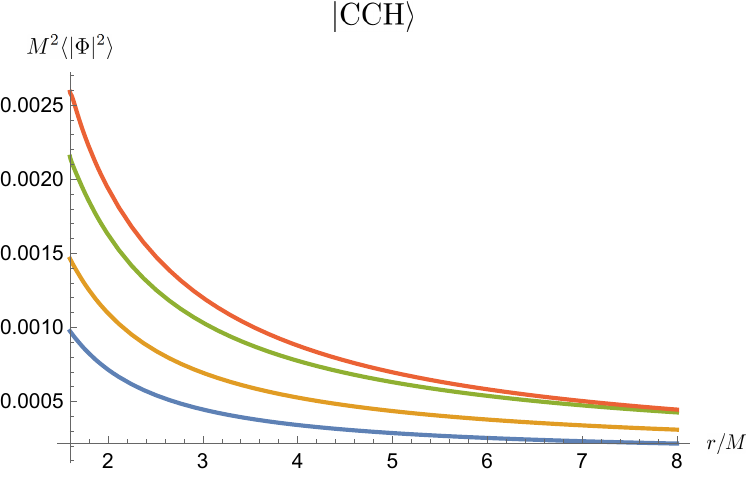}
        \caption{Renormalized vacuum polarization $\langle |{\hat {\Phi }}|^{2} \rangle $ for a quantum charged scalar field on an RN black hole with $Q=0.8M$ and a selection of values of the scalar field charge $q$. 
        Three quantum states for the scalar field are considered, namely the Boulware $|{\mathrm{B}}\rangle $, Unruh $|{\mathrm{U}}\rangle $ and CCH $|{\mathrm{CCH}}\rangle $ states.}
        \label{fig:SC2}
\end{figure}

\begin{figure*}
\centering
         \includegraphics[width=0.47\textwidth]{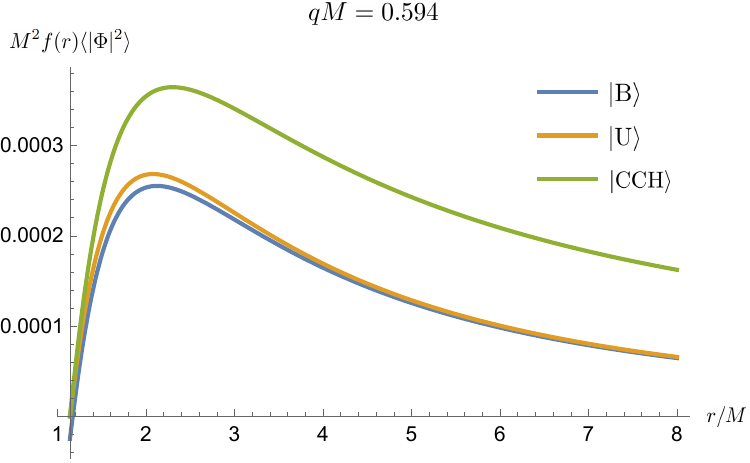}
         \includegraphics[width=0.47\textwidth]{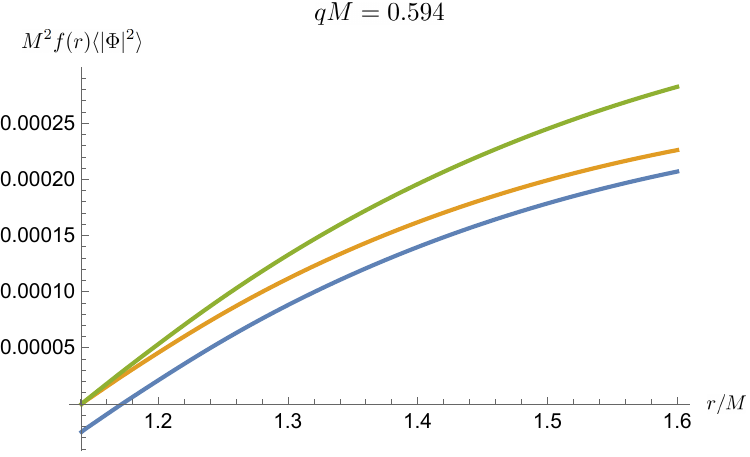}
         \includegraphics[width=0.47\textwidth]{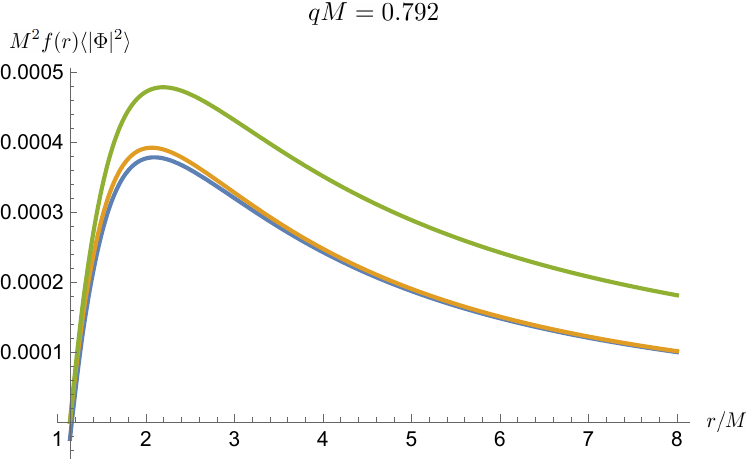}
         \includegraphics[width=0.47\textwidth]{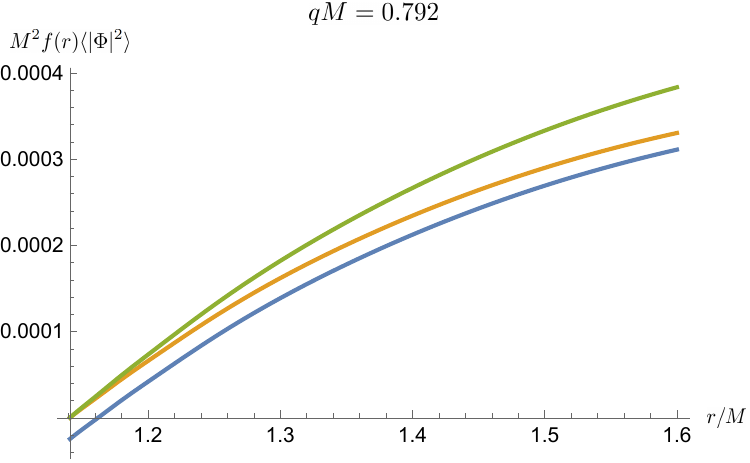}
         \includegraphics[width=0.47\textwidth]{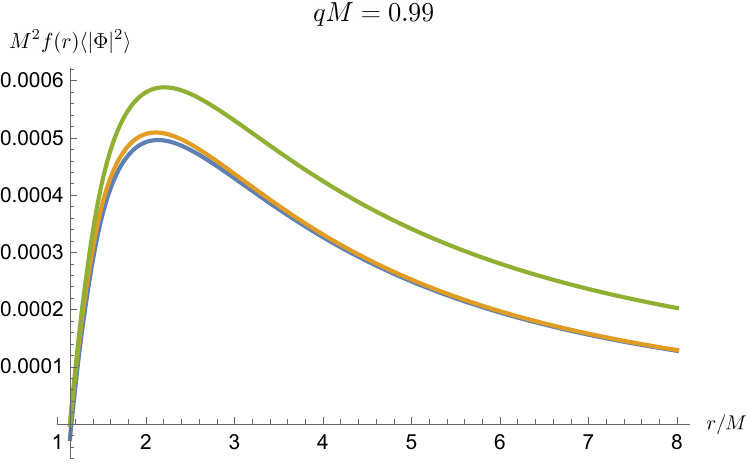}
         \includegraphics[width=0.47\textwidth]{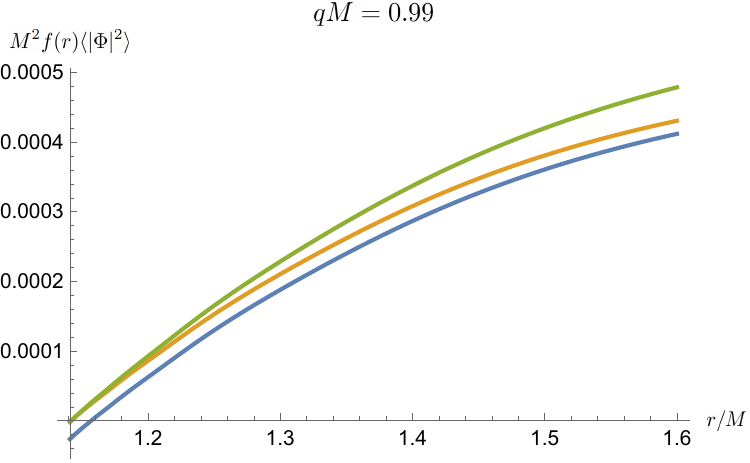}
        \caption{Renormalized vacuum polarization $\langle |{\hat {\Phi }}|^{2} \rangle $ for a quantum charged scalar field on an RN black hole with $Q=0.99M$ and a selection of values of the scalar field charge $q$. 
        Three quantum states for the scalar field are considered, namely the Boulware $|{\mathrm{B}}\rangle $, Unruh $|{\mathrm{U}}\rangle $ and CCH $|{\mathrm{CCH}}\rangle $ states.
        The plots on the right-hand-side show the region close to the event horizon.}
        \label{fig:SC3}
\end{figure*}

\begin{figure}
     \centering
         \includegraphics[width=0.47\textwidth]{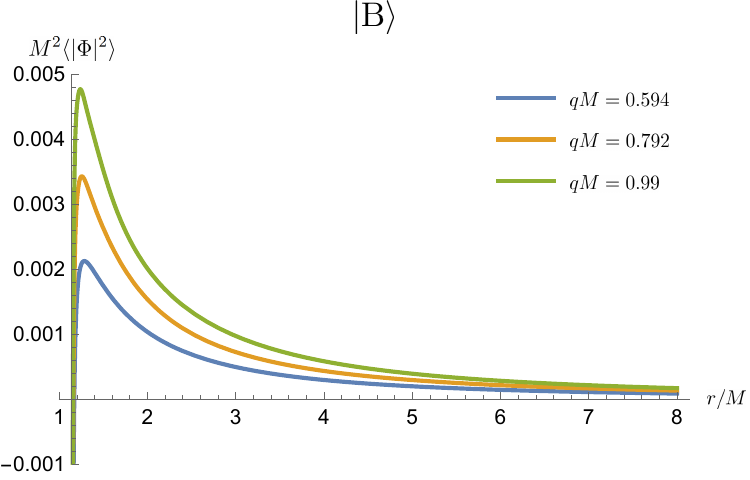}
         \includegraphics[width=0.47\textwidth]{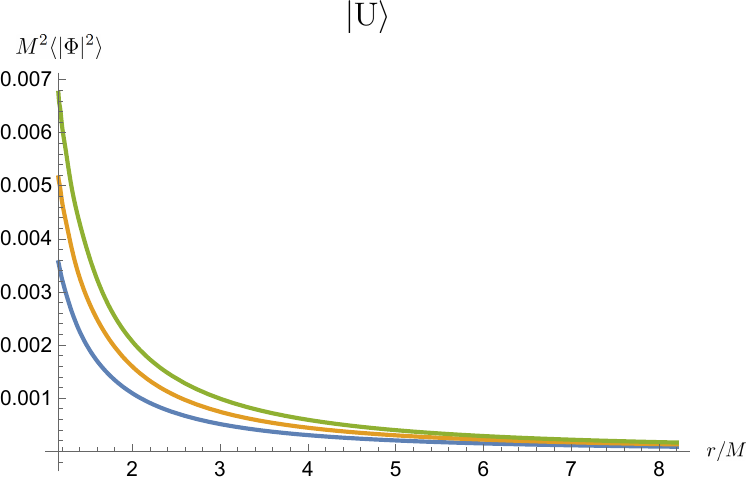}
         \includegraphics[width=0.47\textwidth]{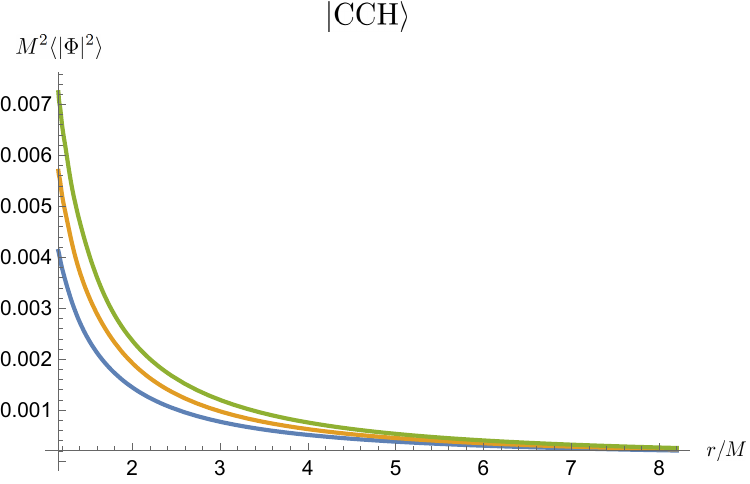}
        \caption{Renormalized vacuum polarization $\langle |{\hat {\Phi }}|^{2} \rangle $ for a quantum charged scalar field on an RN black hole with $Q=0.99M$ and a selection of values of the scalar field charge $q$. 
        Three quantum states for the scalar field are considered, namely the Boulware $|{\mathrm{B}}\rangle $, Unruh $|{\mathrm{U}}\rangle $ and CCH $|{\mathrm{CCH}}\rangle $ states.}
        \label{fig:SC4}
\end{figure}

We present our numerical results for the renormalized vacuum polarization $\langle |{\hat {\Phi }}|^{2} \rangle $ in Figs.~\ref{fig:SC1}--\ref{fig:SC4}, for the black hole and charged scalar field parameters given at the start of Sec.~\ref{sec:results}.

We begin, in Figs.~\ref{fig:SC1}--\ref{fig:SC2}, by presenting $\langle {|\hat {\Phi }}|^{2} \rangle $ in the three quantum states $|{\mathrm{B}}\rangle $, $|{\mathrm{U}}\rangle $  and $|{\mathrm{CCH}}\rangle $, when the black hole charge is $Q=0.8M$.
Fig.~\ref{fig:SC1} shows how the renormalized vacuum polarization depends on the quantum state for fixed scalar field charge, while Fig.~\ref{fig:SC2} shows the effect of varying the scalar field charge on the expectation value in a particular state.

In Fig.~\ref{fig:SC1} we have multiplied the vacuum polarization by the metric function $f(r)$ (\ref{eq:frRN}), so that we are plotting quantities which are regular on the horizon for all three quantum states.
We see that, in the Unruh and CCH states, the quantity $f(r)\langle {|\hat {\Phi }}|^{2} \rangle $ vanishes on the horizon, while for the Boulware state it takes a nonzero, finite, value. 
This implies that the vacuum polarization diverges on the horizon for the Boulware state but is regular on the horizon in both the Unruh and CCH states, as can be seen in Fig.~\ref{fig:SC2}.
Far from the black hole, Fig.~\ref{fig:SC1} shows that the values of the vacuum polarization in the Boulware and Unruh states are very similar (and the difference between them at fixed radial coordinate $r$ decreases as $q$ increases), while the vacuum polarization in the CCH state takes a very different (and much larger) value. 
This suggests that the dominant contribution to the vacuum polarization in the Unruh state, far from the black hole, is coming from the superradiant modes, while, for the CCH state, the contribution to the vacuum polarization from the ``in'' modes is $\sim 4$ times larger than the contribution from the ``up'' modes.

From Fig.~\ref{fig:SC2}, we see that the renormalized vacuum polarization in both the Unruh and CCH states is positive everywhere on and outside the event horizon, and is a monotonically decreasing function of the radial coordinate $r$.  
The values of the vacuum polarization are also monotonically increasing for fixed $r$ as the scalar field charge $q$ increases. 
In contrast, for the Boulware state the vacuum polarization is negative close to the horizon (and diverges on the horizon).
Near the horizon, it is monotonically increasing, and takes positive values for sufficiently large radial coordinate $r$. 
We find that the vacuum polarization has a maximum outside the horizon, and then is monotonically decreasing as $r$ increases.

Far from the event horizon, the vacuum polarization in the Boulware and Unruh states is tending to zero.
However, for the CCH state, while the vacuum polarization is decreasing as $r$ increases, it appears to be tending to a nonzero constant as $r\rightarrow \infty $. 
This indicates that the CCH state is not a vacuum state far from the black hole, as expected, since it contains an incoming flux of radiation. 

Many of these qualitative features are preserved when we consider a larger value of the black hole charge, namely $Q=0.99M$, in Figs.~\ref{fig:SC3}--\ref{fig:SC4}.
In Fig.~\ref{fig:SC3} we present $f(r)\langle |{\hat {\Phi }}|^{2} \rangle $ for the three states under consideration and a selection of values of the scalar field charge $q$.
The plots on the left-hand-side show all values of the radial coordinate $r$ for which we have performed our computations, while the plots on the right-hand-side show just the region close to the event horizon.
From the latter plots (and those in Fig.~\ref{fig:SC4}), we can see that the renormalized vacuum polarization in the Unruh and CCH states is regular on the event horizon, but that in the Boulware state diverges like $f(r)^{-1}$ as the horizon is approached.
The expectation values of the vacuum polarization in the Unruh and Boulware states are, from Fig.~\ref{fig:SC3}, virtually indistinguishable on the scale plotted, except for a region close to the horizon, while, as in Fig.~\ref{fig:SC1}, the expectation value in the CCH state is significantly larger far from the black hole.

As in Fig.~\ref{fig:SC2}, in Fig.~\ref{fig:SC4} we see that increasing the scalar field charge $q$ increases the expectation value of the vacuum polarization for fixed radial coordinate $r$.
Finally, comparing the results in Figs.~\ref{fig:SC2} and \ref{fig:SC4}, it can be seen that increasing the black hole charge also results in an increase in the vacuum polarization. 

The qualitative features of the vacuum polarization profiles shown in Figs.~\ref{fig:SC1}--\ref{fig:SC4} are the same as those found for a neutral scalar field on a neutral Schwarzschild black hole (see, for example, \cite{Levi:2016esr}).
In that situation, the vacuum polarization is monotonically decreasing in both the Unruh and Hartle-Hawking states, and is regular on the event horizon, while the vacuum polarization profile in the Boulware state also has a similar shape and diverges on the horizon.

In the extremal limit, $Q\rightarrow M$, the temperature $T=\kappa /2\pi $ (where $\kappa $ is the surface gravity (\ref{eq:kappa})) of the black hole vanishes. 
While the Hawking radiation switches off in this limit, the spontaneous emission in the superradiant modes does not.
In this limit the three states, Boulware, Unruh and CCH, become the same (as can be seen from the unrenormalized mode sums (\ref{eq:SCunrenstates})). 
Even though we have used a large value of $Q=0.99M$ in Figs.~\ref{fig:SC3} and \ref{fig:SC4}, while it is clear that the Unruh and Boulware states are nearly identical, the vacuum polarization for the CCH state is still significantly larger than that for the other two states, although the difference between the vacuum polarizations in the CCH and Unruh states is smaller for $Q=0.99M$ in Fig.~\ref{fig:SC3} than for $Q=0.8M$ in Fig.~\ref{fig:SC1}.
It seems that one has to be even closer to the extremal limit in order for the CCH state to become indistinguishable from the Unruh state. 

\subsection{Renormalized charged scalar current}
\label{sec:currentres}

\begin{figure*}
     \centering
         \includegraphics[width=0.47\textwidth]{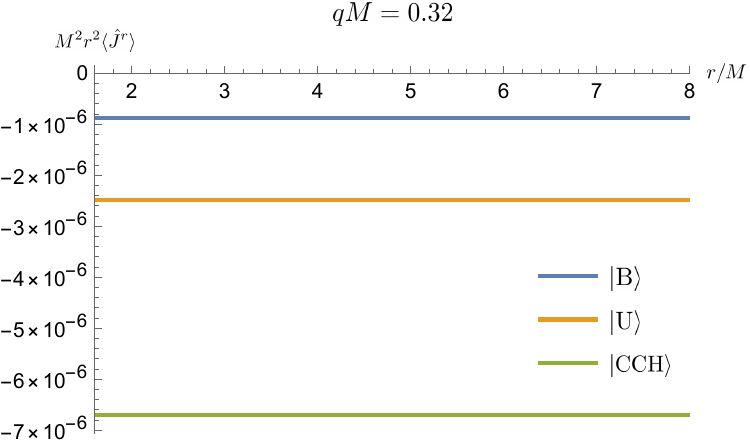}
         \includegraphics[width=0.47\textwidth]{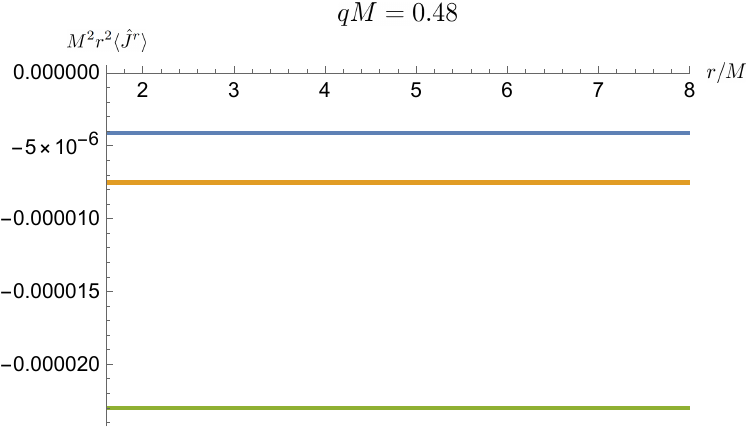}
         \includegraphics[width=0.47\textwidth]{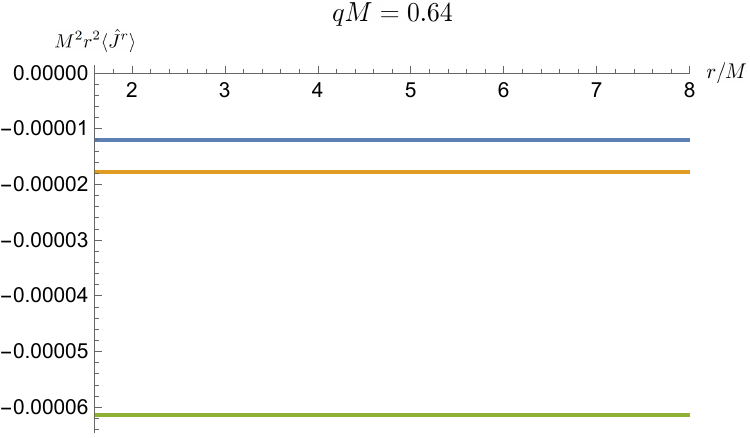}
         \includegraphics[width=0.47\textwidth]{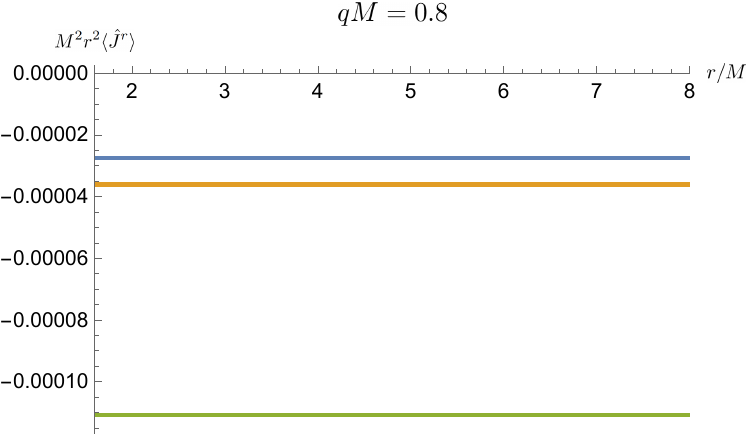}
        \caption{Radial component of the renormalized charge current $\langle {\hat {J}}^{r} \rangle $ multiplied by $r^{2}$, for a quantum charged scalar field on an RN black hole with $Q=0.8M$ and a selection of values of the scalar field charge $q$. 
        Three quantum states for the scalar field are considered, namely the Boulware $|{\mathrm{B}}\rangle $, Unruh $|{\mathrm{U}}\rangle $ and CCH $|{\mathrm{CCH}}\rangle $ states.}
        \label{fig:Jr1}
\end{figure*}

\begin{figure}
     \centering
         \includegraphics[width=0.47\textwidth]{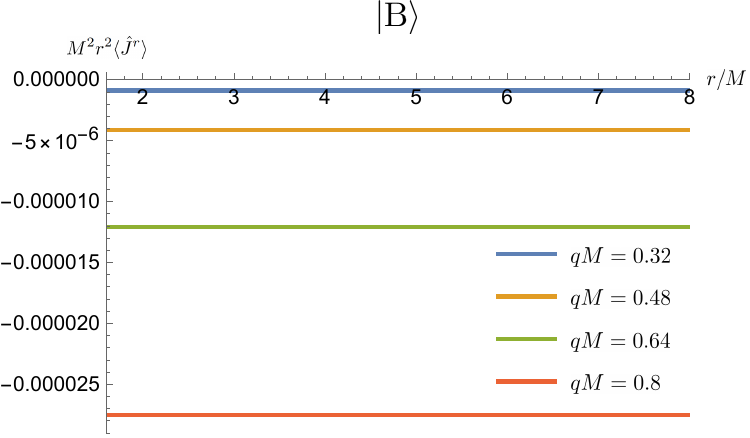}
         \includegraphics[width=0.47\textwidth]{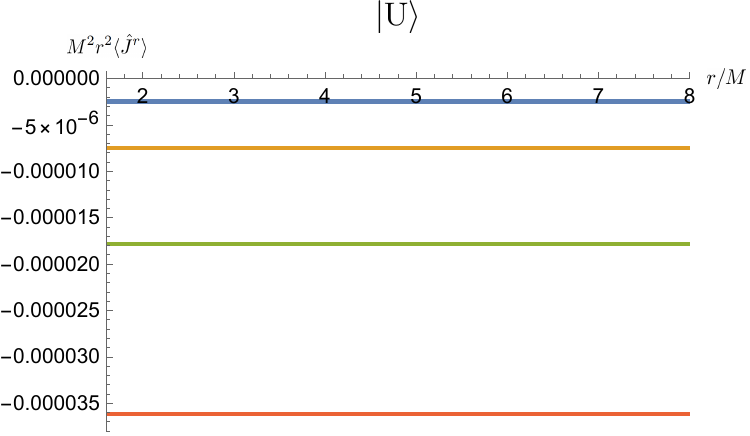}
         \includegraphics[width=0.47\textwidth]{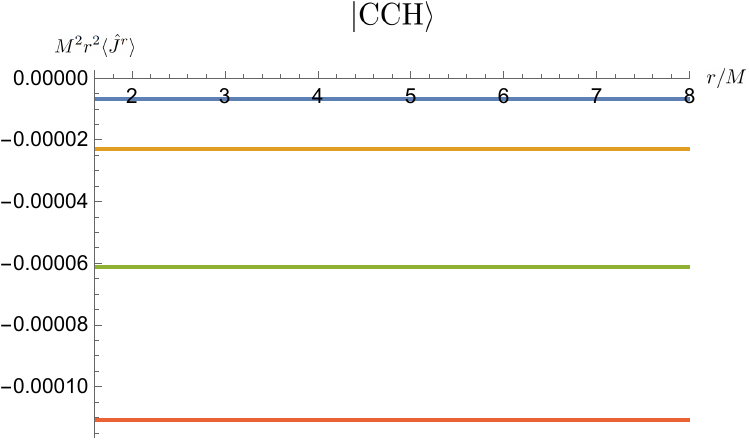}
        \caption{Radial component of the renormalized charge current $r^{2}\langle {\hat {J}}^{r} \rangle $ multiplied by $r^{2}$, for a quantum charged scalar field on an RN black hole with $Q=0.8M$ and a selection of values of the scalar field charge $q$. 
        Three quantum states for the scalar field are considered, namely the Boulware $|{\mathrm{B}}\rangle $, Unruh $|{\mathrm{U}}\rangle $ and CCH $|{\mathrm{CCH}}\rangle $ states.}
        \label{fig:Jr2}
\end{figure}

\begin{figure}
     \centering
         \includegraphics[width=0.47\textwidth]{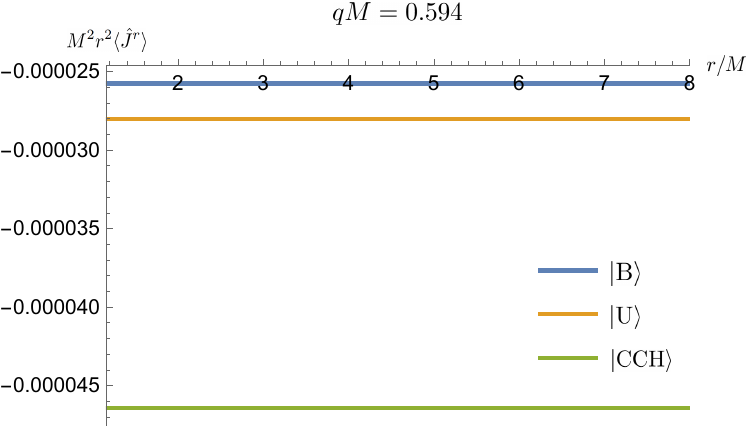}
         \includegraphics[width=0.47\textwidth]{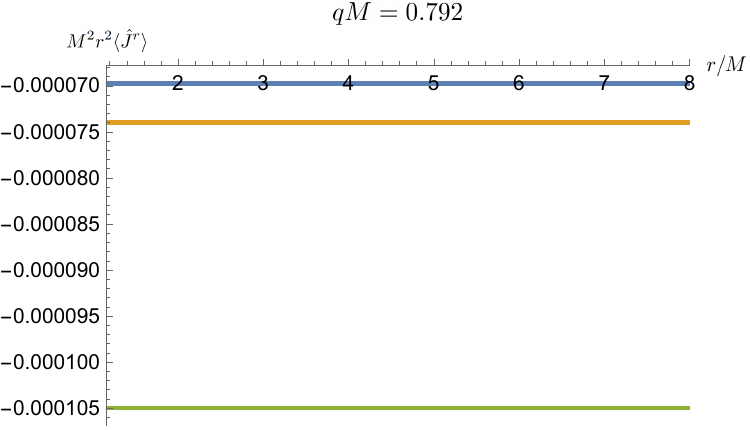}
         \includegraphics[width=0.47\textwidth]{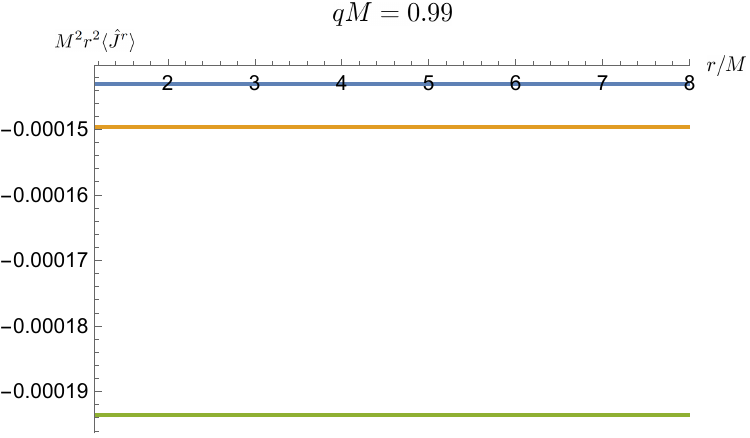}
        \caption{Radial component of the renormalized charge current $\langle {\hat {J}}^{r} \rangle $ multiplied by $r^{2}$, for a quantum charged scalar field on an RN black hole with $Q=0.99M$ and a selection of values of the scalar field charge $q$. 
        Three quantum states for the scalar field are considered, namely the Boulware $|{\mathrm{B}}\rangle $, Unruh $|{\mathrm{U}}\rangle $ and CCH $|{\mathrm{CCH}}\rangle $ states.}
        \label{fig:Jr3}
\end{figure}

\begin{figure}
     \centering
         \includegraphics[width=0.47\textwidth]{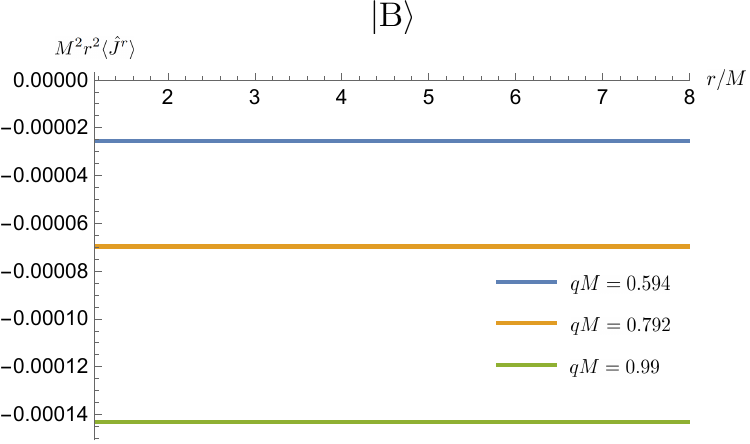}
         \includegraphics[width=0.47\textwidth]{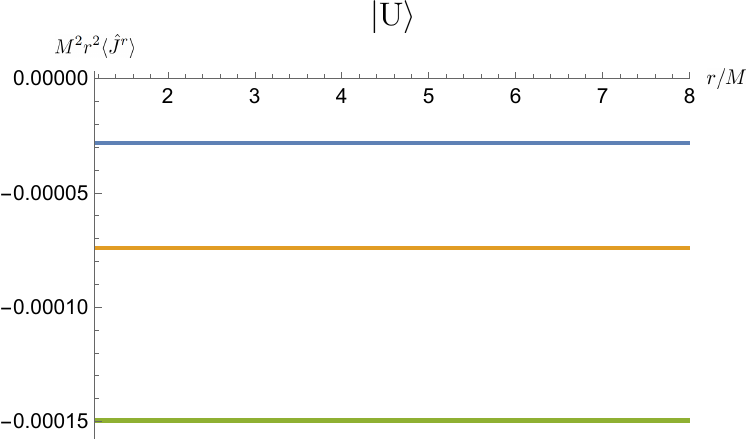}
         \includegraphics[width=0.47\textwidth]{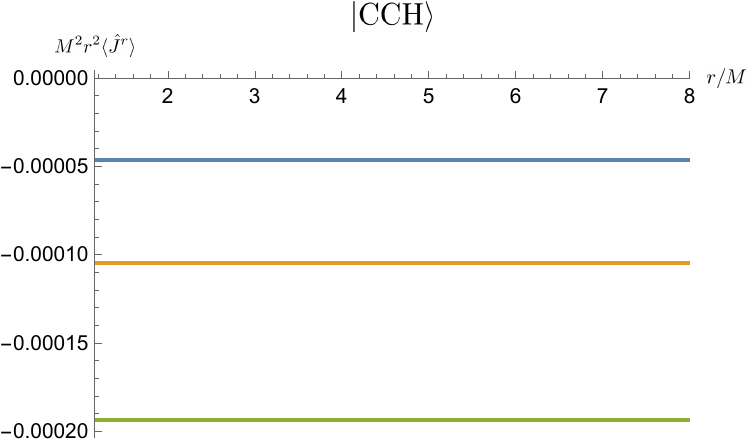}
        \caption{Radial component of the renormalized charge current $\langle {\hat {J}}^{r} \rangle $ multiplied by $r^{2}$, for a quantum charged scalar field on an RN black hole with $Q=0.99M$ and a selection of values of the scalar field charge $q$. 
        Three quantum states for the scalar field are considered, namely the Boulware $|{\mathrm{B}}\rangle $, Unruh $|{\mathrm{U}}\rangle $ and CCH $|{\mathrm{CCH}}\rangle $ states.}
        \label{fig:Jr4}
\end{figure}

\begin{figure*}
     \centering
         \includegraphics[width=0.47\textwidth]{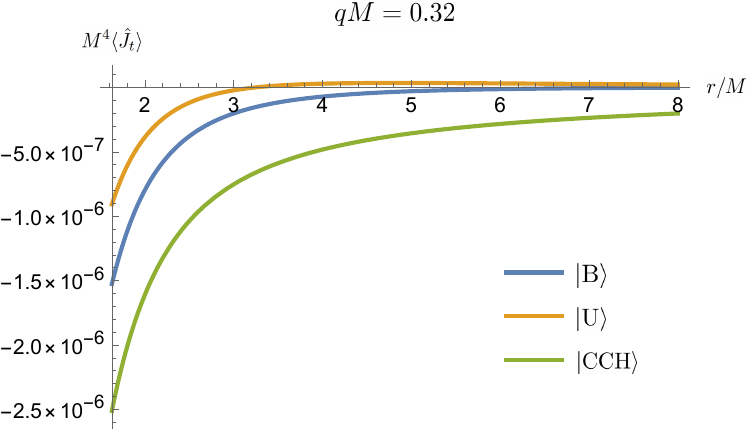}
         \includegraphics[width=0.47\textwidth]{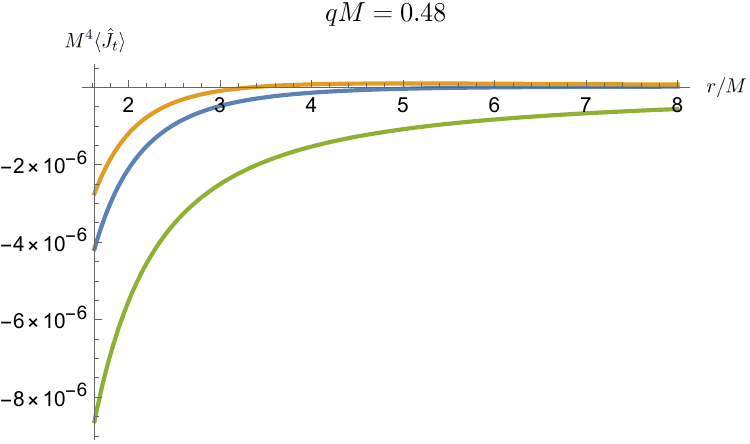}
         \includegraphics[width=0.47\textwidth]{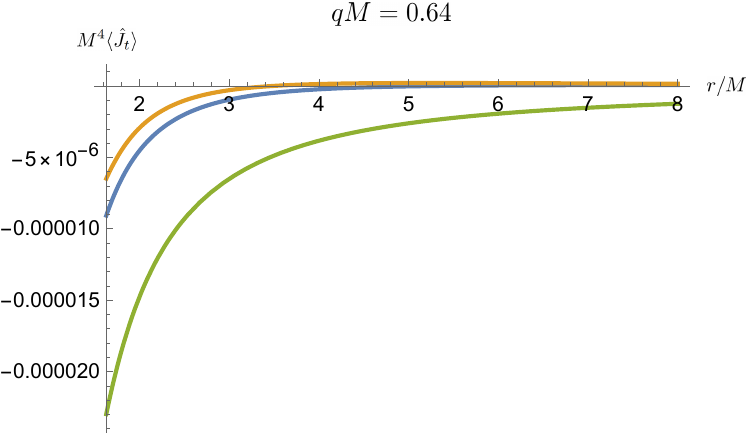}
         \includegraphics[width=0.47\textwidth]{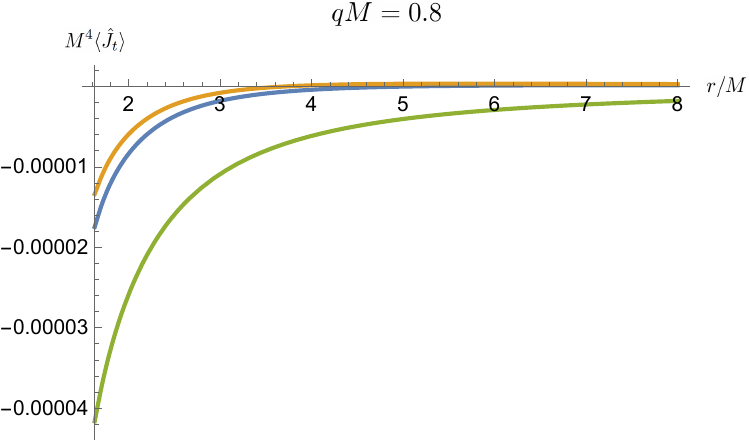}
        \caption{Time component of the renormalized charge current $\langle {\hat {J}}_{t} \rangle $, for a quantum charged scalar field on an RN black hole with $Q=0.8M$ and a selection of values of the scalar field charge $q$. 
        Three quantum states for the scalar field are considered, namely the Boulware $|{\mathrm{B}}\rangle $, Unruh $|{\mathrm{U}}\rangle $ and CCH $|{\mathrm{CCH}}\rangle $ states.}
        \label{fig:Jt1}
\end{figure*}

\begin{figure}
     \centering
         \includegraphics[width=0.47\textwidth]{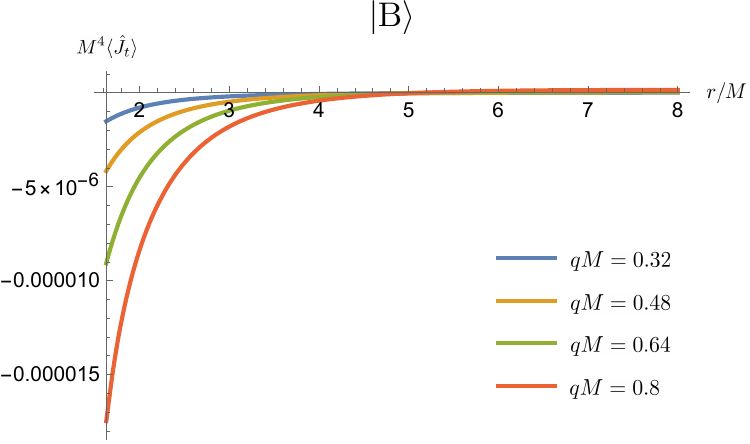}
         \includegraphics[width=0.47\textwidth]{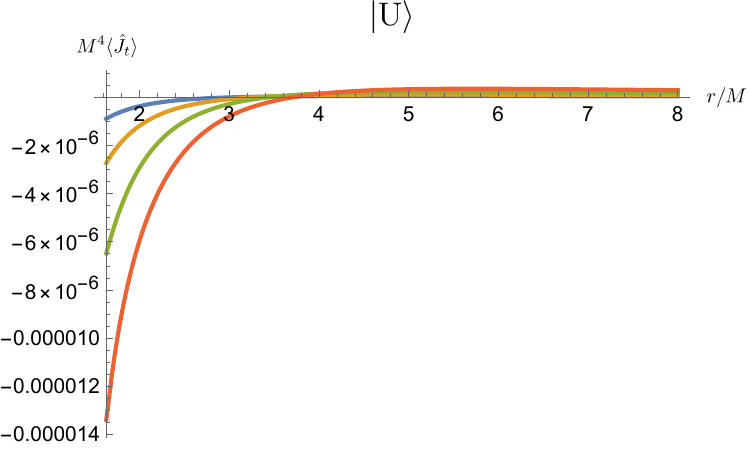}
         \includegraphics[width=0.47\textwidth]{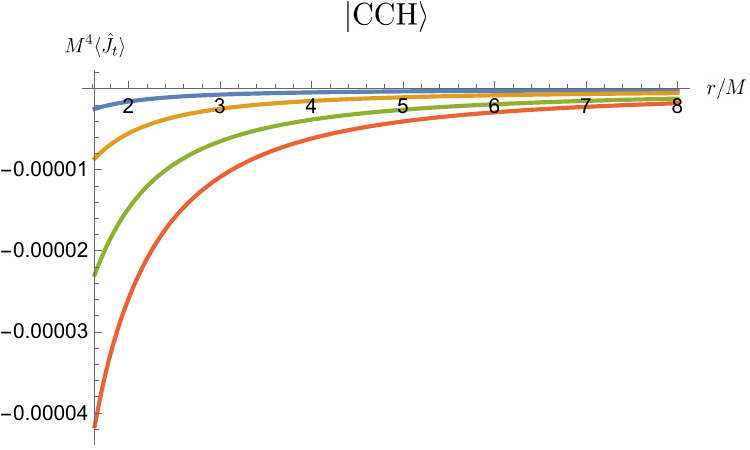}
        \caption{Time component of the renormalized charge current $\langle {\hat {J}}_{t} \rangle $, for a quantum charged scalar field on an RN black hole with $Q=0.8M$ and a selection of values of the scalar field charge $q$. 
        Three quantum states for the scalar field are considered, namely the Boulware $|{\mathrm{B}}\rangle $, Unruh $|{\mathrm{U}}\rangle $ and CCH $|{\mathrm{CCH}}\rangle $ states.}
        \label{fig:Jt2}
\end{figure}

\begin{figure}
     \centering
         \includegraphics[width=0.47\textwidth]{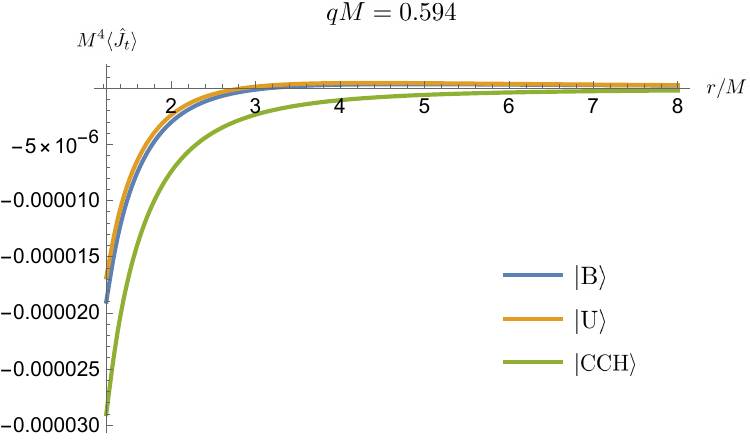}
         \includegraphics[width=0.47\textwidth]{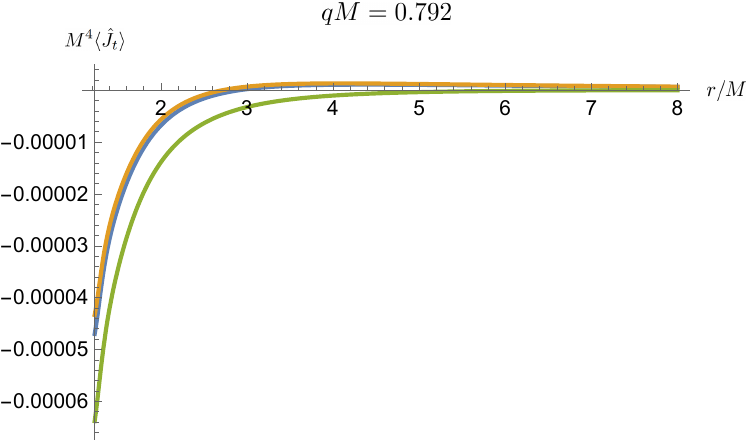}
         \includegraphics[width=0.47\textwidth]{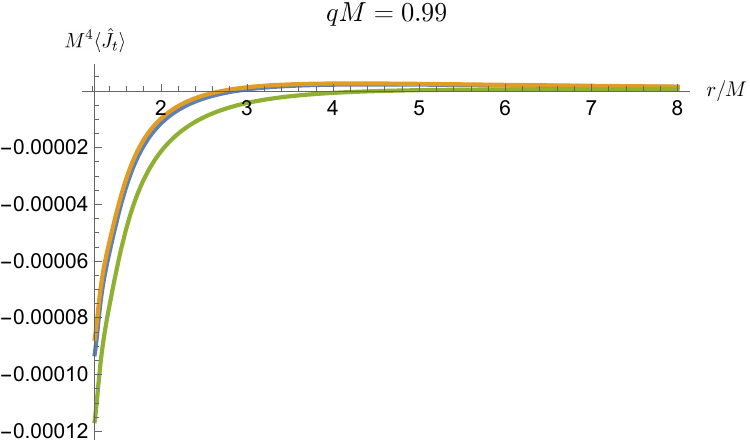}
        \caption{Time component of the renormalized charge current $\langle {\hat {J}}_{t} \rangle $, for a quantum charged scalar field on an RN black hole with $Q=0.99M$ and a selection of values of the scalar field charge $q$. 
        Three quantum states for the scalar field are considered, namely the Boulware $|{\mathrm{B}}\rangle $, Unruh $|{\mathrm{U}}\rangle $ and CCH $|{\mathrm{CCH}}\rangle $ states.}
        \label{fig:Jt3}
\end{figure}

\begin{figure}
     \centering
         \includegraphics[width=0.47\textwidth]{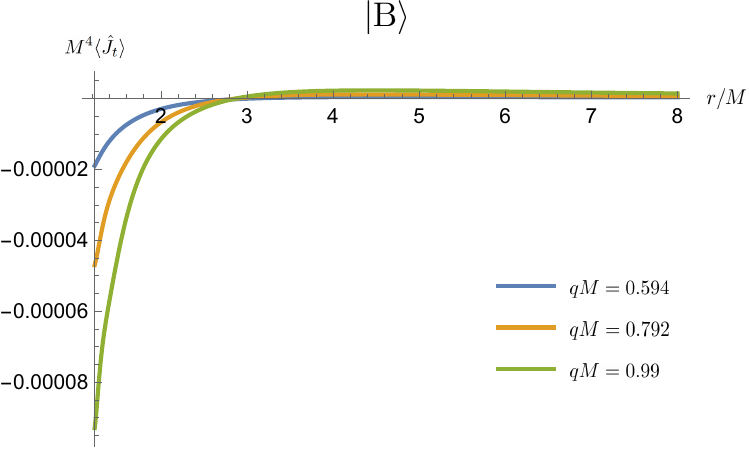}
         \includegraphics[width=0.47\textwidth]{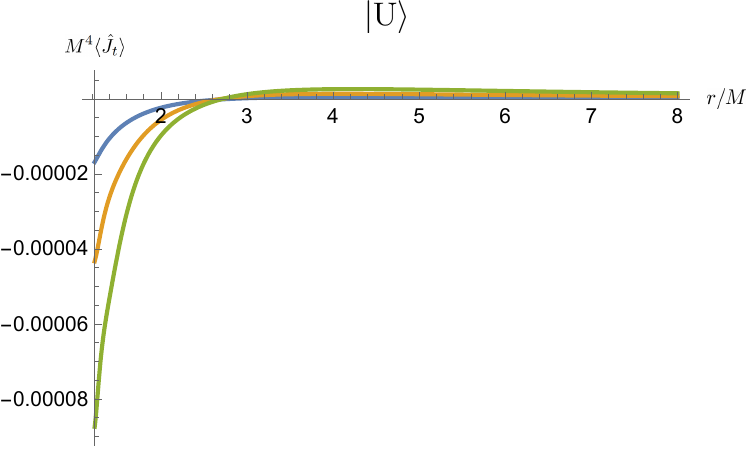}
         \includegraphics[width=0.47\textwidth]{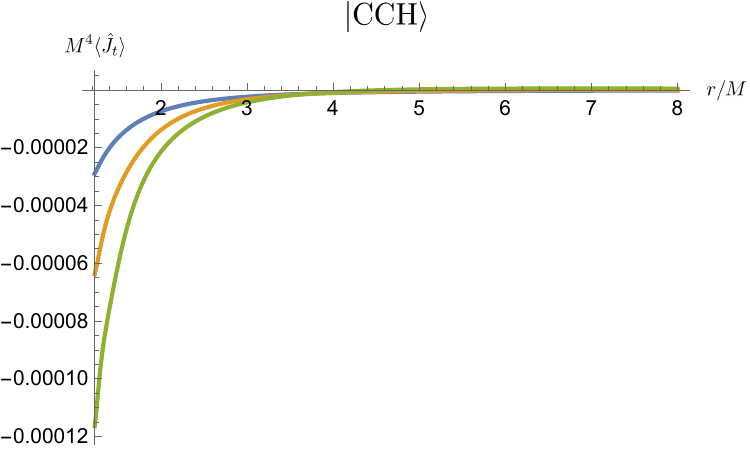}
        \caption{Time component of the renormalized charge current $\langle {\hat {J}}_{t} \rangle $, for a quantum charged scalar field on an RN black hole with $Q=0.99M$ and a selection of values of the scalar field charge $q$. 
        Three quantum states for the scalar field are considered, namely the Boulware $|{\mathrm{B}}\rangle $, Unruh $|{\mathrm{U}}\rangle $ and CCH $|{\mathrm{CCH}}\rangle $ states.}
        \label{fig:Jt4}
\end{figure}

Our results for the renormalized expectation value of the charge current $\langle {\hat {J}}^{\mu }\rangle $ are presented in Figs.~\ref{fig:Jr1}--\ref{fig:Jt4}, for the black hole and charged scalar field parameters given at the start of Sec.~\ref{sec:results}.

We examine first the radial component of the current, $\langle {\hat {J}}^{r}\rangle $, in Figs.~\ref{fig:Jr1}--\ref{fig:Jr4}. 
In each of these plots, we have multiplied the radial component of the current by $r^{2}$.
As expected from (\ref{eq:Jr}), this gives a constant independent of $r$, but dependent on the quantum state under consideration.
We first note that $r^{2}\langle {\hat {J}}^{r}\rangle $  is at least a factor of ten smaller than the vacuum polarization for the same quantum state and black hole and scalar field parameters. 
This was expected, since in the mode sums (\ref{eq:Jrmode}) for this quantity, the modes with positive and negative frequency contribute with different signs, whereas in the vacuum polarization (\ref{eq:SCunrenstates}), the contributions from the positive and negative frequency modes have the same sign.
The charge current is therefore a measure of the {\em {net}} charge in the quantum scalar field.

The constant ${\mathcal {K}}$ in (\ref{eq:Jr}) is the flux of charge from the black hole.
We see that this is nonzero in all three states studied, indicating that none of these quantum states is an equilibrium state.  
We find that ${\mathcal {K}}>0$ for all three states and all values of the black hole and scalar field charges studied. 
This means that, in all three states, the black hole is losing charge (we only consider the situation in which both the black hole charge $q$ and the scalar field charge $Q$ are positive).
This is as expected for the Boulware and Unruh states. 
The former contains an outgoing flux of particles in the superradiant modes, while the latter contains both the superradiant flux and the outgoing Hawking radiation.
For the CCH state, there is an ingoing flux of charge in the ``in'' modes which could potentially increase the charge of the black hole, however overall the black hole is losing charge in this state also.

In Figs.~\ref{fig:Jr1}--\ref{fig:Jr4} we explore how the charge flux ${\mathcal {K}}$ depends on the quantum state, the scalar field charge $q$ and the black hole charge $Q$. 
Fixing the black hole charge to be $Q=0.8M$, we see from Fig.~\ref{fig:Jr1} that the charge flux is greatest in the CCH state, and takes its smallest value in the Boulware state.
The difference between the fluxes in the Unruh and Boulware states decreases as the scalar field charge increases, suggesting that in this limit the superradiant flux is more significant than the Hawking radiation.
The flux in the CCH state is at least a factor of two greater than that in the Boulware or Unruh states. 
Fig.~\ref{fig:Jr2} compares the charge flux for different values of the scalar field charge $q$, again for $Q=0.8$. 
As might be expected, increasing the scalar field charge increases the flux of charge from the black hole in each state. 

A similar picture can be seen on increasing the black hole charge to $Q=0.99M$.
In Fig.~\ref{fig:Jr3}, in order to compare the charge current in the three states, the vertical axis does not extend to the origin. 
As for the lower black hole charge, the charge fluxes in the Unruh and Boulware states have similar magnitudes for fixed scalar field charge, and the flux in the CCH state is significantly larger, although it is now less than double the charge flux in the other two states. 
The main conclusion from Fig.~\ref{fig:Jr4}, is that, as observed previously, increasing the scalar field charge increases the flux of charge from the black hole.
The analytic expressions in the very recent paper \cite{Alberti:2025mpg} for the flux of charge   in the Unruh state when the RN black hole is near-extremal are valid only in the small charge regime $|qQ| \ll 1$, and thus our numerical results do not agree with those analytic expressions.
 
The time component of the expectation value of the current, $\langle {\hat {J}}^{t}\rangle $, corresponds to the charge density in the quantum scalar field, and is shown in Figs.~\ref{fig:Jt1}--\ref{fig:Jt4}. 
We plot $\langle {\hat {J}}_{t}\rangle = -f(r) \langle {\hat {J}}^{t}\rangle $ since $\langle {\hat {J}}_{t}\rangle$ is regular on the event horizon in all three quantum states under consideration.
We note that $\langle {\hat {J}}_{t}\rangle $ is negative for all three quantum states and all values of the black hole and scalar field charges that we consider.
Therefore we find that $\langle {\hat {J}}^{t}\rangle $ is positive. 

In Fig.~\ref{fig:Jt1} we compare the time component of the current $\langle {\hat {J}}_{t}\rangle $ in the three quantum states of interest for a black hole with charge $Q=0.8M$ and a selection of values of the scalar field charge $q$.
In every case $\langle {\hat {J}}_{t}\rangle $ has its maximum magnitude at the event horizon, and, in a neighbourhood of the horizon, it is increasing (decreasing in magnitude) as the radial coordinate increases. 
We see that the quantum scalar field generates a ``cloud'' of charge, with a greater charge density near the horizon of the black hole.
At a fixed value of the radial coordinate $r$, the charge density is largest for the CCH state.
Perhaps surprisingly, it is smallest for the Unruh (rather than Boulware) states, even though there is a greater charge flux in the Unruh state compared with the Boulware state. 
The difference between the Unruh and Boulware states decreases as the scalar field charge $q$ increases. 
For large values of the radial coordinate $r$, for the Unruh and Boulware states, it seems to be the case that $\langle {\hat {J}}_{t}\rangle \rightarrow 0 $ as $r\rightarrow \infty $, and the ``cloud'' of charge in the quantum field is localized in the vicinity of the event horizon. 
For the CCH state, while $\langle {\hat {J}}_{t}\rangle $  is decreasing in magnitude as $r$ increases, there is no indication that it tends to zero as $r\rightarrow \infty $.
This is further evidence that the CCH state is nonempty far from the black hole.
In this state the ``cloud'' of charge in the quantum scalar field extends to infinity. 

Turning now to Fig.~\ref{fig:Jt2}, we see that, for each particular quantum state, the time component $\langle {\hat {J}}_{t}\rangle $ increases in magnitude as the scalar field charge increases. 
For the Unruh state,  $\langle {\hat {J}}_{t}\rangle $ is very small (but positive) when $r/M$ is larger than $\approx 5$, and hence the charge density in the quantum scalar field is negative far from the black hole in these two quantum states.
A similar change in the sign of $\langle {\hat {J}}_{t}\rangle $ far from the event horizon was observed in Ref.~\cite{Klein:2021les} for a charged quantum scalar field on a Reissner-Nordstr\"om-de Sitter black hole, in the Unruh state. 
The sign of $\langle {\hat {J}}_{t}\rangle $ in the Boulware state also changes for sufficiently large $r$ and larger values of $q$ (such as $qM=0.8$ in Fig.~\ref{fig:Jt2}).

The effect of increasing the black hole charge $Q$ to $0.99M$ on the expectation value $\langle {\hat {J}}_{t}\rangle $ can be seen in Fig.~\ref{fig:Jt3}. 
As seen for the vacuum polarization in Fig.~\ref{fig:SC3}, in this case the time component of the current for the Unruh and Boulware states is almost indistinguishable, and when the scalar field charge is also large, $qM=0.99$, the expectation value in the CCH state takes values close to those in the other two states.
The effect of the Hawking radiation is therefore becoming less important than the superradiant emission, as expected as the temperature of the black hole becomes small. 
Fig.~\ref{fig:Jt4} shows that, for this larger value of the black hole charge, increasing the scalar field charge $q$ increases the magnitude of $\langle {\hat {J}}_{t}\rangle $ for each fixed quantum state.
As in Fig.~\ref{fig:Jt2}, the expectation value $\langle {\hat {J}}_{t}\rangle $ is very  small but positive for large $r$, when we consider the states $|{\mathrm{B}}\rangle $ and $|{\mathrm{U}}\rangle $.
For the state $|{\mathrm{CCH}}\rangle $, it remains negative for all $r$ considered. 

The remaining question is whether the expectation value of the current $\langle {\hat {J}}^{\mu }\rangle $ is regular across the event horizon of the black hole.
To address this issue, since the usual Schwarzschild coordinates $(t,r,\theta ,\varphi )$ are not regular at the horizon, we consider the ingoing and outgoing Eddington-Finkelstein coordinates, $(v,r,\theta, \varphi )$ and $(u,r,\theta ,\varphi )$ respectively, where $u$ and $v$ are defined in (\ref{eq:EF}). 
Making the change  to one of these sets of coordinates, the radial component of the charge current $\langle {\hat {J}}^{r}\rangle $ is unchanged, and the time component $\langle {\hat {J}}^{t}\rangle $ is replaced by either $\langle {\hat {J}}^{u}\rangle $ or $\langle {\hat {J}}^{v}\rangle $ as applicable, where
\begin{subequations}
    \begin{align}
        \langle {\hat {J}}^{u}\rangle  & = \langle {\hat {J}}^{t}\rangle - f(r)^{-1}\langle {\hat {J}}^{r}\rangle   = -\frac{1}{f(r)} \left[ \langle {\hat {J}}_{t}\rangle + \langle {\hat {J}}^{r}\rangle  \right] ,
        \\
        \langle {\hat {J}}^{v}\rangle  & = \langle {\hat {J}}^{t}\rangle + f(r)^{-1} \langle {\hat {J}}^{r}\rangle  
        = -\frac{1}{f(r)} \left[ \langle {\hat {J}}_{t}\rangle - \langle {\hat {J}}^{r}\rangle  \right] ,
    \end{align}
\end{subequations}
where $f(r)$ is the metric function (\ref{eq:frRN}).
For all three quantum states studied, and the values of the black hole and scalar field charges that we consider, we find that $\langle {\hat {J}}^{r}\rangle $ is negative and finite (and nonzero) on the event horizon at $r=r_{+}$.
We also find that $\langle {\hat {J}}_{t}\rangle $ is also negative and finite  (and nonzero) on the horizon.
Therefore, it must be the case that $\langle {\hat {J}}^{u}\rangle $ diverges like $f(r)^{-1}$ as $r\rightarrow r_{+}$.
We deduce that all three quantum states are divergent on the past horizon of the black hole.

However, we cannot immediately rule out regularity on the future horizon of the black hole. 
In order for $\langle {\hat {J}}^{v}\rangle $ to be regular on the future horizon, it must be the case that $\langle {\hat {J}}_{t}\rangle =\langle {\hat {J}}^{r}\rangle $ at $r=r_{+}$.
Close examination of our numerical results reveals that, to within the accuracy of our numerical computations, this appears to be the case in the Unruh and CCH states, but not for the Boulware state. 
Therefore the Unruh and CCH states are regular at the future (but not the past) horizon, while the Boulware state diverges on both the future and past horizons.
This is in agreement with the calculation of the renormalized charge current on a Reissner-Nordstr\"om-de Sitter black hole \cite{Klein:2021les}, which is also regular on the future event horizon when the field is in the Unruh state (Ref.~\cite{Klein:2021les} did not consider the Boulware or CCH states).
 
\section{Backreaction}
\label{sec:backreaction}

We now examine the backreaction of the charged scalar current on the electromagnetic field, governed by the semiclassical Maxwell equations (\ref{eq:SCmaxwell}). 
Adopting the approach of \cite{Flanagan:1996gw}, and temporarily reinstating the reduced Planck constant $\hbar $, we consider a perturbative (in $\hbar $) expansion in both the metric and electromagnetic field. 
Since none of the three states we consider in this paper is an equilibrium state,
we assume that the configuration remains spherically symmetric, but the perturbations in the metric and electromagnetic field can depend on time $t$ as well as the radial coordinate $r$.
We consider only the ${\mathcal {O}}(\hbar )$ corrections to the metric and electromagnetic field. 

For the metric, we make the following ansatz to ${\mathcal {O}}(\hbar )$ \cite{Klein:2021ctt}:
\begin{multline}
    ds^{2} = -\left[ f(r) + \delta f (t,r) \right]     dt^{2} 
    \\
    + \left[ f(r) + \delta f (t,r) \right] ^{-1} \, dr^{2} + \left[ r^{2} + 2r\, \delta R(t,r)\right] \, d\Omega ^{2},
    \label{eq:pertmetric}
\end{multline}
where $f(r)$ is the background RN metric function (\ref{eq:frRN}), and $\delta f(t,r)$, $\delta R(t,r)$ are perturbations which we assume are ${\mathcal {O}}(\hbar )$.
Similarly, we take the perturbed electromagnetic potential to have the form
\begin{equation}
    A = \left[ - \frac{Q}{r} + \delta A_{t}(t,r) \right] \, dt ,
\end{equation}
where the first term is the background electrostatic potential (\ref{eqn:gaugeField}) and the second term is a perturbation of order ${\mathcal {O}}(\hbar )$.
The only nonzero components of the perturbed electromagnetic field are then, to ${\mathcal {O}}(\hbar )$, 
\begin{equation}
   \delta F^{tr} (t,r) =-\delta F^{rt}(t,r)  = \partial _{r} \left[ \delta A_{t}(t,r) \right] .
   \label{eq:Ftrpert}
\end{equation}
The electromagnetic field strength $F^{\mu \nu }$ is a tensor, however, with the metric ansatz (\ref{eq:pertmetric}), the perturbation of the field strength \cite{Klein:2021ctt,Klein:2023rwg} is
\begin{equation}
    \delta {\sqrt {-F_{\mu \nu }F^{\mu \nu }}} = {\sqrt{2}} \, \delta F^{tr}(t,r) 
    \label{eq:pertEM}
\end{equation}
and this is a scalar physical quantity.

We now consider the ${\mathcal {O}}(\hbar )$ terms in the semiclassical Maxwell equations (\ref{eq:SCmaxwell}), where the expectation value on the right-hand-side is ${\mathcal {O}}(\hbar )$ and evaluated on the background (nonperturbed) metric.
For all three states we study, the expectation values $\langle {\hat {J}}^{\theta } \rangle $ and $\langle {\hat {J}}^{\varphi} \rangle$ vanish identically.
With our ansatz for the perturbed electromagnetic potential, the $\nu = \theta $ and $\nu = \varphi$ components of (\ref{eq:SCmaxwell}) are then trivially satisfied at order ${\mathcal {O}}(\hbar )$.
This leaves the $\nu = t$ and $\nu = r$ components, which take the following form, where we retain only the ${\mathcal {O}}(\hbar )$ terms:
\begin{subequations}
\label{eq:hbarMaxwell}
\begin{align}
    4\pi \langle {\hat {J}}^{t} \rangle  
    & = \frac{1}{r^{2}}\partial _{r} \left[ r^2 \delta F^{rt}(t,r) \right]
     -\frac{2Q}{r^{2}}\partial _{r}\left[ \frac{1}{r}\delta R (t,r) \right] , 
     \label{eq:Maxwellt}
    \\
    4\pi \langle {\hat {J}}^{r} \rangle 
    & = \partial _{t}\left[ \delta F^{tr} (t,r) \right] 
 +     \frac{2Q}{r^{3}}  \partial _{t} \left[\delta R(t,r) \right] .
 \label{eq:Maxwellr}
 \end{align}
\end{subequations}
We note that (\ref{eq:hbarMaxwell}) contain the metric perturbation $\delta R(t,r)$ as well as $\delta F^{tr}(t,r)=-\delta F^{rt}(t,r)$, and that the expectation values on the left-hand-side depend on the radial coordinate $r$ only.

Since the radial component of the renormalized charged scalar current $\langle {\hat {J}}^{r} \rangle $ is given by (\ref{eq:Jr}) (where ${\mathcal {K}}$ is state-dependent), the second perturbed Maxwell equation (\ref{eq:Maxwellr}) simplifies to:
\begin{subequations}
\label{eq:pertMaxwell}
\begin{equation}
    \partial _{t} \left[  r^{2}\delta F^{tr} (t,r)  + \frac{2Q}{r}\delta R(t,r)  \right]
    = -4\pi {\mathcal {K}},
\end{equation}
while the first perturbed Maxwell equation (\ref{eq:Maxwellt}) can be written as
\begin{equation}
    \partial _{r} \left[  r^{2}\delta F^{tr} (t,r)  + \frac{2Q}{r}\delta R(t,r)  \right]
    = -4\pi r^{2} \langle {\hat {J}}^{t} \rangle .
\end{equation}
\end{subequations}
We can readily integrate (\ref{eq:pertMaxwell}) to give
\begin{subequations}
 \label{eq:MaxSol}   
\begin{equation}
 \delta F^{tr} (t,r)  + \frac{2Q}{r^{3}}\delta R(t,r)   = -\frac{4\pi}{r^{2}} \left[  {\mathcal {F}}(t) 
 + {\mathcal {G}}(r) \right] , 
   \label{eq:MaxSol1}
\end{equation}
where 
\begin{equation}
    {\mathcal {F}}(t) = {\mathcal {K}}t, \qquad 
    {\mathcal {G}}(r) = \int _{r'=r_{1}}^{r} r'^{2} \langle {\hat {J}}^{t} \rangle \, dr' ,
    \label{eq:calFG}
\end{equation}
\end{subequations}
and $r_{1}$ is an arbitrary constant of integration.
We now examine each of the terms on the right-hand-side of (\ref{eq:MaxSol1}) in turn.

The first term on the right-hand-side corresponds to a Coulomb electric field with a time-varying charge $Q(t)$.
The rate of change of charge $\partial _{t}Q(t)\propto -{\mathcal {K}}<0$ for all three quantum states considered in this work.
Therefore, since we take $Q(t)>0$, the black hole is always losing electric charge.
For fixed values of the background black hole and scalar field charges, the rate of discharge is greatest in the CCH state.
The rate of discharge also increases as either the scalar field charge $q$ or the initial black hole charge $Q$ increase. 

The second term on the right-hand-side of (\ref{eq:MaxSol1}) has, to the best of our knowledge, not been considered previously in the literature.
Since $\langle {\hat {J}}^{t}\rangle = -f(r)^{-1}\langle {\hat {J}}_{t}\rangle$ does not have a closed-form dependence on the radial coordinate $r$, we study the quantity ${\mathcal {G}}(r)$ (\ref{eq:calFG}) numerically.
To do this, we need to perform a definite integral, fixing the lower limit of integration to be a constant $r_{1}$.
Changing $r_{1}$ simply corresponds to a constant shift in the time-dependent function ${\mathcal {F}}(t)$.

The most natural choice would be to set $r_{1}=r_{+}$, the event horizon radius.
However, we see from Figs.~\ref{fig:Jt1}--\ref{fig:Jt4} that $\langle {\hat {J}}_{t}\rangle $ takes a finite nonzero value on the horizon for all three states considered in this paper, and hence $\langle {\hat {J}}^{t}\rangle = -f(r)^{-1}\langle {\hat {J}}_{t}\rangle$ diverges like $f(r)^{-1}\sim (r-r_{+})^{-1}$ as $r\rightarrow r_{+}$ and the event horizon is approached. 
Therefore, on integrating with respect to $r$, there is a logarithmic singularity in  $\int r^{2}\langle {\hat {J}}^{t} \rangle \, dr$ as $r\rightarrow r_{+}$.
Since the expectation value of $\langle {\hat {J}}_{t}\rangle$ on the horizon is very small, typically $\sim 10^{-5}/M^4$, the coefficient of the logarithmic singularity is also very small.
Nonetheless, at first glance this seems to be an unsatisfactory result, since we are working in a semiclassical framework in which quantum corrections to the metric and electromagnetic field are assumed to be small.
However, on the left-hand-side of (\ref{eq:MaxSol1}), we have a combination of electromagnetic field and metric perturbations, namely $\delta F^{tr}(t,r)$ and $\delta R(t,r)$.
From (\ref{eq:pertEM}), the electromagnetic field strength perturbation $\delta F^{tr}(t,r)$ is a physical quantity which we would expect to be regular on the event horizon when we consider the backreaction of a quantum state which is itself regular on the event horizon.
However, the metric function $\delta R(t,r)$ is not directly measureable, and there is no reason {\emph {a priori}} to rule out a logarithmic singularity in this quantity as the event horizon is approached.
Furthermore, the location of the event horizon is likely to be shifted by the backreaction of the quantum field on the space-time geometry. 
To explore whether the metric perturbation $\delta R(t,r)$ can indeed have a logarithmic singularity, and find the shifted location of the event horizon, it would be necessary to consider the semiclassical Einstein equations giving the backreaction of the quantum field on the space-time geometry.
The semiclassical Einstein equations involve the renormalized expectation value of the stress-energy tensor operator, the computation of which is beyond the scope of this present work. 
 
\begin{figure*}
     \centering
         \includegraphics[width=0.47\textwidth]{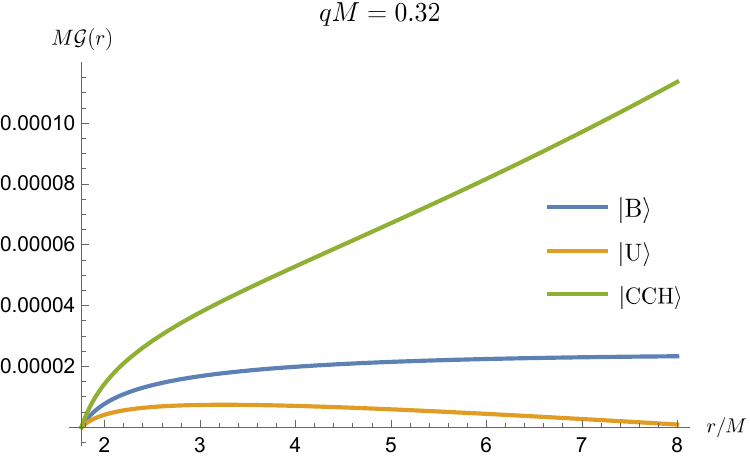}
         \includegraphics[width=0.47\textwidth]{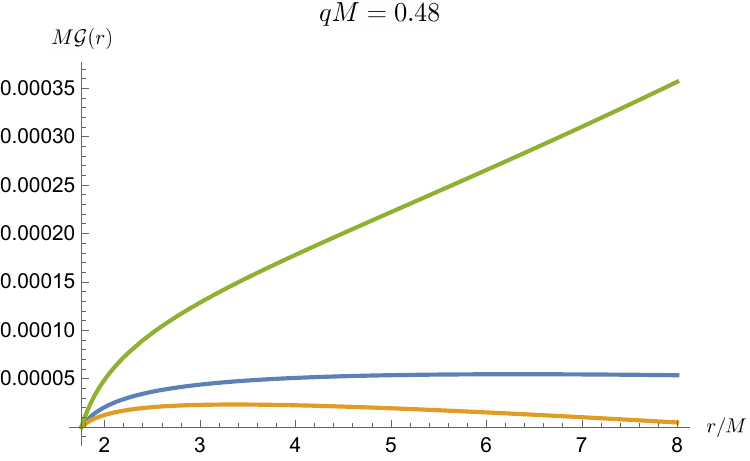}
         \includegraphics[width=0.47\textwidth]{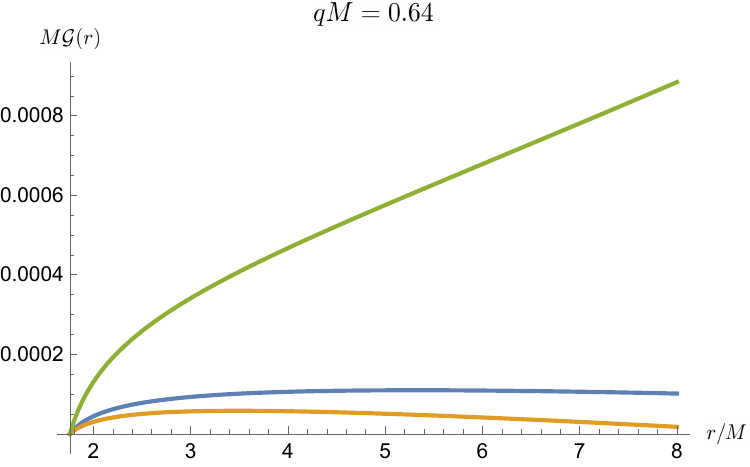}
         \includegraphics[width=0.47\textwidth]{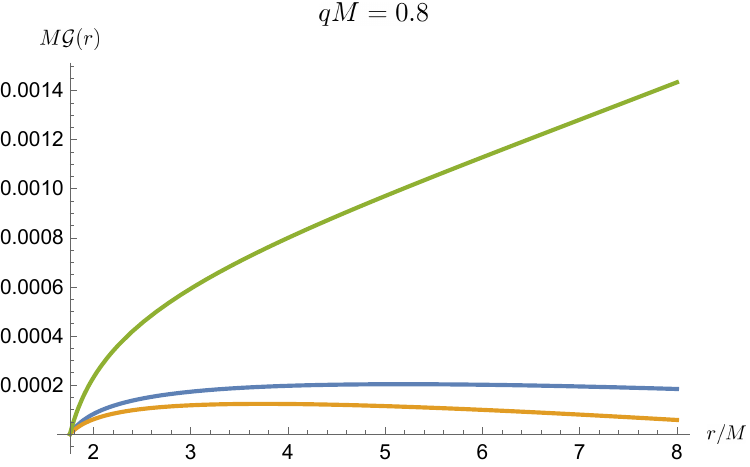}
        \caption{${\mathcal {G}}(r)$ (\ref{eq:calFG}) for a quantum charged scalar field on an RN black hole with $Q=0.8M$ and a selection of values of the scalar field charge $q$. 
        Three quantum states for the scalar field are considered, namely the Boulware $|{\mathrm{B}}\rangle $, Unruh $|{\mathrm{U}}\rangle $ and CCH $|{\mathrm{CCH}}\rangle $ states.
        The lower limit of integration is taken to be $r_{1}=11r_{+}/10$, where $r_{+}=8M/5$ is the event horizon radius.}
        \label{fig:int1}
\end{figure*}

\begin{figure}
     \centering
         \includegraphics[width=0.47\textwidth]{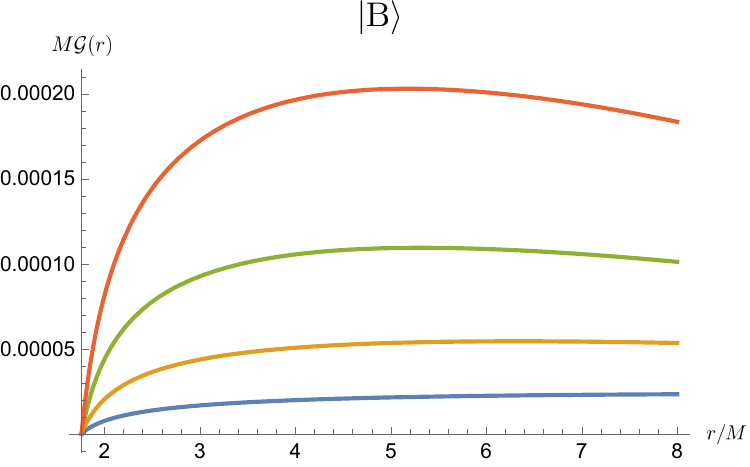}
         \includegraphics[width=0.47\textwidth]{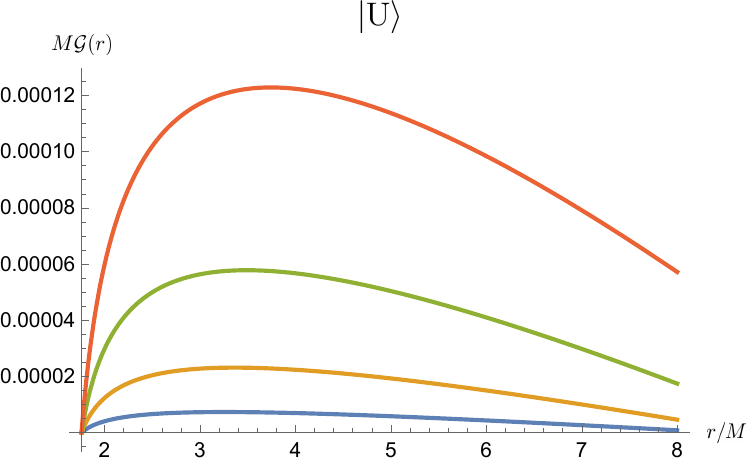}
         \includegraphics[width=0.47\textwidth]{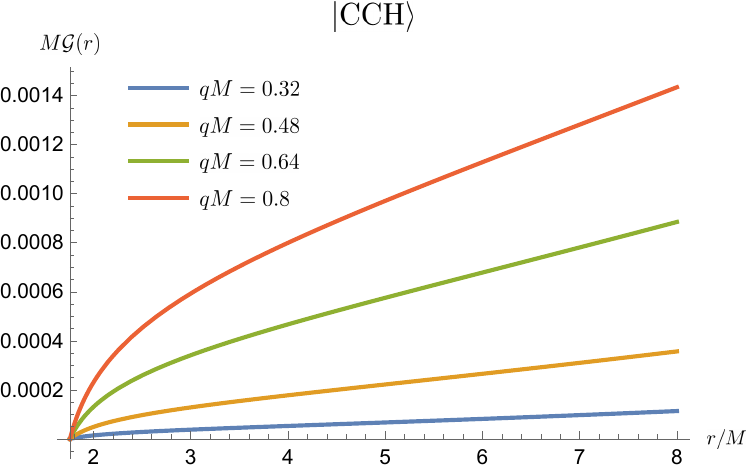}
        \caption{${\mathcal {G}}(r)$ (\ref{eq:calFG}) for a quantum charged scalar field on an RN black hole with $Q=0.8M$ and a selection of values of the scalar field charge $q$. 
        Three quantum states for the scalar field are considered, namely the Boulware $|{\mathrm{B}}\rangle $, Unruh $|{\mathrm{U}}\rangle $ and CCH $|{\mathrm{CCH}}\rangle $ states.
        The lower limit of integration is taken to be $r_{1}=11r_{+}/10$, where $r_{+}=8M/5$ is the event horizon radius.}
        \label{fig:int2}
\end{figure}

\begin{figure}
     \centering
         \includegraphics[width=0.47\textwidth]{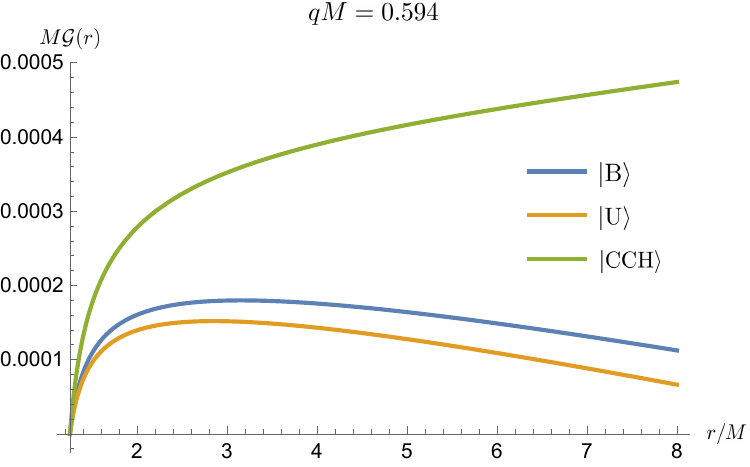}
         \includegraphics[width=0.47\textwidth]{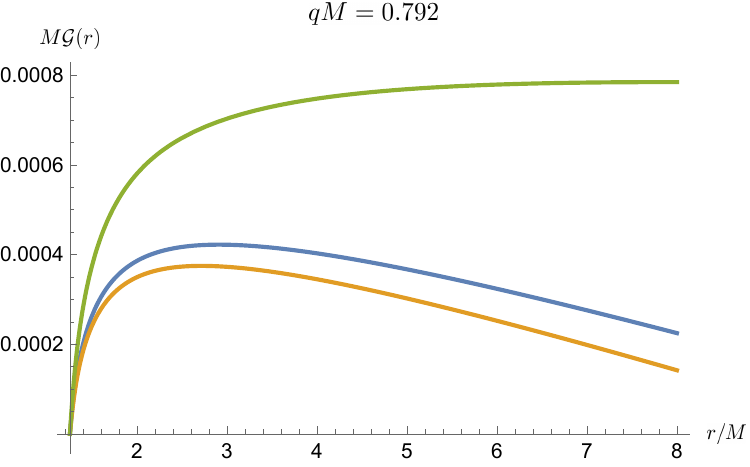}
         \includegraphics[width=0.47\textwidth]{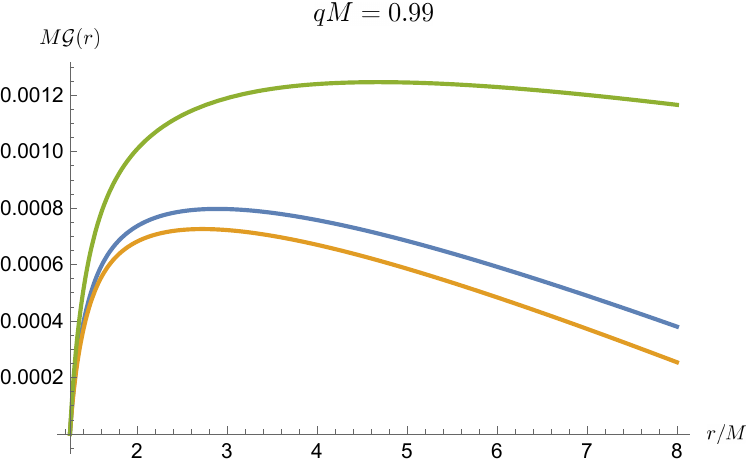}
        \caption{${\mathcal {G}}(r)$ (\ref{eq:calFG}) for a quantum charged scalar field on an RN black hole with $Q=0.99M$ and a selection of values of the scalar field charge $q$. 
        Three quantum states for the scalar field are considered, namely the Boulware $|{\mathrm{B}}\rangle $, Unruh $|{\mathrm{U}}\rangle $ and CCH $|{\mathrm{CCH}}\rangle $ states.
        The lower limit of integration is taken to be $r_{1}=11r_{+}/10$, where $r_{+}\approx 1.141M$ is the event horizon radius.}
        \label{fig:int3}
\end{figure}

\begin{figure}
     \centering
         \includegraphics[width=0.47\textwidth]{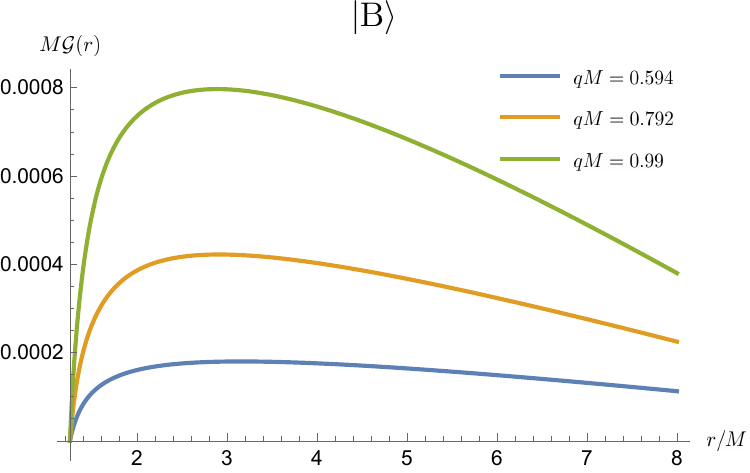}
         \includegraphics[width=0.47\textwidth]{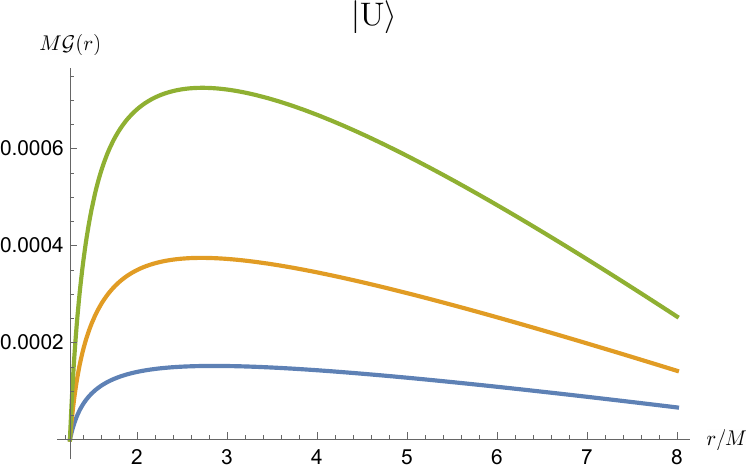}
         \includegraphics[width=0.47\textwidth]{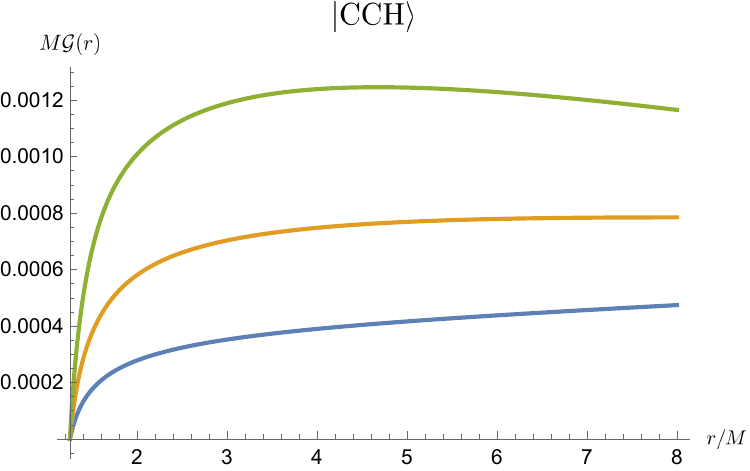}
        \caption{${\mathcal {G}}(r)$ (\ref{eq:calFG}) for a quantum charged scalar field on an RN black hole with $Q=0.99M$ and a selection of values of the scalar field charge $q$. 
        Three quantum states for the scalar field are considered, namely the Boulware $|{\mathrm{B}}\rangle $, Unruh $|{\mathrm{U}}\rangle $ and CCH $|{\mathrm{CCH}}\rangle $ states.
        The lower limit of integration is taken to be $r_{1}=11r_{+}/10$, where $r_{+}\approx 1.141M$ is the event horizon radius.}
        \label{fig:int4}
\end{figure}

To explore the properties of ${\mathcal {G}}(r)$ (\ref{eq:calFG}), we therefore need to make a choice of $r_{1}$.
In Figs.~\ref{fig:int1}--\ref{fig:int4} we show this integral as a function of $r$, for the renormalized scalar current displayed in Figs.~\ref{fig:Jt1}--\ref{fig:Jt4} respectively, setting
$r_{1}=11r_{+}/10$. 
Changing the value of $r_{1}$ corresponds to adding a constant to ${\mathcal {G}}(r)$.

In Figs.~\ref{fig:int1} and \ref{fig:int2} we show ${\mathcal {G}}(r)$ as a function of $r$ for a black hole with charge $Q=0.8M$ and a selection of values of the scalar field charge $q$. 
Fig.~\ref{fig:int1} compares the three states of interest at fixed scalar field charge, while Fig.~\ref{fig:int2} examines the effect of changing the scalar field charge on ${\mathcal {G}}(r)$.
Looking first at Fig.~\ref{fig:int1}, away from the event horizon, it is striking that, at fixed radius, ${\mathcal {G}}(r)$ is larger in the Boulware state than it is in the Unruh state. 
The difference between these two states narrows as the scalar field charge increases.
This follows from our results in Fig.~\ref{fig:Jt1}, as the charge density is greater in the Boulware than in the Unruh states. 
In all three states, ${\mathcal {G}}(r)$ increases as the scalar field charge increases. 

In Fig.~\ref{fig:int2}, we see that, for the Unruh state, ${\mathcal {G}}(r)$ is decreasing as the radial coordinate $r$ increases far from the black hole (this is also the case for the Boulware state for larger values of the scalar field charge $q$), due to the change in sign of the charge density in these states far from the black hole (see comments on Fig.~\ref{fig:Jt2}). 
Since ${\mathcal {G}}(r)$ is multiplied by $r^{-2}$ in the perturbed electromagnetic field strength (\ref{eq:MaxSol1}), far from the black hole, this term will be subleading compared to the Coulomb field arising from the function ${\mathcal {F}}(t)$ in (\ref{eq:MaxSol1}).  

The properties of the function ${\mathcal {G}}(r)$ for the CCH state, at least for the values of the radial coordinate $r$ studied, are very different far from the black hole.
In particular, from Figs.~\ref{fig:int1} and \ref{fig:int2} we see that ${\mathcal {G}}(r)$ is increasing approximately linearly as the radial coordinate $r$ increases. 
From Fig.~\ref{fig:int2}, the rate of increase also increases as the scalar field charge increases.
This implies that the term ${\mathcal {G}}(r)$ will dominate over the Coulomb term ${\mathcal {F}}(t)$ in (\ref{eq:MaxSol1}) far from the black hole.
Therefore, in the CCH state, at large $r$ the dominant contribution to the perturbed electromagnetic field strength $\delta F^{tr}(t,r)$ is coming from the charged scalar ``cloud'' rather than the flux of charge.

The perturbation of the total charge in the space-time is given by an integral over the sphere at infinity ${\mathbb{S}^{2}_{\infty }}$:
\begin{equation}
    \delta Q =  \frac{1}{4\pi } \int _{\mathbb{S}^{2}_{\infty }} \left[  r^{2} \delta F^{tr} + \frac{2Q}{r}\delta R (t,r) \right] \sin \theta \, d\theta \, d\varphi  .
    \label{eq:chargepert}
\end{equation}
Substituting in from (\ref{eq:MaxSol1}) gives
\begin{equation}
    \delta Q =-  4 \pi \left\{ {\mathcal {F}}(t) + \lim _{r\rightarrow \infty }\left[  {\mathcal {G}}(r) \right] \right\}.
    \label{eq:chargepert1}
\end{equation}
The solution of the perturbed Maxwell equations (\ref{eq:MaxSol}) gives this charge perturbation only up to an arbitrary constant (corresponding to the choice of the lower limit $r_{1}$ in ${\mathcal {G}}(r)$). 
One natural choice of $r_{1}$ is, as discussed above, the event horizon radius $r_{+}$. 
Another natural choice is to set $r_{1}=\infty $, in which case the second term in (\ref{eq:chargepert1}) vanishes,
and the charge perturbation $\delta Q$ is entirely due to the Coulomb term ${\mathcal {F}}(t)$.

For the Unruh state (and also for the Boulware state for some values of the scalar field charge $q$ depicted in Figs.~\ref{fig:int1} and \ref{fig:int2}), the data available suggest that  $\lim _{r\rightarrow \infty }\left[  {\mathcal {G}}(r) \right] = 0$. 
For those values of the scalar charge $q$ for which ${\mathcal {G}}(r)$ has a finite (but nonzero) limit as $r\rightarrow \infty $, we can set this limit to be zero by an alternative choice of the integration constant. 
In both these cases the total net charge in the scalar ``cloud'' vanishes.
However, for the CCH state, if the trend shown in Figs.~\ref{fig:int1}--\ref{fig:int2} continues for larger values of the radial coordinate $r$, then this suggests that $\lim _{r\rightarrow \infty }\left[  {\mathcal {G}}(r) \right] \rightarrow \infty $
and the total scalar ``cloud'' charge is infinite, due to the infinite volume of space-time. 

The effect of increasing the black hole charge $Q$ on the scalar ``cloud'' can be seen in Figs.~\ref{fig:int3} and \ref{fig:int4}. 
As in Fig.~\ref{fig:int1}, in Fig.~\ref{fig:int3} we see that ${\mathcal {G}}(r)$ is greater for the Boulware state than it is for the Unruh state, with the difference between these two states being comparatively small. 
For both these states, ${\mathcal {G}}(r)$ is decreasing as $r$ increases far from the black hole.
The behaviour of ${\mathcal {G}}(r)$ far from the black hole in the CCH state is however qualitatively different from that seen for the lower value of the black hole charge in Fig.~\ref{fig:int2}.
In particular, the rate of increase of ${\mathcal {G}}(r)$ as a function is $r$ is much smaller in Fig.~\ref{fig:int4} than in Fig.~\ref{fig:int2}. 
For very large scalar field charge $qM=0.99$, we see that ${\mathcal {G}}(r)$ is in fact decreasing as $r$ increases far from the black hole (it may be that we would find similar behaviour for other values of $Q$ and $q$ if we extended our computations to larger values of $r$).
When $Q=0.99M$, the black hole is near-extremal and, in this limit, the Hawking radiation (which vanishes in the extremal limit) becomes subleading compared to the superradiant emission.
This means that the CCH state approaches the Boulware and Unruh states, and the scalar ``cloud'' becomes dominated by the superradiant flux.

\section{Conclusions}
\label{sec:conc}

In this work we have extended the ``pragmatic mode-sum'' methodology of \cite{Levi:2015eea} to compute the renormalized expectation values of the charge current $\langle {\hat {J}}^{\mu } \rangle $ for a massless quantum charged scalar field on a charged RN black hole background.
Since the field is massless, it is subject to classical charge superradiance. 
We have considered three possible states for the scalar field (all of which are ``past'' states in the terminology of \cite{Balakumar:2022yvx}), namely the Boulware, Unruh and CCH states. 
The construction of all three states in the presence of charge superradiance is uncontroversial. 

To calculate $\langle {\hat {J}}^{\mu } \rangle $, we employ the $t$-splitting variant of the ``pragmatic mode-sum'' approach of Ref.~\cite{Levi:2015eea}.
One difference from the work of \cite{Levi:2015eea} is that we use the Hadamard parametrix in our renormalization prescription rather than DeWitt-Schwinger renormalization. 
This choice does not affect the singularity structure of the Green function or the physical interpretation of our results.
As in a previous computation of the renormalized charge current on a Reissner-Nordstr\"om-de Sitter black hole \cite{Klein:2021les}, we work on the Lorentzian black hole geometry.
In \cite{Klein:2021les}, a different choice of gauge for the background electrostatic potential is made at each space-time point, which has the advantage that only finite renormalization terms are required to find the expectation value of the current, but at the expense of having to compute a separate set of field modes for each space-time point considered.
Here we fix the electrostatic potential gauge from the outset, so that only one set of field modes is needed for each value of the scalar and black hole charges that we study.
The disadvantage of our choice of gauge is that the time component of the charge current  requires renormalization. 

Finding the renormalized vacuum polarization is a straightforward extension of the method of \cite{Levi:2015eea} in the neutral scalar field case; the only difference is a small modification in the finite renormalization terms due to the background electrostatic potential.
The renormalized current is the sum of two terms; one is proportional to the gauge potential multiplied by the vacuum polarization, while the second involves a derivative of the scalar field operator. 
The radial component of the charged scalar current does not require renormalization \cite{Balakumar:2022yvx}, and therefore we only need to renormalize the quantity in the charge current which involves a time derivative of the quantum scalar field.
The key new aspect of our methodology is the application of the ``pragmatic mode-sum'' prescription to this quantity.  

We find that none of the three states considered is an equilibrium state; all three have a nonzero flux of charge from the black hole, given by the radial component of the charge current.
For the Boulware state, this flux corresponds to quantum emission in the classically superradiant modes.
For the Unruh state, in addition to the superradiant flux, there is also the thermal Hawking radiation.
The CCH state has an ingoing flux of charged particles as well as the outgoing Hawking radiation and superradiant flux, but overall there is a net outwards flux.
In all three states, the quantum scalar field is discharging the black hole.  

Since we have a massless scalar field which is subject to classical superradiance, this discharge process continues in the extremal limit, when the temperature of the black hole tends to zero and the Hawking radiation is absent. 
In this limit the three states we study coincide.  
Even though we have considered a black hole having a charge/mass ratio of $0.99$, to probe the extremal limit would require studying black holes whose charge/mass ratios are closer to unity, which is very challenging numerically. 
Furthermore, it has recently been argued \cite{Brown:2024ajk,Mohan:2024rtn} that the semiclassical approach we employ in this work breaks down in the extremal limit, and nonperturbative effects become dominant.  

The time component of the charge current gives the charge density in the quantum scalar field, resulting in the formation of a quantum scalar ``cloud'' of charge surrounding the black hole.
By solving the semiclassical Maxwell equations to leading order in $\hbar $, we have explored the backreaction of this scalar cloud on the electromagnetic field. 
For the Unruh and Boulware states, the charge density tends to zero far from the black hole, so that the total charge contained in the scalar ``cloud'' is finite, and the dominant contribution to the electromagnetic field far from the black hole appears to be the change in the Coulomb term due to the discharge of the black hole. 
In contrast, the total scalar ``cloud'' charge in the CCH state appears to be potentially unbounded (at least for some of the scalar field and black hole charges considered) since this state is nonempty at infinity and the space-time has an infinite volume.

Solving the semiclassical Maxwell equations gives a combination of the perturbed electromagnetic field and a metric perturbation. 
For the quantum states considered in this paper, we find that this combination has a logarithmic divergence at the event horizon, suggesting that the semiclassical approximation employed here (that is, assuming that quantum corrections are small and working with linearized equations) may break down near the horizon.
This logarithmic divergence is a generic feature, and will occur for all nonequilibrium states of a charged scalar field on a charged black hole background, even if the scalar field is sufficiently massive that charge superradiance is absent.
However, the logarithmic divergence is not present for equilibrium states, such as the Hartle-Hawking state considered in \cite{Breen:2024ggu}, which is a thermal equilibrium state that can be constructed for a massive scalar field when there is no charge superradiance.

The metric perturbation which arises in the semiclassical Maxwell equations is not a directly observable quantity, and it may be that all physical quantities (such as the electromagnetic field strength) are regular everywhere outside the black hole.
To address this question, it is necessary to solve the semiclassical Einstein equations for the perturbed metric and electromagnetic field.
Similarly, in this paper we have considered only the rate of change of charge of the black hole, whereas an adiabatic approach provides evidence that the evolution of the charge/mass ratio of the black hole during its evaporation exhibits complicated behaviour \cite{Hiscock:1990ex,Ong:2019vnv}.
Finding the rate of change of the black hole mass also requires solving the semiclassical Einstein equations. 
However, the semiclassical Einstein equations have, as their source, the renormalized expectation value of the stress-energy tensor operator, whose computation is beyond the scope of the present work.
We therefore postpone further consideration of the backreaction problem to future research.

\bigskip

\begin{acknowledgments}
We thank Cormac Breen and Peter Taylor for helpful discussions.
We acknowledge IT Services at The University of Sheffield for the provision of services for High Performance Computing.
The work of E.W.~is supported by STFC grant number ST/X000621/1.
\end{acknowledgments}

\section*{Data availability}
The data that support the findings of this article are openly available \cite{Figshare}.

\vfill

\bibliography{pmr}

\end{document}